\documentclass[preprint]{elsarticle}

\usepackage{booktabs} 
\usepackage{graphicx}
\usepackage{epsfig}
\usepackage{amsmath}
\usepackage{amssymb}
\usepackage{comment}
\usepackage{multirow}
\usepackage{wrapfig}
\usepackage{caption}
\usepackage{mathrsfs}
\usepackage{enumerate}
\usepackage{array}
\usepackage{color}
\usepackage{url}
\usepackage{cite}
\usepackage{amsthm}

\usepackage{algorithmic}
\usepackage[noend,linesnumbered,algoruled]{algorithm2e}
\usepackage{epstopdf}
\usepackage{hyperref}
\hypersetup{
  colorlinks   = true, 
  urlcolor     = blue, 
  linkcolor    = blue, 
  citecolor   = blue 
}
\usepackage{multicol}

\newcommand{\hproof}[1]{\noindent {\em Proof:} \
#1 \boxtheorem\\ \vspace{2mm}}

\newcommand{\vectt}[1]{\bar{#1}}

\newcommand{\mc}[1]{\mathcal{ #1}}
\newcommand{\mf}[1]{\mathfrak{ #1}}

\newcommand{\nit}[1]{{\it #1}}

\newcommand{\boxtheorem}{\hfill $\blacksquare$\vspace{2mm}}
\newcommand{\ignore}[1]{}
\newcommand{\qchase}[2]{$\langle #1,#2\rangle$-chase}
\newcommand{\qmchase}[3]{\langle #1,#2\rangle\mbox{-}\nit{chase}(#3)}

\newcommand{\red}[1]{\textcolor{red}{#1}}
\newcommand{\blue}[1]{\textcolor{black}{#1}}

\newcommand{\comlb}[1]{{\vspace{2mm}\noindent \bf \red{COMM(LEO):}}~ #1 \hfill {\bf
    END.}\\}
\newcommand{\commos}[1]{{\vspace{2mm}\noindent \bf \blue{COMM(MOSTAFA):}}~ #1 \hfill {\bf
    END.}\\}
\newcommand{\da}{Datalog~}

\newcommand{\dpm}{{Datalog}$^\pm$}
\newcommand{\dl}{DL}
\newcommand{\edb}{EDB}
\newcommand{\de}{{Datalog}$^\exists$}
\newcommand{\dplus}{{Datalog}$^+$}
\newcommand{\m}{\hspace{0.5mm}}
\newcommand{\ms}{{\sf MagicD}$^+$}
\newcommand{\nc}{{\em nc}}
\newcommand{\ncs}{{\em ncs}}
\newcommand{\tgd}{{\em tgd}}
\newcommand{\tgds}{{\em tgds}}
\newcommand{\edg}{EDG}
\newcommand{\egd}{{\em egd}}
\newcommand{\egds}{{\em egds}}
\newcommand{\SCh}{\nit{SCh}}
\newcommand{\sch}{\nit{sch}}
\newcommand{\ssch}{\nit{syn-sch}}
\newcommand{\wsch}{\nit{WSCh}}
\newcommand{\jwsch}{\nit{JWSCh}}
\newcommand{\stk}{\nit{Sticky}}
\newcommand{\ws}{\nit{WS}}
\newcommand{\wa}{\nit{WA}}
\newcommand{\jws}{\nit{JWS}}
\newcommand{\ja}{\nit{JA}}

\newcommand{\GSCh}{\nit{GSCh}}
\newcommand{\schqa}{${\sf SChQA}^\mc{S}$}

\newcommand{\sips}{sips}
\newcommand{\qa}{QA}
\newcommand{\cq}{CQ}
\newcommand{\cqs}{CQs}
\newcommand{\bcq}{BCQ}
\newcommand{\constants}{\ensuremath{\Gamma^C}}
\newcommand{\nulls}{\ensuremath{\Gamma^N}}

\newcommand{\schema}{\ensuremath{\mc{R}}}
\newcommand{\fp}[1]{\nit{FinPos}(#1)}

\newcommand{\myrightarrow}[1]{\rightarrow_{#1}}
\newcommand{\myRightarrow}[1]{\ll_{#1}}
\newcommand{\select}{\mc{S}}
\newcommand{\rank}{\Pi}
\newcommand{\finiteRank}{\rank_F}
\newcommand{\infiniteRank}{\rank_\infty}
\newcommand{\finiteExists}{\rank^\exists_F}

\newcommand{\fo}{FO}

\newcommand{\exptime}{{\sc exptime}}
\newcommand{\ptime}{{\sc ptime}}
\newcommand{\acz}{{\sc ac}$_0$}
\makeatletter
\newcommand{\setword}[2]{%
  \phantomsection
  #1\def\@currentlabel{\unexpanded{#1}}\label{#2}%
}
\theoremstyle{definition}
\newtheorem{example}{Example}[section]
\newtheorem{theorem}{Theorem}[section]
\newtheorem{definition}{Definition}
\newtheorem{proposition}{Proposition}
\newtheorem{corollary}{Corollary}
\newtheorem{lemma}{Lemma}

\journal{ }

\begin{document}

\begin{frontmatter}
\title{Extending Sticky-\dpm \ via Finite-Position  Selection Functions:  \ Tractability, Algorithms, and Optimization}

\author[add1]{Leopoldo Bertossi}
\ead{leopoldo.bertossi@uai.cl}
\author[add2]{Mostafa Milani}
\ead{mostafa.milani@uwo.ca}

\address[add1]{Universidad Adolfo Ib\'a\~nez,  and  IMFD, Santiago, Chile}
\address[add2]{Department of Computer Science, The University of Western Ontario, London, Canada}

\begin{abstract} {\em Weakly-Sticky} (\ws\m) \dpm \ is an expressive member of the family of \dpm \ program classes that is defined  on the basis of the conditions of {\em stickiness} and {\em weak-acyclicity}. Conjunctive query answering (QA) over the \ws \ programs has been investigated, and its tractability in data complexity has been established. However, the design and implementation of practical QA algorithms and their optimizations have been open. In order to fill this gap, we first study \stk \ and \ws \ programs from the point of view of the behavior of the chase procedure. We  extend the stickiness property of the chase to that of {\em generalized stickiness of the chase} (\GSCh) modulo an oracle that {\em selects} (and provides) the predicate positions where finitely values appear during the chase. Stickiness modulo a {\em selection function} $\select$ that provides only a subset of those positions defines \sch$(\select)$, a semantic subclass of \GSCh. Program classes with selection functions include \stk \ and \ws, and another syntactic class that we introduce and characterize, namely \jws, of  {\em jointly-weakly-sticky} programs, which contains \ws. The selection functions for these last three classes are computable, and no external, possibly non-computable oracle is needed. We propose a  bottom-up QA algorithm for programs in the class \sch($\select$), for a general selection $\select$. \ As a particular case, we obtain a polynomial-time QA algorithm for \jws \ and weakly-sticky programs.  Unlike \ws, \jws \ turns out to be closed under  magic-sets  query optimization. As a consequence,  both  the generic polynomial-time QA algorithm and its magic-set optimization can be particularized and applied to \ws.
\end{abstract}
\end{frontmatter}

\section{Introduction}\label{sec:intr}
Ontology-based data access (OBDA) \citep{poggi} allows to access data, usually stored in a relational database, through a conceptual layer that takes the form of an ontology. Queries can be expressed in terms of the ontology language, but are answered by eventually requesting data from the extensional data source underneath. Common languages of choice for representing ontologies are certain syntactic classes  of {\em description logic} (\dl) \citep{artale} and, more recently, of {\em \dpm} \citep{cali09,cali12,cali12jws}. Those classes  are expected to be both sufficiently expressive and computationally well-behaved in relation to query answering (QA) for conjunctive queries (CQs). In this work we use \dpm.

\dpm \ extends the  \da relational query language~\citep{ceri} by allowing: (a)  Existentially quantified variables ($\exists$-variables) in rule heads, and so extending classical Datalog rules. These new and old rules represent {\em tuple-generating dependencies} (tgds) \citep{abiteboul}. (b) Constraints in the form of rules with  equality atoms or an always  false propositional atom $\bot$. The former represent ``equality-generating dependencies" (egds) and the latter,  ``negative constraints" \citep{cali12}. The ``$+$" in \dpm\ stands for those extensions, while the ``$-$" reflects syntactic restrictions on programs for better computational properties.

\dpm \ is expressive enough to represent in logical and declarative terms useful ontologies, in particular those that capture and extend the common
conceptual data models \citep{cali12is} and Semantic Web data \citep{arenas}.
The rules of a \dpm \ program can be seen as forming an ontology  on top of an extensional database (EDB), $D$, which may be {\em incomplete}. In particular, the ontology: (a)   provides a ``query layer" for $D$,
enabling  OBDA, and (b) specifies the completion of $D$ through program rules that can be enforced to generate new data.  Several approaches and techniques have been proposed for QA under \dl\  \citep{artale,poggi} and  \dpm\  \citep{cali09} ontologies.
\ignore{ Since the conceptual layer provided by the ontology can unify possibly heterogeneous data sources, through a shared meaning of the terms of a  vocabulary in common,
OBDA has been applied in several areas, such as virtual data integration \citep{lenzerini02} and the semantic web with RDF data \citep{arenas}.} 

{\em In the rest of this work we assume that \dpm \ programs contain only  tgds, plus extensional data, but no constraints.\footnote{The conditions and results on the integration of tgds and constraints found in \citep{cali12} also apply to our work. More details can be found in Section \ref{sec:dpm}.} When programs are subject to syntactic restrictions, we talk about \dpm \ programs, whereas when no
conditions are assumed or applied, we sometimes talk about \dplus \ programs, also called \de \ programs}~\citep{baget09, cali09, leone, krotzsch}. \ {\em Queries are always conjunctive; and whenever otherwise stated, every complexity claim refers to data complexity, that is, time complexity in terms of the size of the EDB $D$} \citep{abiteboul}.

From the semantic and computational point of view, the completion of the EDB $D$ is achieved through the so-called {\em chase procedure} (usually simply called ``the chase") that, starting  from the data in $D$, iteratively enforces the rules in the ontology.
That is, when a rule body (the antecedent) becomes true in the instance constructed so far, but not the head (the consequent), a new tuple is generated to make the rule true (as an implication).  This process may propagate existing values to the same or other {\em positions} (or arguments) in predicates; or create new values (nulls) corresponding to existentially quantified variables in rule heads. The following example informally illustrates this process and the notions involved (cf. Section \ref{sec:preliminaries} for details).

\begin{example} \label{ex:intr} Consider a \dplus \ program consisting of the set $\mc{P}$ of rules below and the EDB $D=\{R(a,b)\}$:

\vspace{-7mm}
\begin{align}
R(x,y) \! ~&\rightarrow~ \! \exists z \ R(y,z).\label{eq:s1}\\
R(x,y),R(y,z) \! ~&\rightarrow~ \! S(x,y,z). \hspace{1cm}\label{eq:s2}
\end{align}
\vspace{-4mm}

The program's schema has a binary (i.e. two-argument) predicate, $R$, with positions $R[1],R[2]$, and a ternary predicate $S$, with positions $S[1],S[2],S[3]$. The {\em join variable} $y$ in rule (\ref{eq:s2}), i.e. repeated in its body, appears in positions $R[2]$ and $R[1]$.
\ The initial instance $D$ makes the antecedent of  rule (\ref{eq:s1}) true, but not its head. So, a new tuple, $R(b,\zeta_1)$, is generated by the chase. Now the body of rule (\ref{eq:s2}) becomes true, and its head has to be made true, generating a tuple $S(a,b,\zeta_1)$. Continuing in this way, the extension of $D$ produced by the chase includes the following tuples (among infinitely many  others due to further rule enforcements): $R(b,\zeta_1),$ $S(a,b,\zeta_1),R(\zeta_1,\zeta_2),S(b,\zeta_2,\zeta_1)$. Notice that $S(a,b,\zeta_1)$ and $S(b,\zeta_1,\zeta_2)$ are obtained by replacing the join variable $y$  by $b$ and $\zeta_1$, resp. \boxtheorem\end{example}

The result of the chase, as a possibly infinite instance for the program's schema, is also called ``the chase", and extends the instance $D$ that contains the extensional data.
This chase
gives the semantics to the \dpm \ ontology, by providing an intended model, and can be used, at least in principle and conceptually, for QA in the sense that a query could be posed directly to the materialized chase instance. However, this may not be the best way to
go about QA, and computationally better alternatives have to be explored.

 Actually, when the chase may be infinite, (conjunctive) QA may be undecidable \citep{johnson}. However, for some classes of programs that may produce an  infinite chase, QA is still computable (decidable), and even tractable in the size of $D$. In fact,  syntactically restricted subclasses of \dplus \ programs have been identified and characterized for which
QA is decidable, among them: {\em linear}, {\em guarded} and {\em weakly-guarded}, {\em sticky} and {\em weakly-sticky} (\ws) \dpm\ \citep{cali09,cali12} (cf. also Section \ref{sec:programs}).

Sticky \dpm \ is a  class of programs characterized by  syntactic restrictions on join variables. \ws \ \dpm \ extends \stk \ \dpm \ by also capturing the well-known class of {\em weakly-acyclic programs}, which is defined through the syntactic notions of {\em finite-} and {\em infinite-rank} position~\citep{fagin}. Accordingly, \ws \ \dpm \ is characterized by restrictions on certain join variables occurring
in infinite-rank positions. A non-deterministic QA algorithm  for \ws \ \dpm \ was presented in \citep{cali12}, and was used to establish  that QA can be done in polynomial-time. However, this algorithm was not proposed for practical purposes, but only theoretical ones.

\begin{figure}[h]
\centering
\includegraphics[width=6.0cm]{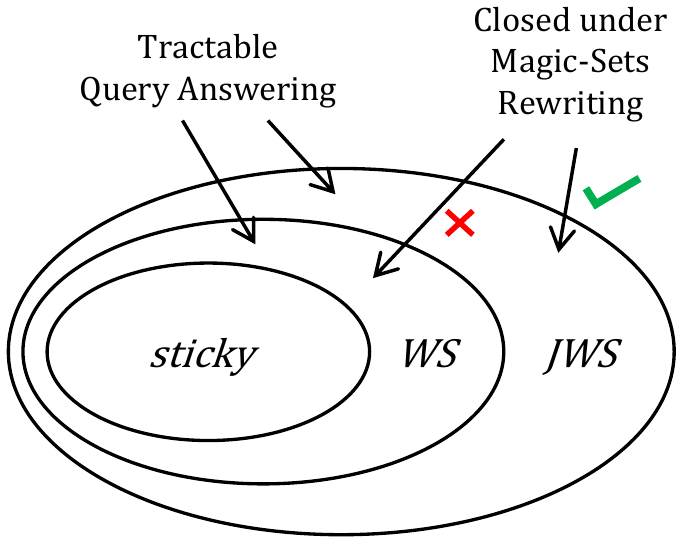}\vspace{-2mm}
\caption{The existing and new settings} \vspace{-2mm}
\label{fig:goals}
\end{figure}

Accordingly, the initial motivation for this work is that of providing a practical, polynomial-time QA algorithm for \ws \ \dpm\!, including the optimization via magic-sets (MS). This is interesting {\em per se}, but is also practically relevant, because
 \ws \ \dpm\! has found natural and interesting applications to the extraction of quality data from possibly dirty databases, as shown in our previous work~\citep{bertossi17,milani15}. This task is accomplished through QA.  However, in order to achieve these goal for \ws \ \dpm\!, we have to go beyond this class: \ws \ \dpm\! \ is not closed under MS. In this direction, we investigate in more abstract terms classes of programs that extend \stk \ and \ws \ \dpm\!, and are defined in terms of the stickiness property of the chase for values identified by a {\em selection function}, among those that appear in body joins and in {\em finite positions} in the  program   (more details below in this introduction). \
 More concretely, but still in high-level terms, our main goals and results in this work are as follows:

\begin{itemize}
\item [(A)] We introduce the generic class of \sch$(\select)$-\dpm\! programs, where $\select$ identifies some of the finite positions in a program body, i.e. those that take finitely many values during the chase. This is a semantic class in that the behavior of a program in it depends on the program's extensional data.  \ws \ \dpm\! is a syntactic subclass of one of those semantic classes.


\item [(B)] We investigate tractability of QA for \sch$(\select)$-\dpm\! \ modulo the availability or computability of the selection function $\select$; and propose a generic, bottom-up, chase-based QA algorithm for this class that calls $\select$ as an oracle or a subroutine. This QA algorithm applies the classical chase procedure, but with a novel termination condition, as needed for QA (cf. Section~\ref{sec:alg}).


\item[(C)] We introduce the class of {\em jointly-weakly-sticky} \ (\jws) \dpm\ programs, as a particular program class that extends \ws \ \dpm\!, is determined by a computable selection function, and is closed under magic-set-based program rewriting (cf. Figure \ref{fig:goals}).

    \item[(D)]We  show that the generic QA algorithm in (B) becomes a deterministic and polynomial-time  for  \ \jws.
\ Since the chase may be infinite, depending on the query, only a finite, small and query-dependent initial portion of the chase is generated and queried.
\item[(E)] We propose a {\em magic-sets} optimization algorithm,  \ms, of the QA algorithm for  \ws\ and \jws \ programs. We show that the query-dependent  rewriting of  program in \jws\ \dpm also belongs to \jws\ \dpm. This algorithm is based on~\citep{alviano12-datalog}.
\end{itemize}

\ignore{\comlb{This was in the previous paragraph. It looks to specific, better move it to the section where it belongs: "Extending classical magic-sets for Datalog \citep{\ignore{bancilhon,beeri-ms,}ceri},
  \ms prevents existential variables from getting bounded, a reasonable adjustment that essentially preserves the semantics of existential rules  during the rewriting."}
\commos{NEW: We have similar explanation there too. I add this as a comment there anyway.}}

In relation to item (A) above,
   we consider both semantic and syntactic  classes of \dpm. \ By a {\em semantic class of programs} we refer to one whose programs {\em also include their EDBs}, and with that EDB the chase exhibits a certain behavior and has some special properties. A syntactic class characterizes its members, i.e. programs, in terms of a condition that is computable or decidable on the basis of the program's rules alone, without involving an EDB (e.g. \stk \ and \ws \ \dpm \ are syntactic classes). A particularly prominent semantic condition (or class of programs that satisfy it) is that of {\em stickiness of the chase} (in short, the {\em sch-property}):

\begin{eqnarray}\hspace{-7mm}\nit{SCh}\hspace{-3mm}&:&\hspace{-3mm}\mbox{\em A program $\mc{P} \cup D$ belongs to
 the \SCh \ class if, due to the enforcement}\nonumber\\
  &&\hspace{-3mm}\mbox{\em    of a rule during the chase, a value replaces a join variable in a rule} \nonumber\\
  &&\hspace{-3mm}\mbox{\em  body, then that value is propagated through all the possible subsequent} \nonumber\\
  &&\hspace{-3mm}\mbox{\em  chase steps, i.e. the value ``sticks".} \ \hfill \label{eq:sch}
\end{eqnarray}

 \begin{example} \label{example:sticky0} (ex. \ref{ex:intr} cont.) Consider programs $\mc{P}$ and $\mc{P}'$ below, both with EDB $D=\{R(a,b)\}$.
\[ \arraycolsep=0pt
\begin{array}{rcl c rcl}
& \mc{P}&&&&\mc{P}'&\\
R(x,y) &~\rightarrow&~ \exists z\;R(y,z), &\hspace*{1.0cm}& R(x,y) &~\rightarrow&~ \exists z\;R(y,z),\\
R(x,y),R(y,z) &~\rightarrow&~ S(x,y,z).&\hspace*{1.0cm}&R(x,y),R(y,z) &~\rightarrow&~ S(x,y,z),\\
&&&\hspace*{1.0cm}&S(x,y,z) &~\rightarrow&~ P(x,z).
\end{array}
\]

\vspace{2mm}
\begin{center}
\includegraphics[width=6cm]{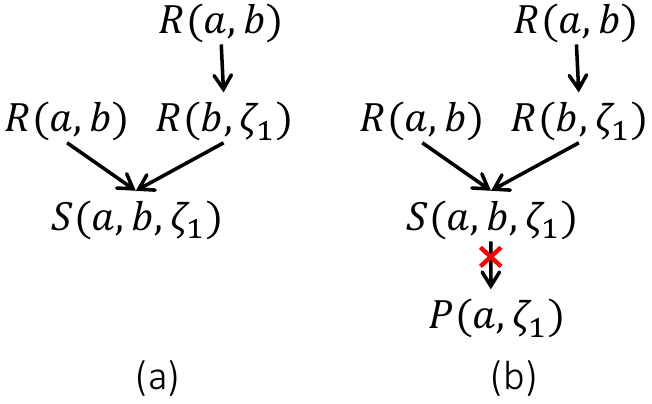}
\end{center}
\vspace{-3mm}
\captionof{figure}{The {\em sch-property}.}\label{fig:chase0}

\ignore{\begin{minipage}[t]{0.55\textwidth}
$\mc{P}$ has the {\em sch-property}, as a portion of its chase in Figure~\ref{fig:chase0} \blue{(a)} shows. $\mc{P}'$ does not have the {\em sch-property}, as shown in the same figure \blue{(b)}: value $b$ is not propagated to $P(a,\zeta_1)$. \boxtheorem
\hspace{0.25cm} Stickiness of the chase defines a {\em semantic class} of programs in the sense that they involve an EDB. This class contains every \stk \ \dpm \ program \citep{cali12}; in this case for every EDB. So, in this case, a purely syntactic property of the program guarantees stickiness.
\end{minipage}
\hspace{-1mm}
\begin{minipage}[t]{0.45\textwidth}
\begin{center}
\vspace{-0.1cm}
\includegraphics[width=5.2cm]{sch}
\end{center}
\vspace{-5mm}
\captionof{figure}{The {\em sch-property}.}\label{fig:chase0}\vspace{-5mm}
\end{minipage}
}

\vspace{2mm}$\mc{P}$ has the {\em sch-property}, as a portion of its chase in Figure~\ref{fig:chase0}(a) shows. $\mc{P}'$ does not have the {\em sch-property}, as shown in Figure~\ref{fig:chase0}(b): value $b$ is not propagated to $P(a,\zeta_1)$. \boxtheorem
\end{example}

Stickiness of the chase defines a {\em semantic class} of programs in the sense that they involve an EDB. This class, {\nit SCh}, contains every \stk \ \dpm \ program \citep{cali12} accompanied by any EDB, as long as the latter is schema-compatible with the program. So, in this case, a purely syntactic property of the program, independent from the EDB, guarantees stickiness.
\ For a program, stickiness  of the chase, i.e. membership of \nit{SCh}, guarantees tractability of QA, because \cq s on such a program can be answered on an initial portion of the chase that has a fixed depth that is independent from the EDB (but depends only on the program and the query), and has a size that  is polynomially bounded by the size of the \edb \ \citep{cali12}.

\ignore{
\comlb{I think this discussion may be too much and too long for the introduction. Please, if it can be reused (after possibly adapting it) in the algorithmic section. I not checking or doing anything in relation to what follows. }
\commos{I modified it and moved it to the beginning of Section~\ref{sec:qa-ch}.}
}

The class of \ws \ \dpm \ programs we start from is defined in such a way it is guaranteed that values not appearing in any {\em finite-rank position} in a body join  are propagated all the way up through the chase, for every EDB. Without going into the technical details about finite-rank positions for the moment, let's just say that they are all {\em finite positions} of the program, where, for a program $\mc{P} \cup D$, a  {\em a position is finite}  if and only if finitely many different values may appear in it during the chase.\footnote{Since there is always a finite number of constants in the EDB of a program, and no constants are created during the chase, the possible creation of infinitely many values at a position is due to the introduction of nulls.}   \ Accordingly, if we  denote with $\fp{\mc{P} \cup D}$ the set of finite positions of a program $\mc{P} \cup D$ consisting of a set of rules $\mc{P}$ and extensional database $D$, the set of finite-rank positions of $\mc{P}$ is contained in $\fp{\mc{P} \cup D}$ (for every $D$). (There may be positions in $\fp{\mc{P} \cup D}$ that are not finite-rank positions of $\mc{P}$ though.)

 We can see that the definition of \ws \ \dpm: (a) is based on a very particular way of choosing finite positions of the program; and (b) is crafted to guarantee that join values not appearing in those finite positions have the propagation property in relation to the chase. So, \ws \ \dpm \ is in essence determined by a {\em selection function} (of finite positions), which is denoted by $\select^\nit{rank}$, and turns out to be syntactic in the sense that in can be computed from the program $\mc{P}$, independently from  $D$ (cf. Section \ref{sec:ws}).

This idea can be generalized in a very natural manner
by replacing in (\ref{eq:sch}),  the condition ``join variable" by the stronger one requiring  ``join variable not appearing in any of the finite positions selected by $\select$", where $\select$ is an abstract {\em selection function} $\select$ that identifies a set of finite positions, say the ``\!$\select$-finite positions", i.e. $\select(\mc{P} \cup D) \subseteq \fp{\mc{P} \cup D}$, a possibly proper inclusion.
\begin{eqnarray}
\mbox{\sch}(\select)\hspace{-3mm}&:&\hspace{-3mm} \mbox{\em A program } \mc{P} \cup D \ \mbox{\em belongs to
 the } \mbox{\sch}(\select) \mbox{\em class if, due to the enforcement}\nonumber\\ &&\hspace{-3mm}\mbox{\em of a rule during the chase, a value replaces a join variable in the rule}\nonumber \\ &&\hspace{-3mm}\mbox{\em body that does not appear in any position in }\mc{S}(\mc{P} \cup D), \mbox{\em then that value} \nonumber \\&&\hspace{-3mm}\mbox{\em is propagated through all the possible subsequent chase steps.} \label{eq:ssch}
 \end{eqnarray}
Since the condition on the join variables is stronger than that for \sch, the new property defines a possibly larger semantic class  of programs (the positions that have to be checked for value propagation may be a subset of those to check for \nit{SCh}). For this  class of programs \sch$(\mc{S})$ that enjoy the {\em $\mc{S}$-stickiness property of the chase}, it holds \ $\nit{SCh} \subseteq \mbox{\sch}(\mc{S})$.

We can define a whole range of program classes by considering different selection functions. There are two extreme cases. On one side, if $\mc{S}$ returns the empty set of finite positions, we reobtain \nit{Sch} in (\ref{eq:sch}). At the other extreme, if $\mc{S}$ selects all the finite positions, i.e. $\mc{S}(\mc{P} \cup D) := \fp{\mc{P} \cup D}$, then we obtain the class \nit{GSCh} of programs  with the {\em generalized-stickiness property of the chase}: {\em A program } $\mc{P} \cup D$ {\em belongs to
 the \nit{GSCh} class if, due to the enforcement of a rule during the chase, a value replaces a join variable in the rule body that  does not appear in any  position in} $\fp{\mc{P} \cup D}$, {\em then that value is propagated through all the possible subsequent chase steps.}

Clearly the {\em GSCh} class contains the \SCh \ class, and all the other classes \sch$(\mc{S})$.  We should notice that, given a Datalog$^+$ program  $\mc{P} \cup D$,  computing (deciding) $\fp{\mc{P} \cup D}$ is unsolvable (undecidable) \citep{deutsch}. Accordingly, it is also undecidable if a Datalog$^+$ program belongs to the {\em GSCh} class. The same may happen with some of the other \sch$(\mc{S})$ classes.

As another particular case of (\ref{eq:ssch}), we obtain the semantic class \sch($\mc{S}^\nit{rank}$) related to \ws \ \dpm \ by using the syntactic selection function $\mc{S}^\nit{rank}$ that characterizes the finite-rank positions (cf. Section \ref{sec:ws}). Although \ws \ \dpm \ is a syntactic class (membership does not depend on the EDBs), \sch($\mc{S}^\nit{rank}$) is  still semantic, because -even with a syntactic $\mc{S}$- the stickiness property may depend on the EDB. However, every program in the syntactic \ws \ \dpm \ class  belongs to \sch($\mc{S}^\nit{rank}$), for every EDB.

\begin{figure}[h]
\centering
\includegraphics[width=8.5cm]{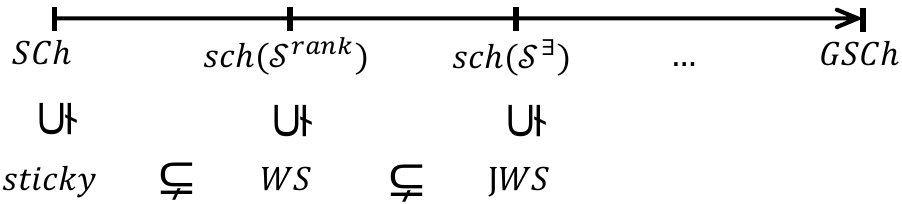}\vspace{-2mm}
\caption{Some program classes in the range} \vspace{-2mm}
\label{fig:someClasses}
\end{figure}

In Section \ref{sec:jws} we will introduce another syntactic selection   function, $\mc{S}^\exists$, that will lead to the semantic class \sch$(\mc{S}^\exists)$. The associated syntactic class will be that of \jws \ \dpm. $\mc{S}^\exists$ that is inspired by the {\em existential dependency graph} of a program \citep{krotzsch}. Since $\mc{S}^\nit{rank}(\mc{P} \cup D) \subseteq    \mc{S}^\exists(\mc{P} \cup D) \subseteq \fp{\mc{P} \cup D}$, it holds \sch($\mc{S}^\nit{rank}$) $\subseteq$ \sch($\mc{S}^{\exists}$). The programs in the associated syntactic class \jws \ \dpm \ will all belong to \sch($\mc{S}^{\exists}$), for every EDB. \
The containment relationships between the syntactic and semantic classes discussed so far are shown in Figure \ref{fig:someClasses}, with containment from left to right. \ We define the semantic classes in Section~\ref{sec:stk}, generalizing sticky \dpm \ on the basis of the classical chase.

\ignore{
The program classes \sch($\mc{S}^\nit{rank}$) and \sch$(\mc{S}^\exists)$ have as purely syntactic subclasses those of \ws\ and \jws \ \dpm \ programs, resp. (cf. Sections \ref{sec:programs} and \ref{sec:jws}). They are syntactic in the sense that a program belongs to the class if it passes a syntactic test that does not depend on the data. As a consequence, programs in  \ws\ and \jws \ belong to \sch($\mc{S}^\nit{rank}$) and \sch$(\mc{S}^\exists)$, resp., with any EDB associated to them. (Cf. Figure \ref{fig:someClasses}, with containment from left to right.)}

We propose a general QA algorithm for the \sch$(\mc{S})$ classes (cf. Section \ref{sec:qa-ch}). It assumes that the positions identified by the selection $\mc{S}$ are {\em computationally accessible}, which may happen through an efficient, data-independent computation as in the case of syntactic classes above, or through an oracle that just returns them (say, in constant time) when  $\mc{S}$ is not computable (and the finiteness of the positions it returns may depend on the data).

More precisely, the algorithm relies on the stickiness property of the chase for the program class at hand; and becomes of
polynomial-time in data complexity, modulo access to \sch$(\mc{S})$, on the assumption that $\select$ returns positions where polynomially many values appear during the chase. This is the case, in  particular, for the semantic classes \sch($\mc{S}^\nit{rank}$) and \sch$(\mc{S}^\exists)$. Therefore, we obtain polynomial-time \qa \ algorithms for their syntactic subclasses, \ws \ and \jws \ \dpm, resp.

In general terms, the just described approach to QA works as follows:
Given a query over a program in \sch$(\select)$, the $\select$-stickiness property of the program restricts the number of values that replace the join variables in non-$\select$-finite positions, because those values are   propagated all the way to the query atom, which has a  fixed arity. On the other hand, the join variables in $\select$-finite positions can only be replaced with finitely many values. As a result, the depth of the proof-schema, which depends on these join values, is also limited by the size of the query and the number of values in $\select$-finite positions. This guarantees the decidability of \qa \ for programs in \sch$(\select)$. Furthermore, if the number of those join values is polynomial in the size of the \edb, the depth of proof-schema will be polynomial in \edb, \ and \qa \ becomes tractable.



The paper is structured as follows: Section~\ref{sec:preliminaries} is a review of some basics of the database theory, the chase procedure, and \dpm \ program classes. Section~\ref{sec:stk} contains our semantic and syntactic generalizations of stickiness using selection functions. \ignore{We formalize selection functions in Section~\ref{sec:selection-func}, and introduce the semantic and the syntactic classes based on the notion of stickiness in Sections~\ref{sec:selection} and \ref{sec:selectionSynt}, respectively. The \jws \ class of programs is presented in Section~\ref{sec:jws}.} Section~\ref{sec:qa-ch} and Section~\ref{sec:mg} contain the QA algorithm and \ms\!. In this paper we use mainly intuitive and informal introductions of concepts and techniques, illustrated by examples. This work extends and is build upon our earlier work on \qa \ for extensions of \ws \ \dpm \ programs~\citep{milani15amw,milani16rr}.

\section{Preliminaries and Background} \label{sec:preliminaries}

In this section, we briefly review relational databases and the \dpm \ program classes.

\subsection{Relational Databases}\label{sec:relational}

We consider relational schemas $\schema$ with two disjoint domains:  \constants, with possibly infinitely many {\em constants}, and \nulls, of infinitely many {\em labeled nulls}. $\schema$ also contains predicates of fixed finite arities. If $P$ is an $n$-ary predicate (i.e. with $n$ arguments) and $1\leq i \leq n$, $P[i]$ denotes its $i$-th position. $\schema$ gives rise to a language $\mf{L}(\schema)$ of first-order (FO)  predicate logic with equality ($=$).  Variables are usually denoted with $x, y, z, ...$, and finite sequences thereof by $\bar{x}, ...$. Constants are usually denoted with $a, b, c, ...$; and nulls are denoted with $\zeta, \zeta_1, ...$. An {\em atom} is of the form $P(t_1, \ldots, t_n)$, with $P$ an $n$-ary predicate  and $t_1, \ldots, t_n$ {\em terms}, i.e. constants, nulls, or variables.
 The atom is {\em ground} (aka. a tuple) if it contains no variables. An {\em instance} $I$ for schema $\schema$ is a possibly infinite set of ground atoms.  The {\em active domain} of an instance $I$, denoted ${\it Adom}(I)$, is the set of constants and nulls that appear in atoms of $I$. Instances can be used as interpretation structures for language $\mf{L}(\schema)$. A {\em database instance} is a finite instance that contains no  nulls.

 A {\em homomorphism} from instance $I$ to instance $I'$ for the same schema is a structure-preserving mapping, ${\it Adom}(I) \!\rightarrow \!{\it Adom}(I')$, such that: (a) $t \in \constants$ implies $h(t)=t$, and (b) for every ground atom $P(\vectt{t}) \in I$, it holds $P(h(\vectt{t})) \in I'$. ($h(\bar{t})$ is defined componentwise.) 

A {\em conjunctive query} (\cq) is a \fo \ formula,  $\mc{Q}(\vectt{x})$, of the form:

\vspace{-3mm}
\begin{align}\label{frm:cq}\exists  \vectt{y}\;(P_1(\vectt{x}_1)\wedge \dots \wedge P_n(\vectt{x}_n)),\end{align}
\vspace{-3mm}

\noindent with (distinct) free variables $\vectt{x} := (\bigcup \vectt{x}_i) \smallsetminus \vectt{y}$. If $\mc{Q}$ has $m$ (free) variables, for an instance $I$, $\vectt{t} \in (\constants \cup \nulls)^m$ \ is an {\em answer} to $\mc{Q}$ if $I \models \mc{Q}[\vectt{t}]$, meaning that  $Q[\vectt{t}]$ becomes true in $I$  when the variables in $\vectt{x}$ are componentwise replaced by the values in $\vectt{t}$. $\mc{Q}(I)$ denotes the set of answers to $\mc{Q}$ in $I$. $\mc{Q}$ is a {\em boolean conjunctive query} (\bcq) when $\vectt{x}$ is empty; and if it is {\em true} in $I$,  $\mc{Q}(I) := \{\nit{true}\}$. Otherwise, $\mc{Q}(I) = \emptyset$, and we say it is
{\em false}.

A {\em tuple-generating dependency} (\tgd), also called  a {\em rule}, is an implicitly universally quantified  sentence of $\mf{L}(\schema)$ of the form:

\vspace{-4mm}
\begin{align}\sigma\!: \ \ P_1(\vectt{x}_1), \ldots, P_n(\vectt{x}_n) \ \rightarrow \ \exists \vectt{y} \ P(\vectt{x},\vectt{y}), \label{frm:tgd}\end{align}
\vspace{-4mm}

\noindent with $\bar{x} \subseteq \bigcup_i \bar{x}_i$, and the dots in the antecedent standing for conjunctions. The variables in~$\vectt{y}$ (that could be empty) are the {\em existential variables}. We assume $\vectt{y}\cap \cup \vectt{x}_i=\emptyset$. With ${\it head}(\sigma)$ and ${\it body}(\sigma)$ we denote the atom in the consequent and the set of atoms in the antecedent of $\sigma$, respectively. A \tgd \ may contain constants from \constants \ in predicate positions.

A {\em constraint} is an {\em equality-generating dependency} (\egd) or a {\em negative constraint} (\nc), which are also sentences of $\mf{L}(\schema)$, respectively, of the forms:

\vspace{-3mm}
\begin{align}
P_1(\vectt{x}_1), \ldots, P_n(\vectt{x}_n) \ \rightarrow \  x=x',\label{frm:egd}\\
P_1(\vectt{x}_1), \ldots, P_n(\vectt{x}_n) \ \rightarrow \ \bot,\label{frm:nc}
\end{align}
\vspace{-3mm}

\noindent where $x,x' \in \bigcup_i \bar{x}_i$, and $\bot$ is a symbol that denotes the Boolean constant (propositional variable) that is always false. Satisfaction of constraints by an instance is  as in \fo \ logic. {\em Tgds}, \egds, and \ncs \ are particular kinds of relational {\em integrity constraints} (ICs)~\citep{abiteboul}. In particular,  \egds \ include {\em key constraints} and {\em functional dependencies} (FDs). ICs also include {\em inclusion dependencies} (IDs) that are subsumed by \tgds.

Relational databases work under the {\em closed world assumption} (CWA)~\citep{abiteboul}: ground atoms not belonging to a database instance are assumed to be false. As a result of this form of data-completeness assumption, an IC is always true or false when checked for satisfaction on a database instance, never undetermined. As we will see below, if instances are allowed to be incomplete or open, i.e. with undetermined or missing ground atoms, ICs can be used, by enforcing them, to generate new tuples.

\da \ is a declarative query language for relational databases that is based on the logic programming paradigm, and allows to define recursive views~\citep{abiteboul,ceri}. A \da \ program $\mc{P}$ for schema $\schema$ is a finite set of non-existential rules, i.e. as in (\ref{frm:tgd}) above but without $\exists$-variables. Some of the predicates in $\mc{P}$ are {\em extensional}, i.e. they do not appear in rule heads, and the extensions  for them (i.e. their sets of tuples) are given by a complete database instance $D$ (for the extensional subschema of $\schema$), which is called {\em the extensional database} \ (EDB).\footnote{That is, the {\em closed-world assumption} (CWA) applies to the extensional atoms in $D$: If a ground atom for the extensional subschema is not explicitly a member of $D$, it is assumed to be {\em false}.} The other, {\em intentional}, predicates are defined by rules that have them in  their heads. For Datalog programs, we may assume, without loss of generality, that intentional predicates appear only in rules, but not in the EDB.

 The {\em minimal-model semantics} of a \da \ program with respect to (wrt.) an extensional database instance $D$ is given by a fixed-point semantics: the extensions of the intentional predicates are obtained by, starting from $D$, iteratively enforcing the rules and creating tuples for the intentional predicates, i.e. whenever a ground (or instantiated) rule body becomes true in the extension obtained so far, but not the head, the corresponding ground head atom is added to the extension under computation. If the set of initial ground atoms is finite, the process reaches a fixed-point after a number of steps that is polynomially bounded in the size of $D$.

A \cq \ as in~(\ref{frm:cq}) can be expressed as a \da \ rule of the form:

\vspace{-4mm}
\begin{align}P_1(\vectt{x}_1),...,P_n(\vectt{x}_n)\rightarrow \nit{ans}_\mc{Q}(\vectt{x}),\end{align}
\vspace{-4mm}

\noindent where $\nit{ans}_\mc{Q}(\cdot)$ \ignore{ \notin \schema$} is an auxiliary, answer-collecting predicate.  The answers to query $\mc{Q}$ form the extension of  predicate $\nit{ans}_\mc{Q}(\cdot)$ in the minimal model. When $\mc{Q}$  is a \bcq,  $\nit{ans}_{\mc{Q}}$ is a propositional atom; and $\mc{Q}$ is true in the undelying instance exactly when the atom $\nit{ans}_\mc{Q}$ belongs to the minimal model of the program.

\begin{example} \label{ex:datalog} A \da \ program $\mc{P}$ containing the rules

\vspace{-3mm}
\begin{align*}
P(x,y)~\rightarrow~R(x,y),\\
P(x,y),R(y,z)~\rightarrow~R(x,z)
\end{align*}
\vspace{-3mm}

\noindent recursively defines, on top of the extensional relation $P$, the intentional predicate $R$ as the transitive closure of  $P$. For $D=\{P(a,b),P(b,d)\}$ as extensional database, the extension of $R$ can be computed by iteratively adding tuples enforcing the program rules, which results in $\{R(a,b),R(b,d),R(a,d)\}$.

The CQ $\mc{Q}(x)\!: \ R(x,b) \wedge R(x,d)$ can be expressed by the rule \ $R(x,b), R(x,d)$  $\rightarrow \ \nit{ans}_{\mc{Q}}(x)$. The set of answer is the computed extension for $\nit{ans}_{\mc{Q}}(x)$, namely $\{a\}$.
\boxtheorem
\end{example}

\subsection{Datalog$^\pm$}\label{sec:dpm}

\dpm \ is an extension of Datalog. The ``$+$" stands for the extension, and the ``$-$", for some syntactic restrictions on the program that guarantee some good computational properties. We will refer to some of those restrictions in Section \ref{sec:programs}. Accordingly, until then we will consider \dplus \ programs.

A \dplus \ program may contain, in addition to (non-existential) Datalog rules, existential rules of the form (\ref{frm:tgd}), constraints of the forms (\ref{frm:egd}) and (\ref{frm:nc}), and a finite extensional database $D$ that may be incomplete and contains ground atoms for the extensional predicates, i.e. those that do not appear in rule heads, and  possibly also for the intensional predicates, i.e.  those appearing in rule heads. \ {\em We will usually denote with $\mc{P}$ the set of rules and constraints, and with $D$ the extensional database (EDB). Accordingly a program is of the form $\mc{P} \cup D$. When no possible confusion arises, we simply refer to $\mc{P}$ as the ``program".} \ A program has an associated schema formed by the predicates in it. \ The set of positions (for the predicates) in a program $\mc{P}$ is denoted with $\nit{Pos}(\mc{P})$. \ We may safely assume all the predicates in an EDB for program $\mc{P}$ also appear in $\mc{P}$.

\ignore{\comlb{Notice that I got rid of the $I$; we do not need it. Check possible occurrences of I as an EDB to delete. I keep $I$ below  for instances obtained through the chase.}
\\\commos{I agree. I will keep in mind while reading and working on the rest of the paper.}
}

The semantics of a \dplus \ program $\mc{P} \cup D$ is model-theoretic, and given by the class $\nit{Mod}(\mc{P},D)$ of all instances $D'$ for the program's schema that extend $D$ and make $\mc{P}$ true.  In particular, given a an $n$-ary CQ $\mc{Q}(\bar{x})$, $\bar{t} \in (\constants \cup \nulls)^n$ is an {\em answer} wrt. $\mc{P}$ and $D$ iff $D' \models \mc{Q}[\bar{t}]$ for every $D' \in \nit{Mod}(\mc{P},D)$. This is {\em certain answer} semantics that requests truth in {\em all} models. Without any restrictions on the program, and even for programs without constraints, {\em conjunctive query answering} (CQA) may be undecidable \citep{beeri-icalp}.

CQA appeals to all possible models of the program. However, the {\em chase procedure} \citep{maier} can be used to generate a single instance that represents the class $\nit{Mod}(\mc{P},D)$ for this purpose. We show it by means of an example containing only \tgds.

\begin{example} \label{ex:chaseNEW} Consider a program $\mc{P}$ with the set of rules \ $\sigma\!:  R(x,y) \ \rightarrow \ \exists z \ R(y,z)$, \ and \
$\sigma'\!:  R(x,y),R(y,z) \ \rightarrow \ S(x,y,z)$,
\ignore{\vspace{-4mm}
\begin{align*}
\sigma&:&\hspace{-2cm}R(x,y) \! ~&\rightarrow~ \! \exists z \ R(y,z),\\
\sigma'&:&\hspace{-2cm}R(x,y),R(y,z) \! ~&\rightarrow~ \! S(x,y,z),
\end{align*}  }
and an extensional instance $D=\{R(a,b)\}$, providing an incomplete extension for the program's schema.
With the $I_0:=D$, the pair $(\sigma, \theta_1)$, with (value) {\em assignment} (for variables) $\theta_1\!: \ x\mapsto a, y\mapsto b$, is {\em applicable}: $\theta_1(\nit{body}(\sigma))=\{R(a,b)\} \subseteq I_0$. The chase {\em enforces} $\sigma$ by inserting a new tuple $R(b,\zeta_1)$ into $I_0$, with $\zeta_1$ a {\em fresh} null, i.e. not in $D_0$, resulting in instance $I_1 = \{R(a,b),R(b,\zeta_1)\}$. This chase step is denoted as \ $I_0 \myrightarrow{(\sigma, \theta_1)} I_1$.


Now, $(\sigma',\theta_2)$, with $\theta_2\!: \ x\mapsto a, y\mapsto b, z\mapsto \zeta_1$, is applicable in $I_1$, because $\theta_2(\nit{body}(\sigma'))$ $=$ $\{R(a,b),R(b,\zeta_1)\} \subseteq I_1$. The chase adds $S(a,b,\zeta_1)$ into $I_1$, resulting in instance $I_2$.
The chase continues, without stopping, creating an infinite instance, usually called {\em the chase} (instance):

$${\it chase}(\mc{P},D) = \{R(a,b), R(b,\zeta_1), S(a,b,\zeta_1), R(\zeta_1,\zeta_2), R(\zeta_2,\zeta_3), S(b,\zeta_1,\zeta_2), \ldots\}.$$

A query over $\mc{P} \cup D$ can be answered on a finite, initial portion of the infinite chase instance. For example, only after adding $R(b,\zeta_1)$, we can return {\em true} as the answer to the \bcq \ $\mc{Q}:\exists x\;R(b,x)$. For certain syntactic classes of programs, such as the ``sticky" program  in this example (c.f. Section~\ref{sec:programs}), it is also possible to return a negative answer to BCQ. For example, we can return {\em false} as the answer to $\mc{Q}':\exists x,y\;S(x,y,a)$ after a finite number of chase steps, confirming that constant $a$ will never occur in position $S[3]$.  For a given program (with the right properties), the size of the initial, finite portion of the chase for QA depends on the query. For example, to answer $\mc{Q}'':\exists x,y,z,w,u\;(S(x,y,z) \wedge S(y,z,w) \wedge S(z,w,u)$, actually positively, we need to generate more atoms than those needed to answer $\mc{Q}$, to map them to the three atoms in $\mc{Q}''$.
\boxtheorem
\end{example}

Some natural questions arise from Example~\ref{ex:chaseNEW}, among them: Assuming the program has good properties in relation to the chase (say it belongs to one of the classes we investigate in this work), (a) How far do we have to finitely develop the chase to correctly answer a given CQ? \ (b) How does it depend on the query? \ (c) Having that finite portion of the chase, possibly materialized, what other CQs can be answered on that portion? \ We address these question in Section~\ref{sec:qa-ch}.

In a nutshell, we use a modified version of the chase that includes two special ingredients, namely {\em homomorphism checking} along the chase, and {\em ``freezing'' of some nulls}, i.e. treating them as constants. The latter technique was introduced in~\citep{leone} for a different program class. The application of these two elements depend on the query, and produces a finite portion of the (classical) chase that is large enough to correctly answer the query.

Depending on the programs and instances, the chase may be finite or infinite; and different orders of chase steps may result in different sequences and instances. However, it is possible to define a {\em canonical chase procedure} that determines  a  canonical sequence of chase steps, and consequently, a canonical chase instance~\citep{cali13}. 



Given a program $\mc{P}$ and EDB $D$, the chase (instance) is a \emph{universal model}~\citep{fagin}: For every  $I \in \nit{Mod}(\mc{P},D)$, there is a homomorphism from the chase into $I$. For this reason, the (certain) answers to a \cq \ $\mc{Q}$ under $\mc{P}$ and $D$ can be computed by evaluating $\mc{Q}$ over the chase instance (discarding the answers containing nulls) \citep{fagin}.

If the program $\mc{P} \cup D$ has \ncs, they are expected to be satisfied by the chase. That is, the BCQ  associated to the \nc \ (\ref{frm:nc}), i.e. $\mc{Q}_\eta\!: \ \exists\vectt{x}_1 \cdots \vectt{x}_n(P_1(\vectt{x}_1)\wedge \cdots \wedge P_n(\vectt{x}_n)$, obtained from the the body of (\ref{frm:nc}), must be false. If this is not the case, we say $\mc{P}$ is {\em inconsistent}. If $\mc{P}$ has \egds, they are also expected to be satisfied by the chase. However, one can modify the chase in order to enforce the \egds, which may be possible or not. In the latter case, we say the {\em chase fails}. (It is possible to define a canonical chase that involves \egds \ \citep{cali13}.)

\begin{example} \label{ex:non-separableNEW} Consider a program $\mc{P}$ with $D=\{R(a,b)\}$ with two rules and an \egd:

\vspace{-4mm}
\begin{align}
R(x,y) ~&\rightarrow~ \exists z\; \exists w\;S(y,z,w),\label{frm:sep1NEW}\\
S(x,y,y) ~&\rightarrow~ P(x,y),\label{frm:sep2NEW}\\
S(x,y,z) ~&\rightarrow~ y=z.\label{frm:sep3NEW}
\end{align}
The chase of $\mc{P}$ first applies (\ref{frm:sep1NEW}) and results in $I_1=\{R(a,b),S(b,\zeta_1,\zeta_2)\}$. There are no more {\em tgd}/assignment applicable pairs. But, if we enforce the \egd~(\ref{frm:sep3NEW}), equating $\zeta_1$ and $\zeta_2$, we obtain $I_2=\{R(a,b),S(b,\zeta_1,\zeta_1)\}$.  Now, (\ref{frm:sep2NEW}) and $\theta': x\mapsto b, y\mapsto \zeta_1$ are applicable, so we add $P(b,\zeta_1)$ to $I_2$, generating $I_3=\{R(a,b),S(b,\zeta_1,\zeta_1),P(b,\zeta_1)\}$.
The chase terminates (no applicable \tgds \ or \egds), obtaining $\nit{chase}(\mc{P},D) = I_3$. \ Notice that the program consisting only of (\ref{frm:sep1NEW}) and (\ref{frm:sep2NEW}) produces $I_1$ as the chase, which makes the BCQ
$\exists x \exists y P(x,y)$ evaluate to {\em false}. With the program also including the \egd \ (\ref{frm:sep3NEW}) the answer is now {\em true}.

Now consider  program $\mc{P}'$ that is $\mc{P}$ with the extra rule \ $R(x,y) ~\rightarrow~ \exists z\;S(z,x,y)$, which enforced on $I_3$ results in $I_4=\{R(a,b),S(b,\zeta_1,\zeta_1),P(b,\zeta_1),S(\zeta_3,a,b)\}$.
Now (\ref{frm:sep3NEW}) is applied, which creates a chase failure as it tries to equate constants $a$ and $b$. In this case the set of \tgds \ and the \egd \ are mutually inconsistent. \boxtheorem\end{example}

Characterizations of computationally well-behaved classes of \dpm \ programs usually do not consider any kind of \egds \ and \ncs \ in the program, but only the \tgds. However, considering \ncs \ is not complicated for these characterizations since they may have a trivial effect of QA or no effect at all. More precisely, if a program $\mc{P}$ consists of a set of \tgds \ $\mc{P}^R$ and a set of \ncs \ $\mc{P}^C$, then CQA amounts to deciding if $\mc{P}^R \cup \mc{P}^C \cup D \models \mc{Q}$, for which the following result holds.

\begin{proposition} \citep[theo. 6.1]{cali12}\footnote{ \ We haven't found an explicit proof of this claim in the literature. So, we give it here.}
$\mc{P}^R \cup \mc{P}^C \cup D \models \mc{Q}$ \ iff \
(a) $\mc{P}^R \cup D \models \mc{Q}$, \ or \ (b) for some $\eta \in \mc{P}^C$, $\mc{P}^R \cup D \models \mc{Q}_\eta$, where $\mc{Q}_\eta$ is the BCQ obtained as the existential closure of the body of $\eta$.
\boxtheorem\end{proposition}

\hproof{\!\! \ Assume (b) does not hold, then $\nit{Mod}(\mc{P}^R \cup \mc{P}^C\!,D) \neq \emptyset$. We have to show that $\mc{P}^R \cup \mc{P}^C  \cup D \models \mc{Q}$ iff $\mc{P}^R \cup D \models \mc{Q}$. From right to left is obvious. Now, from left to right, assume $\mc{P}^R \cup \mc{P}^C \cup D \models \mc{Q}$, and let $D' \in \nit{Mod}(\mc{P}^R,D)$. We have to show that $I \models \mc{Q}$. Let $I^{\nit{ch}}$ be $\nit{chase}(\mc{P}^R,D)$, for which $I^{\nit{ch}} \models \mc{P}^C$ holds (otherwise, due to the universality of the chase and preservation of CQA under homomorphisms,  $\mc{P}^R \cup \mc{P}^C \cup D$ would be inconsistent). Then, $I^{\nit{ch}} \in \nit{Mod}(\mc{P}^R \cup \mc{P}^C,D)$, and then, by hypothesis, $I^{\nit{ch}} \models \mc{Q}$. Since $I^{\nit{ch}}$ can be homomorphically embedded into $I$, we obtain \ $I \models \mc{Q}$. }

 Case (b) above holds exactly when $\mc{P} \cup D$ is inconsistent, and $\mc{Q}$ becomes trivially true. This shows that CQA evaluation under \ncs \ can be reduced to the same problem without \ncs, and the data complexity of CQA does not change. Furthermore, \ncs \ may have an effect on CQA only if they are mutually inconsistent with the rest of the program, in which case every BCQ becomes trivially true. The presence of \egds \ may have a more dramatic effect of QA, which can become undecidable, and the presence of \egds \ may also change query answers, as in Example \ref{ex:non-separableNEW} (cf. \citep[sec. 2]{bertossi17} for a more detailed discussion). \ As a consequence, {\em we assume in the rest of this work that
 programs do not have egds \ or ncs.}

\subsection{Programs Classes} \label{sec:programs}

\cq \ answering over \dplus \ programs with arbitrary sets of \tgds \ is  undecidable~\citep{beeri-icalp}. Actually, it is undecidable whether the chase terminates, even for a fixed instance~\citep{beeri-icalp,deutsch}. Several sufficient conditions, syntactic~\citep{deutsch,fagin,krotzsch,marnette} or data-dependent~\citep{meier}, that guarantee chase termination have been identified. {\em Weak-acyclicity}~\citep{fagin} and {\em joint-acyclicity}~\citep{krotzsch} are  syntactic conditions that use a static analysis of a dependency graph of predicate positions in the program.

A non-terminating chase does not imply that \cq \ answering is undecidable. Several program classes are identified for which the chase may be infinite, but \qa \ is still decidable. That is the case for {\em linear}, {\em guarded}, {\em sticky}, {\em weakly-sticky \dpm}~\citep{cali09,cali10vldb,cali11sigrec,cali12}, {\em shy \de}~\citep{leone}, and {\em finite expansion sets} (fes), {\em finite unification sets} (fus), {\em bounded-treewidth sets} (bts)~\citep{baget09,baget11,baget11ai}. Each program class defines conditions on the program rules that lead to good computational properties for \qa.  In the following, we focus on \stk \ and \ws \ \dpm \ programs.

\subsubsection{Weakly-acyclic programs} \label{sec:wa}

The {\em dependency graph} (DG) of a program $\mc{P}$ with schema $\schema$ (cf. Figure~\ref{fig:dg}) is a directed graph whose vertices are the positions (of predicates) in $\mc{P}$. The edges are defined as follows: for every $\sigma \in \mc{P}$, and every universally quantified variable ($\forall$-variable)\footnote{ \ Every variable that is not existentially quantified is implicitly universally quantified.} $x$ in ${\it head}(\sigma)$ in position $p$ in ${\it body}(\sigma)$ (among possibly other positions where $x$ appears in ${\it body}(\sigma)$): \ (a) for each occurrence of $x$ in position $p'$ in ${\it head}(\sigma)$, create an edge from $p$ to $p'$, (b) for each $\exists$-variable $z$ in position $p''$ in ${\it head}(\sigma)$, create a {\it special (dashed) edge} from $p$ to $p''$.

The {\it rank of a position} $p$ in the graph, denoted by $\nit{rank}(p)$, is the maximum number of special edges over all (finite or infinite) paths ending at $p$. $\finiteRank(\mc{P})$ and $\Pi_\infty(\mc{P})$ denote the sets of finite-rank and infinite-rank positions in $\mc{P}$, resp. It is possible to prove that finite-rank positions are finite positions, i.e. they belong to $\nit{FinPos}(\mc{P} \cup D)$ for every EDB $D$ \citep{fagin}.
\ A program is {\em weakly-acyclic} (\wa) if all of the positions belong to $\finiteRank(\mc{P})$~\citep{fagin}.

\begin{example}  \label{ex:dg} Let $\mc{P}$ be a program with rules:

\vspace{-4mm}
\begin{align*}
  R(x,y),R(y,z) &~\rightarrow~ R(x,z),\\
  R(x,y) &~\rightarrow~ \exists z \;P(y,z).
\end{align*}
 The DG \ of $\mc{P}$ is shown in Figure~\ref{fig:dg}. Positions $R[1]$, $R[2]$ and $P[1]$ have rank $0$; and $P[2]$, rank $1$. \ $\mc{P}$ is \wa \ since all positions have finite-rank. There is a cycle in the DG of $\mc{P}$, but  it does not involve any special edge.

 Now, let $\mc{P}'$ be \ws \ program with rules:

\vspace{-4mm}
\begin{align*}
  R(x,y) &~\rightarrow~ P(y,x),\\
  P(x,y) &~\rightarrow~ \exists z \;R(y,z).
\end{align*}

\noindent The DG of $\mc{P}'$ is shown in Figure~\ref{fig:dg2}. Positions $R[1]$ and $P[2]$ have rank $0$. The program is not \wa \ since $R[2]$ and $P[1]$ have infinite rank. $\mc{P}'$ is not \ws,  because its DG graph has a cycle with a special edge.\boxtheorem
\end{example}

\begin{multicols}{2}
\begin{center}
\includegraphics[width=3.1cm]{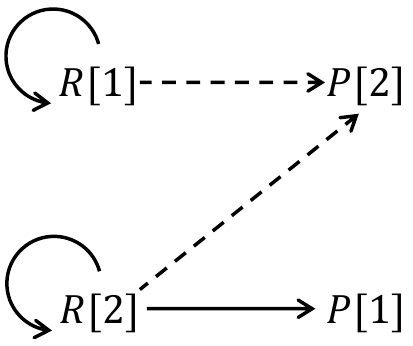}
\vspace{-2mm}
\captionof{figure}{Dependency graph of $\mc{P}$}\label{fig:dg}

\phantom{oo}

\includegraphics[width=2.75cm]{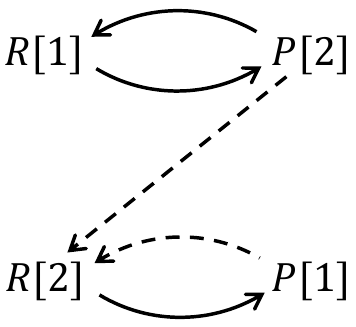}
\vspace{-2mm}
\captionof{figure}{Dependency graph of $\mc{P}'$}\label{fig:dg2}
\end{center}
\end{multicols}

The problem of \bcq \ answering over a \wa \ program is \ptime-complete in data complexity~\citep{fagin}. This is because the chase for these programs stops in polynomial time in the size of the data~\citep{fagin}. The same problem is 2\exptime-complete in combined complexity, i.e. in the size of both program rules and the data~\citep{kolaitis}.

\subsubsection{Jointly-acyclic programs} \label{sec:ja}

The definition of the class of {\em jointly-acyclic} (\ja) programs appeals to the {\em existential dependency graph} (EDG) of a program $\mc{P}$~\citep{krotzsch}, denoted EDG$(\mc{P})$, that we briefly review here.

Assume that program $\mc{P}$ has its rules  standardized apart, i.e. no variable appears in more than one rule. For a variable $x$ in rule $\sigma$, let $B_x$ and $H_x$ be the sets of positions where $x$ occurs in the body, resp. in the head, of $\sigma$. For an $\exists$-variable $z$, the set of {\em target positions} of $z$, denoted by $T_z$, is the smallest set of positions such that: (a) $H_z \subseteq T_z$, and (b) $H_x \subseteq T_z$ for every $\forall$-variable $x$ with $B_x \subseteq T_z$. Roughly speaking, $T_z$ is the set of positions where the invented (fresh) null values for the $\exists$-variable $z$ may appear during the chase.

 EDG$(\mc{P})$ is a directed graph with the $\exists$-variables of $\mc{P}$ as its nodes. There is an edge from $z \in \sigma$ to $z' \in \sigma'$ if there is a body variable $x$ in $\sigma'$ such that $B_x \subseteq T_z$. Intuitively, the edge shows that the values invented by $z$ may appear in the body of $\sigma'$, and cause (null) value invention for $z'$. Therefore, a cycle represents the possibility of inventing infinitely many null values for the $\exists$-variables in the cycle. A program is {\em jointly-acyclic} (\ja) if its EDG \ is acyclic.

\begin{example} \label{ex:edg} Consider a program $\mc{P}$ with the following rules:

\begin{minipage}[t]{0.7\textwidth}
\vspace{-0.2cm}
\begin{align}
\hspace*{-4.75cm}P(x_1,y_1) \! ~&\rightarrow~ \! \exists z_1 \ R(y_1,z_1),\hspace*{-3cm}\label{eq:edg1}\\
\hspace*{-4.75cm}R(x_2,y_2),U(x_2),U(y_2) \! ~&\rightarrow~ \! \exists z_2 \ P(y_2,z_2),\hspace*{-3cm}\label{eq:edg2}\\
\hspace*{-4.75cm}P(x_3,y_3) \! ~&\rightarrow~ \! \exists z_3 \ S(x_3,y_3,z_3).\hspace*{-3cm}\label{eq:edg3}
\end{align}
\end{minipage}\hspace{-7mm}
\begin{minipage}[t]{0.28\textwidth}
\begin{center}
\vspace{-0.1cm}
\includegraphics[width=2.25cm]{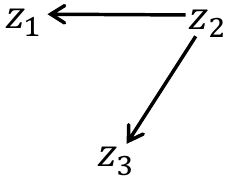}
\end{center}
\vspace{-5mm}
\captionof{figure}{EDG$(\mc{P})$}\label{fig:edg}\vspace{-5mm}
\vspace{0.75cm}
\end{minipage}

$B_{y_1}=\{P[2]\}$ and $H_{y_1}=\{R[1]\}$ are the sets of positions where the variable $y_1$ appears in the body and, resp. the head of rule (\ref{eq:edg1}). Similarly, $B_{x_2}=\{R[1],U[1]\}$, $B_{y_2}=\{R[2],U[1]\}$, and $B_{y_3}=\{P[2]\}$. $T_{z_1}=\{R[2]\}$ and $T_{z_2}=\{P[2],R[1],S[2]\}$ are the sets of target positions of $z_1$ and resp. $z_2$.

In EDG$(\mc{P})$ in Figure~\ref{fig:edg} there is an edge from $z_2$ to $z_1$ since for the body variable $y_1$ in rule (\ref{eq:edg1}), where $z_1$ appears, $B_{y_1} \subseteq T_{z_2}$ holds, which means $y_1$ occurs only in the target positions of $z_2$. Similarly, there is an edge from $z_2$ to $z_3$ since for the body variable $y_3$ in rule (\ref{eq:edg3}), where $z_3$ appears, $B_{y_3} \subseteq T_{z_2}$ holds, which means $y_3$ occurs only in the target positions of $z_2$. There is no edge from $z_1$ to $z_2$ since, in rule (\ref{eq:edg2}), $B_{x_2} \not\subseteq T_{z_1}$ and $B_{y_2} \not\subseteq T_{z_1}$. For a similar reason, there is no self-loop for $z_2$. The graph is acyclic, and then, $\mc{P}$ is \ja.\boxtheorem\end{example}

\ja \ programs have polynomial size (finite) chase wrt. the size of the extensional data, and properly extend \wa \ programs. \bcq \ answering over \ja \ programs is \ptime-complete in data complexity, and 2\exptime-complete is combined complexity~\citep{krotzsch}.

\subsubsection{Sticky programs} \label{sec:sticky} They are characterized through a body variable {\em marking procedure} whose input is the set of rules of a program $\mc{P}$ (the extensional data do not participate in it). The procedure has two steps:

\begin{itemize}
  \item[(a)] {\em Preliminary step}: For each $\sigma \in \mc{P}$ and variable $x$ in ${\it body}(\sigma)$, if there is an atom $A$ in ${\it head}(\sigma)$ where $x$ does not appear, mark each occurrence of $x$ in ${\it body}(\sigma)$.
  \item[(b)] {\em Propagation step}: For each $\sigma \in \mc{P}$, if a marked variable in ${\it body}(\sigma)$ appears in position $p$, then for every $\sigma' \in \mc{P}$ (including $\sigma$), mark  the variables in ${\it body}(\sigma')$ that appear in ${\it head}(\sigma')$ in  position $p$.
\end{itemize}

We say that $\mc{P}$ is {\em sticky} (or belongs to the program class \stk) when, after applying the marking procedure, there is no rule with a marked variable appearing more than once in its body (i.e. not a join variable). Notice that a variable never appears both marked and unmarked in a  same body.






\begin{example} \label{example:sticky} (ex. \ref{example:sticky0} cont.) For program $\mc{P}$ on the left-hand side below, the first rule below already shows marked variable $x$ (with a hat) after the preliminary step. The set of rules on the right-hand side is the final result of the marking procedure applied to  $\mc{P}$:

\vspace{-2mm}
\[ \arraycolsep=0pt
\begin{array}{rcl c rcl}
R(\hat{x},y) &~\rightarrow&~ \exists z\;R(y,z), &\hspace*{1.0cm}& R(\hat{x},\hat{y}) &~\rightarrow&~ \exists z\;R(y,z),\\
R(x,y),R(y,z) &~\rightarrow&~ S(x,y,z),&\hspace*{1.0cm}&R(x,y),R(y,z) &~\rightarrow&~ S(x,y,z).
\end{array}
\]
\vspace{-1mm}

\noindent Variable $y$ in the first rule-body ends up marked after the propagation step: it appears in the rule head, in position $R[1]$, where marked variable $x$ appear in the same rule. Accordingly, $\mc{P}$ is sticky: there is no marked variable that appears more than once in a rule body. \ignore{The chase of $\mc{P} \cup D$ is an infinite instance, $\{R(a,b),R(b,\zeta_1),R(\zeta_1,\zeta_2),S(a,\zeta_1,\zeta_2),R(\zeta_2,\zeta_3),S(\zeta_1,\zeta_2,\zeta_3),...\}$.}

For program $\mc{P}'$, the result of the marking procedure is as follows:

\vspace{-2mm}
\[ \arraycolsep=0pt
\begin{array}{rl}
R(\hat{x},\hat{y}) &~\rightarrow~ \exists z\;R(y,z).\\
R(x,\hat{y}),R(\hat{y},z) &~\rightarrow~ S(x,y,z).\\
S(x,\hat{y},z) &~\rightarrow~ P(x,z).
\end{array}
\]
\vspace{-1mm}

\noindent $\mc{P}'$ is {\em not} sticky: $y$ in the second rule body is marked and occurs twice in it (in $R[2]$ and $R[1]$).
\boxtheorem
\end{example}

The syntactic stickiness condition guarantees that \qa \ can be done in \ptime \ in data complexity; and is \exptime-complete in combined complexity~\citep{cali12}. The chase of a \stk \ program may not terminate, as shown in Example~\ref{example:sticky}. However, a \cq \ can be answered by rewriting it into a \fo \ query, actually a union of CQs, doing {\em backward-chaining} through the rules, and answering the \fo \ query directly on the \edb. The rewriting depends only on the rules and the query; and the size of the rewriting is independent from the \edb~\citep{cali12,gottlob14}.\ignore{\footnote{\ This property is known as \fo \ rewritability that is first introduced for ontology-based query answering and Description Logic (DL) in~\citep{calvanese}.}} 

\ignore{\begin{definition}[first-order rewritability]~\citep{calvanese}\footnote{ \ First-order rewritability for {\em ontology-based query answering} is first introduced in~\citep{calvanese}.} \cq \ answering over a \dplus \ program $\mc{P}$ with extensional database $D$ is {\em first-order (FO) rewritable} if, for every \cq \ $\mc{Q}$, a \fo \ query $\mc{Q}'$ can be constructed such that, $\mc{Q}(\mc{P} \cup D)=\mc{Q}'(D)$.\boxtheorem\end{definition}

\begin{example} \label{ex:fo-re} Let $\mc{P}$ be a program with a database $D=\{P(a,b),U(b)\}$ and the rule:

\vspace{-2mm}
$$P(x,y),U(y) ~\rightarrow~ \exists z\;S(x,y,z).$$

\noindent $\mc{P}$ is \stk \ since it does not have marked variables. Being \stk, it is also \fo \ rewritable. For the \cq \ $\mc{Q}(x):\exists y\;\exists z\; S(x,y,z)$, $\mc{Q}(D)=\emptyset$, since $S$ has no extension in $D$. However, $\mc{Q}(\mc{P} \cup D)=\{a\}$ because $S$ has intensional data obtained through the rule. $\mc{Q}$ can be rewritten into the \fo \ query $\mc{Q}'(x)\!:\!\exists y\exists z\;S(x,y,z)~\vee~\exists t\;(P(x,t)\wedge U(t))$, for which, $\mc{Q}(\mc{P} \cup D)=\mc{Q}'(D)$. Here, $\mc{Q}'$ is obtained by relaxing $\mc{Q}$ and adding $\exists t\;(P(x,t)\wedge U(t))$ that is extracted from the rule body.\boxtheorem\end{example}

\fo \ rewritability is a desirable property as it is well known that the evaluation of \fo \ queries is in the highly tractable class \acz \ (in data complexity)~\citep{vardi}.}

\subsubsection{Weakly-sticky programs} \label{sec:ws}

 They form a syntactic class that extends those of \wa \ and \stk \ programs.  Its characterization does not depend on the extensional data, and uses the notions of finite-rank and marked variable introduced in Sections~\ref{sec:wa} and \ref{sec:sticky}, resp.: A set of rules $\mc{P}$ is {\em weakly-sticky} (\ws) if, for every rule in it and every repeated variable in its body, the variable is either non-marked or appears in a position in $\finiteRank(\mc{P})$.

\begin{example} \label{ex:weakly-sticky}Consider $\mc{P}$ with the set of rules:
\begin{align*}
R(x,y) ~\rightarrow&~ \exists z\; R(y,z),\\
R(x,y),U(y),R(y,z) ~\rightarrow&~ R(x,z),
\end{align*}
\noindent for which $\finiteRank(\mc{P})=\{U[1]\}$ and $\infiniteRank(\mc{P})=\{R[1],R[2]\}$. After applying the marking procedure, every body variable in $\mc{P}$ becomes marked. $\mc{P}$ is \ws \ since the only repeated marked variable is $y$, in the second rule, and it appears in $U[1] \in \finiteRank(\mc{P})$.

Now, let $\mc{P}'$ be the program with the first rule of $\mc{P}$ and the second rule as follows:
\begin{align*}
R(x,y),R(y,z) ~\rightarrow&~ R(x,z).
\end{align*}
\noindent Now, $\finiteRank(\mc{P}')=\emptyset$ and $\infiniteRank(\mc{P}')=\{R[1],R[2]\}$. After applying the marking procedure, every body variable in $\mc{P}'$ is marked. $\mc{P}'$ is {\em not} \ws \ since $y$ in the second rule is repeated, marked and appears in $R[1]$ and $R[2]$, both in $\infiniteRank(\mc{P})$.\boxtheorem\end{example}

The \ws \ condition guarantees tractability of \cq \ answering w.r.t. the size of the \edb~\citep{cali12is}. Intuitively, \ws \ generalizes the syntactic condition of sticky rules by preventing  a repeated, marked variable from appearing only in infinite-rank positions, where it has no bound on the values it can take. However, appearing at least once in a finite position propagates boundedness to its other occurrences. For \ws \ programs, \qa \ can be done by rewriting a \cq \ into a union of CQs, and answering the resulting query over the \edb~\citep{cali12is}. However, unlike \stk \ programs, for \ws \ programs the  rewritten query  and its size may depend on the \edb, but the latter is polynomially bounded by the size of the EDB.

\section{Chase-Based Generalizations of Sticky \dpm} \label{sec:stk}

In Section \ref{sec:intr} we stated our goal of identifying a class of \dpm \ programs that extends \ws, has  a tractable QA problem, and is also closed under magic-set optimization. These desiderata
lead us to analyze more closely the syntactic conditions for \stk \ and \ws \ programs on one side, and, on the other side, value  propagation under the chase for those classes. \ In this section, generalizing from this analysis, we will characterize and use {\em selection functions} $\mc{S}$ that identify sets of finite positions of  programs $\mc{P} \cup D$.

\subsection{Selection Functions}\label{sec:selection-func}

\begin{definition} \label{df:s} (a) A selection function $\mc{S}$ associates every program $\mc{P} \cup D$ with a subset $\mc{S}(\mc{P} \cup D)$ of $\nit{FinPos}(\mc{P} \cup D)$, which is the set of positions that take finitely many values in $\nit{chase}(\mc{P},D)$. (b) The selection functions $\mc{S}^\bot$, \ $\mc{S}^\top$ and $\mc{S}^\nit{rank}$ are defined by: \ $\mc{S}^\bot(\mc{P} \cup D) := \emptyset$, \ $\mc{S}^\top(\mc{P} \cup D) :=\nit{FinPos}(\mc{P} \cup D)$, and
 $\mc{S}^\nit{rank}(\mc{P} \cup D) := \finiteRank(\mc{P})$, the set of finite-rank positions of $\mc{P}$, resp.\boxtheorem
\end{definition}

Notice that $\mc{S}^\top$ is in general non-computable~\citep{deutsch}, but $\mc{S}^\bot$ is clearly computable. $\mc{S}^\nit{rank}$ is a selection function because finitely many values appear in these positions during the chase of the program \citep[Theorem 3.9]{fagin}. It is also computable since the finite-rank positions can be computed from the dependency graph of the program.
 

\begin{definition} \label{def:syn} A selection function $\mc{S}$ is {\em syntactic} iff: \ (a)  there is a computable function $\mc{S}'$ that associates each program $\mc{P}$ with a subset $\mc{S}'(\mc{P})$ of \ $\nit{Pos}(\mc{P})$, such that
 $\mc{S}(\mc{P} \cup D) = \mc{S}'(\mc{P})$, for every $D$; \ and (b) $\mc{S}'(\mc{P}) \subseteq \nit{FinPos}(\mc{P} \cup D)$, for every $D$. \boxtheorem
 \end{definition}

Intuitively, the result of a syntactic selection function depends on the program, but not on the EDB. It soundly returns (some) positions that are finite for a program with any accompanying EDB. Both $\mc{S}^\bot$  and $\mc{S}^\nit{rank}$ are syntactic selection functions, because they only depend on the program, not on the EDB. Particularly, $\mc{S}^\nit{rank}$ only depends on the DG of the program, which is independent of the EDB. We will introduce a computable and syntactic selection function, $\select^\exists$, in Section~\ref{sec:jws}.

\subsection{Semantic Program Classes Defined by Selection Functions}\label{sec:selection}

\ignore{Sticky \dplus \ programs enjoy the {\em``stickiness property"} of the chase ({\em sch-property})~\citep{cali12}. That is, if, due to the application of a rule $\sigma$, when a value replaces a repeated variable in a rule-body, then that value also appears in all the head atoms introduced through the iterative enforcement of applicable rules that starts with $\sigma$'s application. In short, the value is propagated through all possible subsequent chase steps. }

 In order to formally define stickiness properties of the chase, we first recall the {\em chase relation}, $<_{\mc{P},D}$, over the atoms in $\nit{chase}(\mc{P},D)$ \citep[def.~2.1.]{cali12}. Intuitively, $A <_{\mc{P},D} B$ means that $B$ is obtained from $A$ (and possibly other atoms) in a chase step with $\mc{P} \cup D$~.

\begin{definition}[Chase relation] For a \dplus \ program $\mc{P} \cup D$, and atoms $A,B \in \nit{chase}(\mc{P},D)$, $A$ is in {\em chase relation} to $B$, denoted  $A <_{\mc{P},D} B$, if and only if there is a chase step $I_i\myrightarrow{(\sigma_i,\theta_i)}I_{i+1}$  with $\mc{P} \cup D$, such that $A \in \theta_i(\nit{body}(\sigma))$ and $B \in I_{i+1}\setminus I_{i}$. The {\em derivation relation} for $\mc{P} \cup D$, denoted by $\myRightarrow{\mc{P},D}$, is the transitive closure of $<_{\mc{P},D}$. \boxtheorem\end{definition}

\begin{example} (ex.\ignore{~\ref{ex:chs}} \ref{example:sticky0} cont.) \label{ex:rel} According to the chase with $\mc{P} \cup D$ in  Figure~\ref{fig:chase0}, $R(a,b)<_{\mc{P},D} S(a,b,\zeta_1)$, $R(b,\zeta_1) <_{\mc{P},D} S(a,b,\zeta_1)$, and then, $R(a,b)\myRightarrow{\mc{P},D} S(a,b,\zeta_1)$, $R(b,\zeta_1) \myRightarrow{\mc{P},D} S(a,b,\zeta_1)$. However, $S(a,b,\zeta_1) \not\myRightarrow{\mc{P},D} P(a,\zeta_1)$. \ With $\mc{P}' \cup D$: \ $R(a,b)\myRightarrow{\mc{P}',D} S(a,b,\zeta_1)$, $R(b,\zeta_1) \myRightarrow{\mc{P}',D} S(a,b,\zeta_1)$, and $S(a,b,\zeta_1) \myRightarrow{\mc{P}',D} P(a,\zeta_1)$; the last one due to the last rule of $\mc{P}'$.\boxtheorem
\end{example}

We now make precise the definition of the program classes \sch$(\mc{S})$ given in (\ref{eq:ssch}), in Section \ref{sec:intr}.

\begin{definition} \label{df:schs} For a selection function $\select$, a \dplus \ program $\mc{P} \cup D$ has the {\em $\select$-stickiness property} of the chase if and only if, for every chase step $I_i\myrightarrow{(\sigma_i,\theta_i)}I_{i+1}$, the following holds: If a variable $x$ appears more than once in ${\it body}(\sigma_i)$ and {\em not} in $\select(\mc{P} \cup D)$, then $\theta_i(x)$ occurs in the only atom $A$ in $I_{i+1}\setminus I_{i}$,  and in every atom $B$ with $A \myRightarrow{\mc{P}} B$.  \ The class of programs with the $\select$-stickiness property of the chase is denoted with \sch$(\select)$. \boxtheorem\end{definition}

\ignore{\comlb{Mostafa: Did you check the direction of << in other parts of the doc.?}
\commos{Yes I updated it everywhere.}}

\ignore{\comlb{Is the part in red above correct?}\\
\commos{Yes, it should appear not only in A (the head atom) but also any atom B that will be later inferred using A ($A \myRightarrow{\mc{P}} B$).}}


This definition provides semantic classes of programs in that, in general, membership depends on the EDB $D$ associated to the program. With specific selections function we obtain some of the program classes  in
Section \ref{sec:intr}.

\begin{definition} \label{df:schsSpec} (a) \sch$(\select^\bot)$ is the class  of programs with the {\em stickiness-property} of the chase, also denoted with \SCh. \
(b) \sch$(\select^\top)$ is the class of programs with the {\em generalized stickiness-property} of the chase, also denoted with \GSCh. \
(c) \sch$(\select^\nit{rank})$ is the class of programs with the {\em weak stickiness-property} of the chase, also denoted with \wsch.\ignore{ \
(d) \sch$(\select^\exists)$ is the class of programs with the {\em jointly-weakly stickiness-property} of the chase, denoted \jwsch.} \boxtheorem
\end{definition}

Membership for a program $\mc{P} \cup D$ of  the class \GSCh, associated to the uncomputable selection function $\select^\top$ that  returns $\fp{\mc{P} \cup D}$,  is undecidable~\citep{deutsch}.\footnote{Investigating the decidability status of the membership problem for the classes \sch$(\select)$ is outside the scope of this research; at least for the moment.}

\begin{example} \label{ex:chasesticky} (ex.~\ref{example:sticky0} and \ref{ex:rel} cont.) \label{ex:rel2} Clearly, $\mc{P} \cup D \in \SCh$, because the only join variable appears in the rule head. \
Now consider program $\mc{P}' \cup D$. \ All its positions are infinite, i.e. $\Pi_F(\mc{P}') = \fp{\mc{P}' \cup D} = \emptyset$. In fact, it is easy to see that the chase creates infinitely many values in all positions. \ignore{with the first rule invents infinitely many null values in $R[2]$ that are propagated to $R[1]$. The second and the third rules propagate those nulls to every position in $S$ and $P$.} The values in body joins appear only in infinite positions, and have to be checked for stickiness.

It turns out that  $\mc{P}' \cup D \notin \SCh$. In fact, consider the chase step $I_1\myrightarrow{(\sigma,\theta)} I_2$ in which $I_1=\{R(a,b),R(b,\zeta_1)\}$, $I_2=\{R(a,b),R(b,\zeta_1),S(a,b,\zeta)\}$, $\theta: x \mapsto a, y \mapsto b, z \mapsto \zeta_1$, and $\sigma$ is the last rule in $\mc{P}'$. In this chase step, $b$ replaces body variable $x$ that appears twice in the body of $\sigma$. However, $b$ does not continue to appear in the consequent atoms in the next chase steps: $S(a,b,\zeta_1) \myRightarrow{\mc{P}',D} P(a,\zeta_1)$ and $b$ does not appear in $P(a,\zeta_1)$.\boxtheorem
\end{example}

\ignore{
\comlb{Something strange in red above. And you mean null values?}
\commos{I fixed the typo. But about values vs nulls, the stickiness condition is about every value (both nulls and constants), although nulls are the ones that are important for chase termination. Here the example is with value $b$ and not a null value.}
}

The largest of the these \sch$(\select)$ classes is \GSCh, because it imposes the weakest condition on the values that have to be propagated through the chase. More generally, the program class \sch$(\select)$ grows monotonically with $\select$: For selection functions $\select_1$ and $\select_2$ over a same program schema, if $\select_1 \subseteq \select_2$, in the sense that $\select_1(\mc{P} \cup D) \subseteq \select_2(\mc{P} \cup D)$ for every program $\mc{P} \cup D$, then \sch$(\select_1) \subseteq $ \sch$(\select_2)$. This is intuitively clear: the more finite positions are (correctly) identified (and then the less finite positions are treated as infinite), the larger the subclass of \GSCh \ that is identified. \ Accordingly, with \sch$(\select)$ and different selection functions $\mc{S}$ we obtain a range of semantic classes of programs starting with \SCh, ending with \GSCh, as was shown in Figure \ref{fig:someClasses}.

\subsection{Syntactic Program Classes Defined by Selection Functions}\label{sec:selectionSynt}


The semantic classes   \SCh, \wsch, and \jwsch \ in Definition \ref{df:schsSpec} have corresponding syntactic subclasses of programs, which are defined using the same selection functions, plus the marking procedure from Section~\ref{sec:sticky}. \ For \SCh \ and \wsch, they are the classes \stk \ and \ws, introduced in Sections \ref{sec:sticky} and \ref{sec:ws}, resp. For \jwsch, the syntactic class is \jws, of {\em jointly-weakly sticky programs}, which we will introduce in Section \ref{sec:jws}.

In this section, we start in general terms, by defining a range of syntactic program classes \ssch$(\select)$, for syntactic selection functions $\select$. Intuitively, they will correspond to the semantic classes $\sch(\select)$.
\ Given a syntactic selection function $\select$ (as in Definition \ref{def:syn}), the definition of \ssch$(\select)$  follows a pattern similar to that of \ws \ programs: \ (a) it uses the same marking procedure as for \stk \ programs (cf. Section \ref{sec:sticky}), and \ (b) marked join variables are checked for occurrence in positions specified by $\select$.

\begin{definition}\label{df:synt} Given a syntactic selection function $\select$ and a set of rules $\mc{P}$ over the same schema $\schema$, $\mc{P}$ is in \ssch$(\select)$ if and only if, for every rule in it and every repeated variable in its body, the variable is either non-marked or appears in a position in $\select(\mc{P})$. \boxtheorem\end{definition}


By construction, and for example:  \ \stk \ $=$ \ \ssch$(\select^\bot)$, \ws\ $=$ \ \ssch$(\select^\nit{range})$. \ As announced, the semantic class \sch$(\select)$ subsumes the syntactic class \ssch$(\select)$.

\begin{proposition}\label{pr:sch} For every syntactic selection function $\select$ and program $\mc{P}$. If $\mc{P} \in$ \ssch$(\select)$, then, for every EDB $D$ for $\mc{P}$, $\mc{P} \cup D$ is in \sch$(\select)$.\footnote{\ Theorem 3.1 in~\citep{cali12jws} is a special case when $\select=\select^\bot$.}\boxtheorem
\end{proposition}

\hproof{By contradiction, assume that there exists a database $D$ for $\schema$ such that chase of $D$ and $\mc{P}$ does not have the $\select$-stickiness. That means there is a chase step $I_i\myrightarrow{(\sigma_i,\theta_i)}I_{i+1}$ with variables $v$ that occurs more than once in $\nit{body}(\sigma)$, and an atom $A \in I_{i+1}\setminus I_{i}$, for which one of the following holds: $\theta(v) \not \in A$, or there exists $B_1,...,B_k$ such $A <_{\mc{P},D} B_1 <_{\mc{P},D} ... <_{\mc{P},D} B_k$ and $\theta(v) \in B_j, i < k$ but $\theta(v) \not \in B_k$. If $\theta(v) \not \in A$, then $v$ is marked which implies that $\mc{P}$ is not \ssch$(\select)$. Now, assume that $B_k$ is obtained from applying $(\sigma_k,\theta_k)$ with $\theta(v) \in \theta_k(\nit{body}(\sigma_k))$. Clearly, there exists a variable $w$ in $\nit{body}(\sigma_k)$ such that $\theta_k(w) = \theta_i(w)$, but $w$ does not occur in $\nit{head}(\sigma_k)$. Thus, the variable $w$ in $\nit{body}(\sigma_k)$ is marked. Hence, due to the application of the propagation step in the marking procedure, $v$ in $\nit{body}(\sigma_i)$ is marked. This implies that $\mc{P}$ is not sticky, and the claim follows.}

For example, \wsch\ contains the syntactic class \ws \ in the sense that, if $\mc{P} \in$ \ws, then, for every EDB $D$ for $\mc{P}$, \ $\mc{P} \cup D \in$ \wsch. \ Furthermore, the inclusion is proper, i.e. there is a program $\mc{P} \cup D \in$ \wsch\ with $\mc{P} \notin \ws$. \ Similar statements can be made for the classes \stk \ and  \SCh.

\begin{example} \label{ex:sch-non-sticky} The program  $\mc{P}'$ in Example~\ref{example:sticky} is {\em not} (syntactically) sticky: \ $\mc{P'} \notin \stk$. However,  it trivially belongs to \SCh \ with the empty EDB, because   its chase is empty: \
$\mc{P}' \cup \emptyset \ \in \SCh$. \boxtheorem\end{example}

\ignore{
Sticky \dpm \ uses the marking procedure to restrict the repeated body variables and impose the {\em sch-property}. Applying this syntactic restriction only on body variables specified by syntactic selection functions results in syntactic classes that extend sticky \dpm. These syntactic classes are subsumed by the semantic classes defined by the same selection functions; each of these syntactic classes only partially represents its corresponding semantic class. In particular, \SCh \ subsumes sticky \dpm~\citep{cali12}; and \ws \ is a syntactic subclass of \wsch \ (cf. $(g)$ and $(h)$ in Figure~\ref{fig:range-classes}). }

\begin{example} \label{ex:wsch} Consider the program $\mc{P}$ with the set of rules below, to which the marking procedure has been already applied, plus $D=\emptyset$ as EDB.

\vspace{-3mm}
\[ \arraycolsep=0pt
\begin{array}{rl}
R(\hat{x},\hat{y}) &~\rightarrow~ \exists z\;R(y,z),\\
R(x,\hat{y}),R(\hat{y},z) &~\rightarrow~ S(x,y,z),\\
S(x,\hat{y},z) &~\rightarrow~ P(x,z).
\end{array}
\]
\vspace{-1mm}

\noindent $\mc{P} \notin \ws$ \ since $y$ in the second rule is marked and only appears in infinite-rank positions $R[1]$ and $R[2]$. However, $\mc{P} \cup \emptyset \in \wsch$, because the chase is empty.\boxtheorem
\end{example}

It is easy to verify that deciding membership of the syntactic classes of \stk \ and \ws \  can be done in polynomial time  in the program size. Actually, the marking procedure runs in polynomial
time in the size of the program; and the selection functions $\select^\bot$ and $\select^\nit{rank}$ are computable in polynomial time.  More generally, we have:

\begin{proposition} \label{pr:ssch} If a syntactic selection function $\select$ is computable in polynomial time in program size, then membership of \ssch$(\select)$ is decidable in polynomial time in the program size. \boxtheorem\end{proposition}

\subsection{Jointly-Weakly-Sticky Programs} \label{sec:jws}


The class of \jws \ programs to be introduced is based on the syntactic selection function $\select^\exists$ that we introduce in Section \ref{sec:ex-rank}. \jws \ programs are then introduced in Section \ref{sec:jws2}.

\subsubsection{The selection function $\select^\exists$}\label{sec:ex-rank}

The selection function $\select^\exists$ 
appeals to the new notions of $\exists$-rank of a position and the set of finite-existential positions $\finiteExists$. Both are introduced  in Definition~\ref{df:f-ex}. They are similar to the rank of a position, and the set of finite-rank positions, $\finiteRank$, resp., that we reviewed in Section~\ref{sec:wa}.  However, instead of being defined in terms of the {\em dependency graph} (DG) of a program~\citep{fagin} as the latter are, they use the {\em existential dependency graph} (EDG) of a program~\citep{krotzsch}, which was introduced in Section~\ref{sec:ja}.



\begin{definition}[$\exists$-rank and finite-existential positions] \label{df:f-ex}Consider a \dplus \ program with a set of rules $\mc{P}$, and a position $p$ in $\mc{P}$. \ (a) The $\exists$-rank of $p$, denoted by $\exists$-$\nit{rank}(p)$, is the maximum number of nodes in any path in EDG$(\mc{P})$ that ends with some $\exists$-variable $z$ with $p \in T_z$. \ignore{If there is no such a path ending with such $\exists$-variable $z$, $\exists$-$\nit{rank}(p)=1$.} If there is no $\exists$-variable $z$ such that $p\in T_z$, $\exists$-$\nit{rank}(p)=0$. \ (b) The set of {\em finite-existential positions}, denoted by $\finiteExists(\mc{P})$, is the set of positions with finite $\exists$-rank.\boxtheorem
\end{definition}

\begin{example}\label{ex:erank} (ex.~\ref{ex:edg} cont.) The $\exists$-rank of $R[2]$ is $2$ because it is in $T_{z_1}$ and there is a path with nodes $z_2,z_1$ in the EDG in Figure~\ref{fig:edg} that ends with $z_1$. Similarly, the $\exists$-rank of $S[3]$ is $2$, because $S[3] \in T_{z_3}$. The $\exists$-rank of $P[2],R[1],S[2]$ is $1$, because they are in $T_{z_2}$ and the path ending with $z_2$ includes only one variable, i.e. $z_2$. For $S[1]$, $U[1]$ and $P[1]$, their $\exists$-rank is $0$, because there is no $\exists$-variable $z$ such that $S[1] \in T_z$; similarly for $U[1]$ and $P[1]$.\boxtheorem\end{example}





Intuitively, a position in $\finiteExists(\mc{P})$ is not in the target of any $\exists$-variable that may be used to invent infinitely many nulls. Therefore, it specifies a subset of \fp{$\mc{P} \cup D$}, for every EDB $D$ for $\mc{P}$. Accordingly, $\finiteExists(\mc{P})$ determines a syntactic selection function $\select^\exists$, defined by: \  $\select^\exists(\mc{P} \cup D) := \Pi_F^\exists(\mc{P})$.


\ignore{\comlb{The proposition below is not in \citep{krotzsch}, possibly implicitly?}\\
\commos{Yes it is implicit in the definition of existential dependency graphs in~\citep[Sec.~3]{krotzsch}. I mentioned it in the footnote.} }

\begin{proposition} \label{pr:selection} For every program $\mc{P} \cup D$: \ $\finiteRank(\mc{P})\subseteq \finiteExists(\mc{P}) \ \subseteq \fp{\mc{P} \cup D}$.\ignore{\footnote{\ The proposition is mentioned in~\citep[sec.~3 and~4]{krotzsch} but is not proved.}} \boxtheorem\end{proposition}

\hproof{By contradiction, assume there is $p \in \finiteRank(\mc{P})$ with $p \not\in \finiteExists(\mc{P})$. The latter means there is a cycle in EDG$(\mc{P})$ that includes $\exists$-variable $z$ from a rule $\sigma$ and $p \in T_z$. The definition of EDG \ implies that there is a $\forall$-variable $x$ in the body of $\sigma$ for which $B_x \subseteq T_z$. Let $p_z$ and $p_x$ be the two positions where $z$ and $x$ appear in $\sigma$ resp. Then, there is a path in DG$(\mc{P})$ from $p_z$ to $p_x$ and there is also a special edge from $p_x$ to $p_z$ making a cycle including $p_z$ with a special edge. Therefore, $p_z$ has infinite-rank, $p_z \not\in \finiteRank(\mc{P})$. Since $p \in T_z$, we can conclude that $p$ also has infinite-rank, $p \not\in \finiteRank(\mc{P})$, which contradicts the assumption and completes the proof.
\ The second inclusion is also by contradiction. Assume $\exists p \in \finiteExists(\mc{P})$ with $p \not\in \fp{\mc{P} \cup D}$. Then, there is at least one $\exists$-variable $z$ in a rule $\sigma$ that invents infinitely many nulls, and those nulls propagate to $p$. Therefore, $z$ appears in a cycle in \edg \ of $\mc{P}$ and $p$ is in $T_z$, which means $p \not\in \finiteExists(\mc{P})$.}

From Proposition \ref{pr:selection} we immediately obtain:

\begin{corollary} \em For every program $\mc{P} \cup D$: \ $\select^\bot(\mc{P} \cup D) \subseteq \select^\nit{rank}(\mc{P} \cup D) \subseteq \select^\exists(\mc{P} \cup D) \subseteq \select^\top(\mc{P} \cup D)$. \boxtheorem
\end{corollary}

$\select^\exists$ is syntactic, because it only depends on the program, not on the EDB. It is also computable. More precisely, we can decide whether a position $p$  has finite $\exists$-rank by checking whether the $\exists$-variable $z$ the the definition appears in a cycle in the EDG of the program, which can be done in PTIME in  the size of the program.

\subsubsection{\jws \ programs}\label{sec:jws2}

We now introduce the syntactic class \jws \ of programs, and its corresponding semantic class. They  will be particularly relevant in the rest of this work. \ For the next definition we refer to Sections \ref{sec:selectionSynt} and \ref{sec:selection}.

\begin{definition} \label{def:jws} The class \jws \ of {\em join-weakly sticky} programs is \ssch$(\select^\exists)$. The corresponding semantic class, \sch$(\select^\exists)$, contains the programs with the {\em jointly-weakly stickiness-property} of the chase, denoted \jwsch. 
\boxtheorem
\end{definition}

 The inclusion of  \ws \ in \jws \ is an immediate consequence of Proposition~\ref{pr:selection}. It is also strict, as shown in Example \ref{ex:jws}.

\begin{proposition}\label{pr:ws-jws} The class of \ws \ programs is a strict subclass of \jws, i.e. \ws \ $\subsetneqq$ \ \jws. \boxtheorem\end{proposition}

\begin{example} \label{ex:jws} Let $\mc{P}$ be the program below. \ Its DG is shown in Figure~\ref{fig:dg3}, and its EDG has only one node, $z$, without any edge. Then, $\finiteRank(\mc{P})=\{U[1]\}$ and $\finiteExists(\mc{P})=\{U[1],R[1],R[2]\}$.

\begin{minipage}[t]{0.65\textwidth}

\begin{align*}
  \hspace{-1cm}&R(x,y),U(y) ~\rightarrow~ \exists z\; R(y,z).\\
  \hspace{-1cm}&R(x,y),R(y,z) ~\rightarrow~ R(x,z).
\end{align*}

\vspace{3mm}It is easy to check that the marking procedure leaves every body variable marked. \ As a consequence, $\mc{P}$ is not \ws, because marked $y$ in the second rule body does not appear in $U[1]$. However, it is \jws, because all the body positions are finite-existential.\boxtheorem
\end{minipage}\hspace{0.75cm}
\begin{minipage}[t]{0.25\textwidth}
\begin{center}
\vspace{+0.1cm}
\includegraphics[width=3cm]{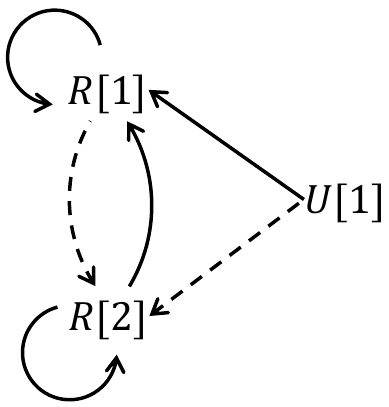}
\end{center}
\vspace{-3mm}
\captionof{figure}{The DG \ of $\mc{P}$}\label{fig:dg3}\vspace{-5mm}
\vspace{0.75cm}
\end{minipage}
\end{example}

\begin{figure}[h]
\centering
\hspace*{-5mm}\includegraphics[width=12cm]{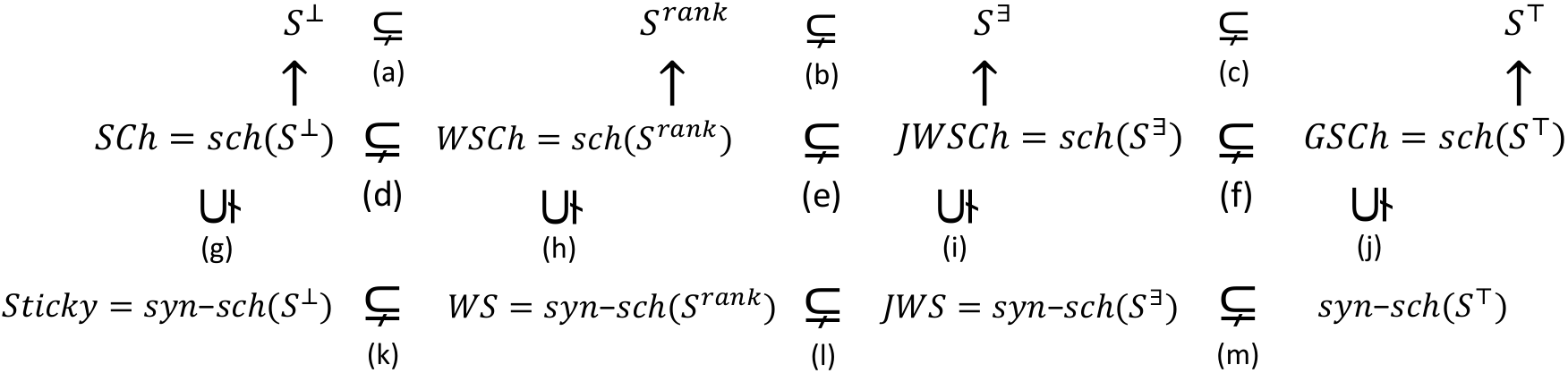}
\caption{Semantic and syntactic program classes, and selection functions}
\label{fig:range-classes}
\end{figure}

By Proposition \ref{pr:sch}, the syntactic class \jws \ has \jwsch \ as a semantic proper super-class. Figure \ref{fig:range-classes} shows the inclusion relationships between the syntactic and semantic program classes in this section. All the inclusions are proper. The non-trivial inequalities in the figure are established through  different examples in this work. Example~\ref{ex:weakly-sticky} shows (d) and (k), while Example~\ref{ex:jws} gives a counter-example to prove (e) and (l). The inequality in (g) is explained in Example~\ref{ex:sch-non-sticky}. Finally, Example~\ref{ex:wsch} proves that the inclusions in (h) and (i) are proper.

\ignore{
\comlb{And the others inequalities in the picture? We might have them, but we need to point to the examples. We do not need to prove or say anything more about the inclusion. So, the remark above: what exactly is shown about
(g) and (j) in that paper? If it is about the inclusion, it is not needed.}

\commos{Assuming this is all about inequalities (not inclusion), I added the part in blue. The figure is also updated. The reference was about inclusion and not the fact that it is a proper inclusion so I removed it as you said.}
}

\ignore{
\begin{figure}[h]
\centering
\includegraphics[width=9.5cm]{axisS-new}\vspace{-2mm}
\caption{Identifying a good syntactic class}\label{fig:quest}
\end{figure}
}

Let us recall that one of the main goals of this work is the identification and characterization of a syntactic class of programs, based on a syntactic and computable selection function, that: \ (a) includes \ws \ programs; \ (b) has tractable QA; \ (c) is closed under magic-set rewriting (c.f. Figure \ref{fig:goals}). \ It turns out that such a syntactic program class is \ssch$(\select^\exists)=$ \ \jws, of {\em jointly-weakly sticky programs}, which we just introduced. \
Tractability of QA for \jws \ and \jwsch \ will be obtained in Section~\ref{sec:qa-ch}; and closure under magic-set rewriting for \jws \ will be established in Section \ref{sec:mg}.

\section{Query Answering for Selection-Based Sticky Classes} \label{sec:qa-ch}

In this section, we present our chase-based, bottom-up QA algorithm, denoted by \schqa \ that is applicable to programs in a semantic class \sch$(\mc{S})$ or in a  syntactic subclass \ssch$(\mc{S})$, where $\mc{S}$ is a fixed selection function. Notice that, in general, $\select$ takes a program $\mc{P}$ and its \edb \ $D$, i.e. $\mc{S}(\mc{P}\cup D)$. However, when $\select$ is syntactic and its result is independent from $D$, we write $\mc{S}(\mc{P})$. Either way, the selection function returns a set of finite positions, which are used by the algorithm, after it ``calls" the selection function.
The algorithm relies on the $\select$-stickiness property of the program. 

Before presenting the algorithm, in Section \ref{sec:alg}, we discuss, in Section \ref{sec:intuitive} and in intuitive terms, the connection between \qa \ and stickiness, for which we  use the notions of {\em proof-tree} and {\em proof-tree schema}. They were introduced in \citep{cali12} to establish the tractability of QA for (semantically or syntactically) sticky programs. In Section \ref{sec:intuitiveS}, that  discussion is extended to the case of $\mc{S}$-sticky programs, providing  the basis for both the QA algorithm under $\mc{S}$-sticky programs, and its proof of correctness. The QA algorithm is presented in Section \ref{sec:alg}.

\subsection{QA and Stickiness}\label{sec:intuitive}

In this section, when we refer to sticky programs, we mean {\em semantically sticky}, in the sense of Definition \ref{df:schsSpec}, that characterizes the class \SCh \ of programs  with the stickiness property of the chase.

In \citep{cali12}, the authors introduce, for a given, possibly open, conjunctive query over a \dplus \ program, the notions of proof-tree and, from the former, that of proof-tree schema. A {\em proof-tree} is a finite tree that shows how an answer to the query is inferred. In it, assuming w.l.o.g. that the query is atomic, the instantiated query atom is placed at the root, the leaves are EDB atoms, and a path goes always from a leaf to the root. A {\em proof-tree schema} (or pattern) represents the general structures of proof-trees for a given query. The authors in~\citep{cali12} show that, for a query over a sticky program, the height of a proof-tree schema has a fixed upper bound that is independent from the EDB. This provides bounds on the number of chase steps needed to reach an answer, which becomes particularly relevant for proving tractability of QA for sticky programs. 

\ignore{\commos{A proof-tree schema does not have a fixed height. For example, when there are multiple proofs, this does not make sense to say the height is a fixed number. I applied  this in the rest of the section. It does not change the discussion much.}
}

In the following we show these notions and discuss them at the light of some examples of sticky programs (cf. \citep{cali12jws} for full details). We do this with the purpose of exploring the extension of those constructs and properties to more relaxed forms of stickiness, as those based on selection functions. 

\begin{figure}
\begin{center}
\includegraphics[width=9.5cm]{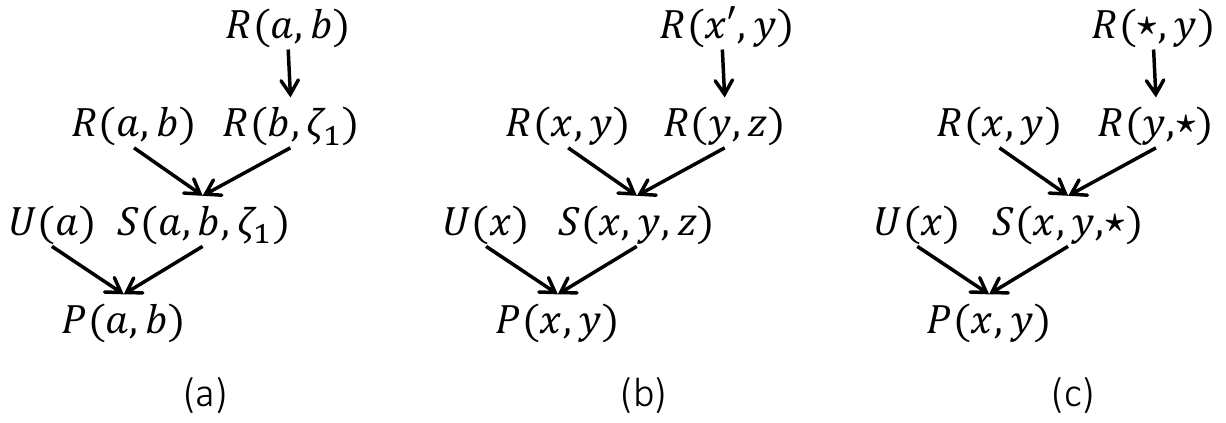}
\end{center}
\vspace{-4mm}
\caption{A proof-tree (a) and its schemas (b) and (c) for $\mc{Q}$.}\label{fig:ptree}
\vspace{-5mm}
\end{figure}

\begin{example} \label{ex:rproof} Consider the program $\mc{P}$ below with \edb \ $D=\{R(a,b),U(a)\}$ and the CQ \
$\mc{Q}: P(x,y)$.

\vspace{-6mm}
\begin{align*}
R(x,y) &\rightarrow \exists z\;R(y,z),\\
R(x,y),R(y,z) &\rightarrow S(x,y,z),\\
U(x),S(x,y,z)&\rightarrow P(x,y).
\end{align*}
It is easy to check that this program is syntactically sticky, and then, also semantically sticky, for the given EDB and any other. 

The query admits answers  w.r.t. $\mc{P} \cup D$, namely $(a,b)$, with a proof-tree for it  shown in Figure \ref{fig:ptree}(a). The derived query atom, $P(a,b)$, appears at the root, which  is an instantiation of the query (a witness for its satisfaction). The
leaves are labeled with atoms in $D$. More than one node in the tree might be labeled with the same atom, and each intermediate node (a ground atom) is generated
via a rule enforcement, and has as children the atoms that participate in a body of that rule, in particular, in a join. \ The proof-tree for $\mc{Q}$ in Figure~\ref{fig:ptree}(a) is one of the possible subtrees of the chase (also conceived as a tree) that reaches the query predicate and answers the query.

Figure~\ref{fig:ptree}(b) shows a {\em proof-tree schema}, that represents how a query atom (or a generic root in a proof-tree) can be inferred via the rules in the program when the leaf nodes in the proof schema are mapped to atoms in an \edb. The proof-tree in Figure~\ref{fig:ptree}(a) is an instance of this proof-tree schema.  \boxtheorem
\end{example}
\vspace{1mm}

Given a program, a query may have different proof-tree schemas, and each proof-tree for an answer is an instantiation of one of them.
Every answer to a query over a program has at least one proof-tree.

The variables and atoms in a proof-tree schema have certain properties we need to discuss. Note that a variable $x$ in (an atom in) a proof-tree schema \red{$\mathcal{T}$} is of either one of two types:\begin{enumerate} \item[(\setword{I}{item:I})] Variable $x$ appears in two  atoms that are not on the same path.\item[(\setword{II}{item:II})] If variable $x$  appears in two different atoms, the atoms  belong to a same path.\end{enumerate} 

\noindent In the proof-tree schema in Figure~\ref{fig:ptree}(b), variables $x$ and $y$ fall in case (\ref{item:I}). Variables $x'$ and $z$ fall in case (\ref{item:II}). 

The first property is that a variable falls under case (\ref{item:I}) only when it appears in a join in the body of a rule that is used to answer the query. In this sense, we sometimes call it ``a join variable". In Figure~\ref{fig:ptree}(b), $y$ and $x$ are join variables, because they  appear in the join between $R(x,y)$ and $R(y,z)$, and, respectively, in  the join between $U(x)$ and $S(x,y,z)$. 

The second property is that there is no pair of atoms $A$ and $B$ in any path in \red{$\mathcal{T}$}, such that $B$ can be transformed into $A$ by locally renaming its variables of type  (\ref{item:II}). For example, in the right-most path in Figure~\ref{fig:ptree}(b), we could not find an atom $R(y,z')$, with $z'$ of type (\ref{item:II}), because it could be transformed into  $R(y,z)$ (in the same path) by renaming $z'$. This property intuitively means a proof-tree schema is a succinct proof for a query answer. For example, if there were $R(y,z)$ and $R(y,z')$ in the same path, we could generate a more succinct proof by removing $R(y,z')$ and every atom between $R(y,z)$ and $R(y,z')$ in the same path, and replacing $z'$ in every other atom with $z$. In other words, we can never find two atoms on a same path of the form $T(\bar{x},\bar{y})$, $T(\bar{x},\bar{z})$ (with the variables occupying the same positions in the predicate), with $\bar{x}$ of type (I), and $\bar{y},\bar{z}$ of type (II).

Using this succinctness property, and the fact that the program's schema is fixed, one can show   that the number of atoms in any path in a proof-tree schema only depends on the number of variables of type (\ref{item:I}) \citep[Lemma 3.3]{cali12}. We can see this by replacing in a proof-tree schema every variable of type  (\ref{item:II}) by a place holder, say $\star$, which is allowed by the fact that these variables do not appear in any other paths, and can be locally replaced. The replacements are shown in Figure Figure~\ref{fig:ptree}(c). \  The maximum number of atoms in a path is bounded above by the all the ways to fill program predicates with variables of type (\ref{item:I}) plus  $\star$.

\ignore{\begin{example} \label{ex:types} (ex.~\ref{ex:rproof} cont.) In the proof-tree schema in Figure~\ref{fig:ptree}(b); variables $x$ and $y$ fall in case (\ref{item:I}). Variables $x'$ and $z$ fall in case (\ref{item:II}).

Now, let us consider the right-most path. In it 
we could not find an atom $R(y,z')$, with $z'$ in case (\ref{item:II}), because it could be transformed into  $R(y,z)$ (in the same path) by renaming $z'$.\boxtheorem\end{example}}

Now, we proceed to analyze variables of type (I) in a proof-tree schema under the assumption that the program is sticky. Notice that the discussion of variables of type (II) above does not make any stickiness assumption.

As mentioned at the beginning of this section, stickiness guarantees that the height of a proof-tree (and a proof-tree schema) for every answer to a \cq \ has an upper bound that is fixed, independent from the \edb. Let us elaborate on this. Stickiness implies that the variables of type (\ref{item:I}) are propagated all the way down to the root, and, as a consequence, the number of type (\ref{item:I}) variables in a proof-tree schema is bounded above by the number of arguments in the query. 

For illustration, in Example~\ref{ex:rproof} and the proof-tree schema in Figure~\ref{fig:ptree}(b); variables $x$ and $y$, both of type (\ref{item:I}), appear in the root atom, which has only two arguments. Therefore, the number of type (\ref{item:I}) variables cannot be larger than two. Since the number of atoms in a path in a proof-tree schema only depends on the number of variables of type (\ref{item:I}), we can conclude that the total number of atoms in any path in the proof-tree schema of a \SCh \ program has a fixed upper bound (which is provided in  \citep[Lemma 3.3]{cali12}).

In Example \ref{ex:dschema}, we will see  a non-sticky program and a query for which  a proof-tree schema has a number of type (\ref{item:I}) variables that depends on the size of the EDB.

As a result of this discussion, we can claim that for an answer to a query over a \SCh \ program, the height of a proof-tree schema has a fixed upper bound. This implies that  QA can be done over an initial portion of the chase of the sticky program, with atoms that are obtained after a fixed number of chase steps. The idea behind the QA algorithm for sticky programs in \SCh \ \ignore{\red{(that actually goes beyond sticky programs, as we will see in the next section)}} consists in exploring a sufficiently large portion of the chase that covers the proof-tree.  \ As we will see in the next section, this kind of analysis of \qa\ on sticky programs and their properties can be generalized to the case of  $\select$-sticky programs.

\subsection{QA and $\mc{S}$-stickiness}\label{sec:intuitiveS}

\ignore{The analysis of \qa\ on sticky programs of the previous section can be generalized to the case of  $\select$-sticky programs, for which the notions of proof-tree and proof-tree schema remain the same. The good properties of proof-tree schemas for sticky programs basically stay the same.} As we saw in the previous section, for sticky programs,  the height of a proof-tree schema has an upper bound that is fixed and  independent from the \edb. It turns out that $\select$-sticky programs enjoy a similar property, with the difference that the upper bound is a fixed number that depends on $\mc{S}$ and the EDB. \ 

In order to show this, consider a proof-tree schema $\mc{T}$ for a query over a program $\mc{P} \cup D$. Given a selection function $\select$, the variables of type~(\ref{item:I}) (of the previous section) in $\mc{T}$ can be divided into two sub-types:
\begin{itemize}
    \item[(\setword{I.1}{subitem:one})] Variables that appear at least once in a position in $\mc{S}(\mc{P} \cup D)$, and
\item[(\setword{I.2}{subitem:two})] Variables that do not appear in these positions.
\end{itemize} 

\begin{figure}
\begin{minipage}[t]{0.48\textwidth}
\begin{center}
\vspace{-0.1cm}
\includegraphics[width=6cm]{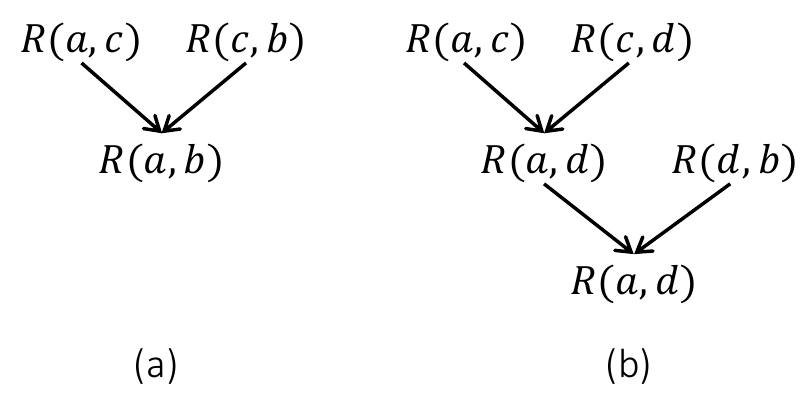}
\vspace{-4.25mm}
\captionof{figure}{The proof-trees of $\mc{Q}$}\label{fig:vtree}\vspace{-5mm}
\vspace{0.75cm}
\end{center}
\end{minipage}
\begin{minipage}[t]{0.48\textwidth}
\begin{center}
\vspace{-0.1cm}
\includegraphics[width=6.2cm]{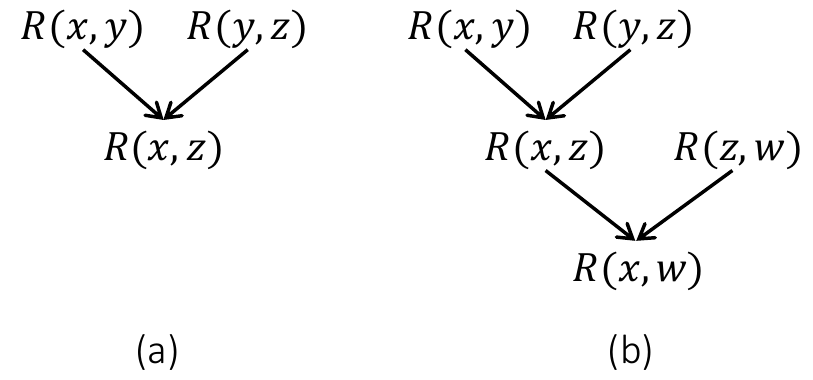}
\vspace{-3mm}
\captionof{figure}{The proof-tree schemas of $\mc{Q}$}\label{fig:vschema}\vspace{-5mm}
\vspace{0.75cm}
\end{center}
\end{minipage}
\end{figure}

\begin{example} \label{ex:subtype} Consider a program $\mc{P} \cup D$, with $\mc{P}$ containing only the rule $R(x,y),R(y,z) \rightarrow R(x,z)$;  $D=\{R(a,c),R(c,d),R(d,b)\}$, and the BCQ query, $\mc{Q}: R(a,b)$, asking if $R(a,b)$ is true. The program is not not in \SCh, because the value $c$ that replaces the join variable $y$ does not appear in $R(a,d)$.

Figures~\ref{fig:vtree}(b) and ~\ref{fig:vschema}(b)  show the proof-trees and the proof-tree schemas for $\mc{Q}$. \ In them, variable $x,w$ are of type (II); and variables $y,z$ are of type (I), because they appear in more than one branch of the trees. The subtypes,
(\ref{subitem:one}) or (\ref{subitem:two}), the latter belong to depend on the selection function $\select$.

Consider $\select=\select^\nit{rank}$. In this case, $\select^\nit{rank}$ contains every position  (of predicates) in $\mc{P}$. Then,  variables $y,z$ in Figure~\ref{fig:vschema}(b) are of sub-type (\ref{subitem:one}), because they appear in $\select^\nit{rank}$-finite positions. In contrast, notice that  for $\select=\select^\bot=\emptyset$, these two variables are of sub-type (\ref{subitem:two}).\boxtheorem\end{example}

For an $\select$-sticky program, the variables of sub-type (\ref{subitem:two}) will appear in the root query atom, and their occurrences are restricted by the query (the same argument as for stickiness above applies to this case). The number of variables of sub-type (\ref{subitem:one}) is limited by the finitely many values in $\select$-finite positions (since the number of values that these variables take is also limited). Therefore, the number of atoms in any path in a proof-tree schema depends on the query, the program's schema and also the number of values that can appear in $\select$-finite positions.  This last number depends on the EDB.


As a consequence, we obtain that for a program in \sch$(\select)$,  the height of a proof-tree schema for a query depends on program's schema, the query, and the number of values in $\select$-finite positions, which in turn depends on the size of the program's \edb. This is illustrated in  Example~\ref{ex:dschema} right below.

\ignore{\comlb{How could it depend on the EDB? In what sense? On the EDB's size? What? I do not understand the ``for different $\select$". The sentence is difficult to parse.}
\commos{The height dependent on the number of values in $\select$-finite positions which depends on the size of \edb. I added the blue sentence and Example~\ref{ex:dschema} to clarify.}
\comlb{It would be good to give a concrete example with a concrete class $\mc{S}$ showing how the schema depends on the EDB. And also the cases i1, i2, ii.}
\commos{I explain this in Example~\ref{ex:dschema}.}  }



\begin{example} \label{ex:dschema}
Consider $\mc{P}$, $D$ and $\mc{Q}$ of Example~\ref{ex:subtype}; and also the EDB $D'=\{R(a,c),R(c,b)\}$. The programs $\mc{P} \cup D$ and $\mc{P} \cup D'$ are not in \SCh, but they are in \sch$(\select^\nit{rank})$. 

The heights of the proof-trees and proof-tree schemas of $\mc{Q}$ w.r.t. $\mc{P} \cup D$ and $\mc{P} \cup D'$ in Figures~\ref{fig:vtree} and ~\ref{fig:vschema} are $3$ and $2$, resp., which means the heights depend on the size of the EDBs.\boxtheorem\end{example}

The discussion in this section shows that, although the chase instance of a program in $\sch(\mc{S})$ may be infinite, QA can be done on a fixed initial portion of it. This is because the height of a proof-tree schema, for any answer, has a fixed upper bound, which may depend on the size of the EDB. In the rest of this section we will make these properties precise. In Section~\ref{sec:alg}, they will be applied to design a QA algorithm for programs in \sch($\select$). It will be based on a query-dependant chase procedure that generates this finite portion, for which the next lemma provided an upper bound. Its proof relies on the considerations we have made so far in this section.

\begin{proposition}\label{prop:ubound} Consider a \cq \ $\mc{Q}$ over a program $\mc{P} \cup D$ in \sch$(\select)$. Let $\mc{T}$ be a proof-tree schema for an answer $t \in \mc{Q}(\mc{P} \cup D)$. An upper bound for the height of $\mc{T}$ is $p\times (s+q+1)^r$, where $p$ is the number of program predicates, $r$ is their maximum arity, $s$ is the number of nulls appearing in  positions in $\select(\mc{P} \cup D)$ during the chase of the program, and $q$ is the number of variables in $\mc{Q}$.\boxtheorem\end{proposition}

\hproof{We find an upper bound on the height of $\mc{T}$ by computing the maximum number of atoms in any path $\beta$ from a leaf node to the root of $\mc{T}$. The number of variables of sub-type (\ref{subitem:two}) in $\beta$ is at most $q$. This is because $\mc{P} \cup D$ is in \sch$(\select)$ which means these variables also appear in $\mc{Q}$. The number of variables of sub-type (\ref{subitem:one}) in $\beta$ is $s$, which is the number of values that these variables take. As discussed in Section~\ref{sec:intuitive}, to count the number of atoms in $\beta$, we can replace every variable of type (\ref{item:II}) by a place holder, $\star$, because these variables do not appear in any other paths. Therefore, the number of possible terms in the atoms in any path of $\mc{T}$ is $s+q+1$. Since there are $p$ predicate names with maximum arity $r$ in $\mc{P}$, we can generate at most $p\times (s+q+1)^r$ atoms with these terms, which will be the upper bound on the length of $\beta$ and also the height of $\mc{T}$.}

Proposition~\ref{prop:ubound} is generic for selection functions $\select$ and specifies an upper bound on the height of the proof-tree schema for programs in a class determined by \sch$(\select)$. The upper-bound depends on $\select$ and $s$. With a more general $\select$, i.e. that returns more positions, the class of programs \sch$(\select)$ is more general and contains more programs. At the same time,  the value of $s$ and the upper-bound $(s+q+1)^r$ increase since $s$ counts values in possibly more positions. This means for programs in a more general class \sch$(\select)$, the height of a proof-tree schema can be larger, and the proof may become more complex. The extreme cases are \sch$(\select^\bot)$ and \sch$(\select^\top)$. In $\SCh=\sch(\select^\bot)$, which is the smallest semantic class, $s=0$, and the upper bound on  proof-tree schemas takes the smallest value. For $\GSCh=\sch(\select^\top)$, which is the most general semantic class, $s$ is maximum, and the upper bound on proof-tree schemas takes the largest possible value. Regarding QA over programs in \sch$(\select)$, this proposition implies that for more general classes of \sch$(\select)$, the chase has to run more steps to cover proofs with larger height.

From the definition of $\mc{S}$-finite position (c.f. Definition  \ref{df:s}),  $s$ in Proposition \ref{prop:ubound} is indeed finite. Neither that definition nor the Proposition give us an upper bound for $s$. For each specific selection function, one has to determine that bound, if possible. However, for some of them we know such a bound. 

\ignore{
\comlb{Please, check the statement below. I ma not sure if we have to talk about the semantic or the syntactic class. As stated, it seems to refer to the semantic class. Should we talk about $\mc{P}$ or $\mc{P} \cup D$? Is the underlined part O.K.? In the next result you are not assuming that the program belongs to the class you are talking about.}\\
\commos{Both Lemmas hols for the general \dplus class. Even if a program has infinite chase, the number of nulls in these positions are bounded, so I removed the semantic class in Lemma~\ref{lemma:new}.}\\
\commos{About $\select^\nit{rank}(\mc{P}\cup D)$ vs $\select^\nit{rank}(\mc{P})$, both are correct, since the selection functions are syntactic. But, since we mentioned at the beginning of Section~\ref{sec:qa-ch} that we use $\select(\mc{P})$ for syntactic functions, I changed to $\select^\nit{rank}(\mc{P})$.}
}

\begin{lemma} \label{lemma:new}
For a \dplus \ program $\mc{P} \cup D$\ignore{\underline{in  $\sch(\select^\nit{rank})$}}, the number  $s$ of nulls appearing in positions in $\select^\nit{rank}(\mc{P})$ is polynomially bounded  in the size of $D$.  Actually, the number of values (constants or nulls) in  positions of $\select^\nit{rank}(\mc{P})$ during the chase is $O(n^{v\times k})$, where $n$ is the number of constants in $D$, $v$ is the maximum number of variables in a rule in $\mc{P}$, and $k$ is the maximum rank of the positions in $\select^\nit{rank}$ (c.f.  Section~\ref{sec:wa}).\boxtheorem
\end{lemma}

The proof of Lemma~\ref{lemma:new} is implicit in that of~\citep[Theorem 3.9]{fagin}, which establishes when $\mc{P}$ is \wa \  that the number of values in the chase of $\mc{P}\cup D$ is $O(n^{v\times k})$. Notice that Lemma~\ref{lemma:new} does not require the program to belong to \wa \ or  $\sch(\select^\nit{rank})$. This is because the lemma is limited to the positions of $\select^\nit{rank}(\mc{P})$, unlike \citep[Theorem 3.9]{fagin} that does not restrict the positions. Also notice that in this lemma we use $\select^\nit{rank}(\mc{P})$, and not $\select^\nit{rank}(\mc{P}\cup D)$, because $\select^\nit{rank}$ is a syntactic selection function that depends only on the program without the EDB. For the same reason, we use $\select^\exists(\mc{P})$ in Lemma~\ref{lemma:ptime} below, where  we establish a similar upper bound for $\select^\exists$. This upper bound depends on $k_\exists$ that is the maximum $\exists$-rank of positions in $\select^\exists$ (c.f. Definition~\ref{df:f-ex}).

\begin{lemma}\label{lemma:ptime} For a \dplus \ program $\mc{P} \cup D$, the number of distinct values (constants or nulls) in $\nit{chase}(\mc{P} \cup D)$ that appear at least once in a position in $\select^\exists(\mc{P})$ is polynomially bounded above by the size of $D$; actually by $O(n^{v\times k_\exists})$, where $n$ is the number of constants in $D$, $v$ is the maximum number of variables in a rule in $\mc{P}$, and $k_\exists$ is the maximum $\exists$-rank of a position in $\mc{S}^\exists(\mc{P})$.\ignore{length of a path in EDG$(\mc{P})$ that ends with an $\exists$-variable in a position in $\mc{S}^\exists$. (We recall that EDG$(\mc{P})$ is the existential dependency graph of $\mc{P}$; c.f.  Section~\ref{sec:ja}.)}\boxtheorem\end{lemma}

\hproof{For the proof, we first partition the positions in $\mc{P}$ into $\Pi_0,\Pi_1,...,\Pi_{k_\exists}$, where $\Pi_i$ is the set of positions with the $\exists$-rank $i$, and $k_\exists$ is the maximum $\exists$-rank, which is bounded by the total number of positions in $\mc{P}$. Let $d_i$ be the number of values that appear in the positions of $\Pi_i$ during the chase of $\mc{P} \cup D$. We prove by induction on $i$ that $d_i$ is polynomial in $n$, i.e. the number of constants in $D$:

\vspace{1mm}
\noindent {\bf Base case:} $d_0$ is $O(n)$ with $n$ because there is only constants from $D$ in positions of $\Pi_0$.

\vspace{1mm}
\noindent {\bf Inductive step:} If for every $j < i$, $d_j$ is a polynomial function $P_j(n)$, then $d_i$ is also a polynomial function $P_i(n)$. To prove this inductive step, consider the following three cases for a value, constant or null, that appears in a position of $\Pi_i$: (a) it is a constant that appears in a position of $\Pi_i$ in an atom in $D$, (b) it is a null or a constant that is copied from a position of $\Pi_j$ to a position in $\Pi_i$, or (c) it is a null that is invented by an $\exists$-variable $z$ in a position in $\Pi_i$. An upper bound for the number of terms in (a) is $n$. By the inductive hypothesis, the number of values in (b) is at most $K_i(n)=P_{i-1}(n)+P_{i-2}(n)+...+P_0(n)$ which is a polynomial function in $n$. For (c), any such variable $z$ appears at the end of at least one path of length $i$ in the \edg \ of $\mc{P}$. Let $\sigma$ be a rule containing such a $\exists$-variable, $z$. The values in $\nit{body}(\sigma)$ are in positions with $\exists$-rank less than $i$. Let $v$ be the maximum number of variables in the body of any rule in $\mc{P}$. Then, $\sigma$ can invent $K_i(n)^v$ new values in the positions of $\Pi_i$ since each variable can be replaced with $K_i(n)$ values from $\Pi_j$. If there are at most $w$ rules in $\mc{P}$ and each rule can have at most $r$ existential variables, which is the maximum arity of the predicates in $\mc{P}$, then there are at most $w\times r\times K_i(n)^v$ distinct values in the positions of $\Pi_i$ which is polynomial in $n$. Putting these together, $d_i$ is at most $n + K_i(n) + w\times r\times K_i(n)^v$.  Applying the recursive definition of $K_i$, we can conclude that $P_i(n)=O(n^{i.v})$, and for $i={k_\exists}$, $d_{k_\exists}$ is a polynomial function $P_{k_\exists}(n)=O(n^{{k_\exists}.v})$.}

\ignore{\comlb{The beginning of what follows here is the same that what we have at the beginning of the proof above. That part could be removed from the proof. However, seeing such a long proof one wonders if the proof is that similar to the one in the other paper. }
\commos{I removed that part from the beginning of the proof. The reason it is long is because I basically repeated the proof with $\mc{S}^\exists(\mc{P})$ rather than $\mc{S}^\nit{rank}(\mc{P})$ and the small changes in the proof. I added a sentence to the blue paragraph below to clarify.}
}

The proof of Lemma~\ref{lemma:ptime} is based on the proof of Theorem 3.9 in ~\citep{fagin}. The main difference is that Lemma~\ref{lemma:ptime}  is about positions in $\mc{S}^\exists(\mc{P})$, whereas the theorem in~\citep{fagin} is about \wa \ programs,  and  positions in $\mc{S}^\nit{rank}(\mc{P})$. We provide the complete proof of Lemma~\ref{lemma:ptime} here to make it clear how $\mc{S}^\exists(\mc{P})$ positions are used in the proof.
\ Notice that, similar to Lemma~\ref{lemma:new}, Lemma~\ref{lemma:ptime} does not require $\mc{P}\cup D$ to be in \wa \ or \sch$(\mc{S})$, and it can be any  \dplus \ program.

From Proposition~\ref{prop:ubound} we conclude that,  when the number of null values in $\select$-finite positions is polynomially bounded above by the size of $D$, then the height of a proof-tree schema is also polynomially bounded above by the size of $D$. Now, from Lemmas \ref{lemma:new}  and  \ref{lemma:ptime}, we conclude that   this is the case for $\select^\nit{rank}$-finite positions~and   $\select^\exists$-finite positions, respectively. For the two associated program classes,  Corollary \ref{cl:ptime} below gives us  explicit upper bounds for the height of proof-tree schemas.

\begin{corollary}\label{cl:ptime}  For a \cq \ $\mc{Q}$ over a program $\mc{P} \cup D$ in \sch$(\select^\nit{rank})$ or  \sch$(\select^\exists)$,  the height of a proof-tree schema for an answer in $\mc{Q}(\mc{P} \cup D)$ is polynomially bounded above by the size of $D$. More precisely, an upper bound is $O(n^{v\times k\times r})$ for $\select=\select^\nit{rank}$, and $O(n^{v\times {k_\exists}\times r})$ for $\select=\select^\exists$. \boxtheorem \end{corollary}

Theorem~\ref{th:decidable} concludes our discussion about the connection between \qa \ and $\select$-stickiness, and it summarizes the results in Proposition~\ref{prop:ubound}, Lemma~\ref{lemma:ptime}, and Corollary~\ref{cl:ptime}. While we state the theorem for semantic program classes \sch($\mc{S}$), the same statement holds for the associated syntactic sub-classes \ssch$(\select)$.

\ignore{\comlb{I have to admit after all this time, I am a bit confused. Isn't it that a selection function computes (or provides) a subset of the finite positions? So, in principle, it could return the empty set. How do we reconcile this with the computability/tractability of QA? Are we assuming $\mc{S}$ is sound and complete? Are S-rank and S-exists sound and complete or only sound? Maybe we should talk on zoom ... See also the beginning of next section.} }

\begin{theorem}\label{th:decidable} Consider \ignore{a \cq \ $\mc{Q}$ over} a program $\mc{P}\cup D$ in \sch($\mc{S}$). The following holds: 

\begin{enumerate}[(a)]
    \item If the selection function $\select$ is computable, then QA over $\mc{P}\cup D$ is decidable.
    \item If the computation of $\mc{S}(\mc{P} \cup D)$ is tractable in the size of $D$, and the number of values (constants or nulls) that appear in the positions in $\mc{S}(\mc{P} \cup D)$ during the chase of $\mc{P} \cup D$ is polynomially bounded above  by the size of $D$, then QA over $\mc{P} \cup D$ is also tractable in the size of $D$.
    \item In particular, when $\select=\select^\exists$ or $\select=\select^\nit{rank}$, QA over $\mc{P} \cup D$  can be done in polynomial time in the size of $D$.
\end{enumerate}
\end{theorem}

\hproof{(a) follows from  Proposition~\ref{prop:ubound} that gives an upper-bound for the height of a proof-tree schema for an answer to a \cq \ over a program in \sch($\select$). This means that the proof can be mapped to a fixed initial portion of the chase. Therefore, QA is decidable for \sch($\select$). Now,  (b) follows from Proposition~\ref{prop:ubound} and the fact that $s$, and then also  the upper bound $p\times (s+q+1)^r$, are polynomially bounded above in the size of \edb. \ignore{\red{The tractability result is obtained from Corollary~\ref{cl:ptime}} that further show \qa is in PTIME for \sch($\select^\exists$) and \sch($\select^\nit{rank}$) since the upper-bound is polynomial in the size of \edb. } Finally,  (c) follows from (b), Corollary~\ref{cl:ptime}, and the fact that computing $\mc{S}(\mc{P} \cup D)$ for $\select=\select^\exists$ or $\select=\select^\nit{rank}$ can be  done in constant time w.r.t. the size of \edb. The last claim holds because $\select^\exists$ and $\select^\nit{rank}$ are syntactic functions, and then,  independent from the \edb.}

In this section, we provided a comprehensive complexity analysis of QA over programs in $\sch(\select)$. We showed $\select$-stickiness for a computable selection function $\mc{S}$ makes QA decidable, and under certain conditions on $\select$, stated in Theorem~\ref{th:decidable}(b), QA becomes tractable. In the next section, we provide a QA algorithm based on the results in this section. It works for the general  $\sch(\select)$ class, while its runtime depends on the selection function $\select$.

\ignore{
\comlb{OLD: People may wonder why we went through all the pain of establishing precise upper bounds if they are not going to be used, e.g. to have a limit to an implemented chase. Let's see if  we can come up later with something.} }

\subsection{The \schqa \ Algorithm} \label{sec:alg}

\schqa \ is a QA algorithm for \dplus \ programs in $\sch(\mc{S})$, where $\mc{S}$ maps programs with their EDBs to sets of finite positions (but not necessarily all finite positions). The algorithm  is parameterized by (or calls as a subroutine) the selection function $\mc{S}$, which can be computed when it is computable or seen as an oracle, otherwise. The algorithm accepts as input a
program $\mc{P} \cup D \in \sch(\mc{S})$ and 
a \cq \ $\mc{Q}$, and returns $\mc{Q}(\mc{P} \cup D)$. The query may contain free variables.

\ignore{
\comlb{Is there an algorithm for each $\mc{P}$ or a generic algorithm for the combo? Asking in relation to data complexity ... Something else: is the query Boolean? If yes (at least for now), we should say it here, and when you talk about the answers below, we should say that we are not talking about the final answers, yes or no, but about the witnesses for existentials in the positive cases.}
\commos{The input contains both $\mc{P},D,\mc{Q}$. I understand that data complexity assumes $D$ is the only input, but I think it will be easier if we present the algorithm in general for $\mc{P},D,\mc{Q}$ and we explain that the data complexity assumes $\mc{P}$ and $\mc{Q}$ are fixed.} 
\commos{It is for general free CQs, not only Boolean. It will generate enough portion of the chase to properly answer a given free CQ.} }

The algorithm runs first what we call the \qchase{\mc{Q}}{\mc{S}} procedure, which is a modified version of the classic chase with $\mc{P} \cup D$, that now generates an initial, finite, and $\mc{Q}$-dependent portion of the (classic) chase instance of $\mc{P} \cup D$. This portion of the chase includes the ground atoms in the proof-trees for the answers to query $\mc{Q}$. Furthermore, \qchase{\mc{Q}}{\mc{S}} differs from the classic chase only in  that it considers a more restrictive condition for the application of a chase step, which guarantees  termination. 
\ After running the \qchase{\mc{Q}}{\mc{S}}, \schqa \ computes the answers to $\mc{Q}$ over this finite portion of the chase, as a regular query posed to a finite instance.

\ignore{
\comlb{Is the last statement correct, or there is more?}
\commos{Yes, is the common term regular or free? I meant here the query is not necessarily Boolean and can return a set of answers rather than just true and false.}  }

In order to define the \qchase{\mc{Q}}{\mc{S}}\ignore{, and then \schqa}, we need  first the notions of {\em homomorphic atoms} and {\em freezing a null}. (C.f. Section \ref{sec:relational} for the definition of homomorphism.)

\begin{definition}[$\Pi$-homomorphism and freezing nulls] Let $\mc{R}$ be a program schema, and $\Pi$ a set of predicate positions. \label{df:freezing}\ (a)
Given two ground atoms $A$ and $B$, i.e. containing only constants or nulls, $A$ is {\em $\Pi$-homomorphic to} $B$ if there is a homomorphism $h:\{A\} \rightarrow \{B\}$\ignore{ (as defined in Section~\ref{sec:preliminaries})} (in particular, the atoms share the predicate and $h(A)=B$), and $h$ is the identity on terms in positions in $\Pi$.   

(b) {\em Freezing a null} $\zeta \in \nulls$ in an \ignore{atom $A$} instance $I$ means replacing every occurrence of $\zeta$ in \ignore{$A$} $I$ with a constant $\zeta^f \in \constants$ (assuming the set of constants is extended with these fresh constants that do not appear anywhere in the initial EDB or the program). \ 
\ignore{(c) Freezing a null $\zeta$ \ignore{that appears in some atom} in an instance $I$ means replacing every occurrence of $\zeta$ in atoms in $I$ with the same constant $\zeta^f$.}
\boxtheorem\end{definition}

Notice that $A$ is {\em homomorphic} to $B$ if $A$ is  $\Pi$-homomorphic to $B$ with $\Pi = \emptyset$ or $\Pi$ does not contain positions of $A$.   Freezing a null in an atom $A$ means freezing the null in instance $\{A\}$.

\begin{example} The ground atom $S(a, \zeta,\zeta)$ is $\{S[1]\}$-homomorphic to $S(a,b,b)$, with  $h=\{ a \mapsto a, \zeta \mapsto b\}$, but it is  not $\{S[2]\}$-homomorphic. \ Atom $S(a,b,b)$ is not $\{S[1]\}$-homomorphic to $S(a, \zeta,\zeta)$. Atom $S(a, \zeta,\zeta)$ is not homomorphic to $S(a,b,c)$.

Freezing the null  $\zeta$ in $S(a, \zeta,\zeta)$ means, in practical terms, treating $\zeta$ in it  as a constant. This may have an impact on possible homomorphisms that involve the atom. For example, after replacing $S(a,\zeta,\zeta)$ by $S(a,\zeta^f,\zeta^f)$, with $\zeta^f$ a constant, $S(a,\zeta^f,\zeta^f)$ is not homomorphic to $S(a,b,b)$ anymore, because $\zeta^f$ and $b$ are (syntactically) different constants.\boxtheorem\end{example}

\begin{definition}[Applicable rule-assignment pair]\label{df:app} Consider a \dplus \ program $\mc{P} \cup D$ and an instance $I\supseteq D$. A  rule-assignment pair $(\sigma,\theta)$, with $\sigma \in \mc{P}$, is $\select(\mc{P} \cup D)$-{\em applicable} over $I$ if: \ (a) $\theta(\nit{body}(\sigma)) \subseteq I$; and (b) there is an assignment $\theta'$ that extends $\theta$, maps the $\exists$-variables of $\sigma$ into nulls that do not appear in $I$ (i.e. they are fresh nulls), and $\theta'(\nit{head}(\sigma))$ is {\em not $\select(\mc{P} \cup D)$-homomorphic} to any atom in $I$. \boxtheorem\end{definition}

 When $\mc{S}$ and $\mc{P} \cup D$ are clear from the context, we will simply say ``the rule is applicable". 
Typically, $I$ will be a finite portion of $\nit{chase}(\mc{P}, D)$. For an instance $I$ and a program $\mc{P}$, we can systematically compute the applicable rule-assignment pairs by first finding $\sigma \in \mc{P}$ for which $\nit{body}(\sigma)$ is satisfied by $I$. That gives an assignment $\theta$ for which $\theta(\nit{body}(\sigma)) \in I$. Next, we construct a $\theta'$ according to  Definition~\ref{df:app}, \ignore{ and we iterate over} and we check for each  atom in $I$ that  there is no $\select(\mc{P} \cup D)$-homomorphism from $\theta'(\nit{head}(\sigma))$.

\begin{example}\label{ex:applicable}Consider a program $\mc{P} \cup D$ with $D=\{P(a,b)\}$ and the rule:
\begin{align}
\sigma: P(x,y) \rightarrow \exists z P(y,z).    
\end{align}
\noindent Also consider  instance $I=D \cup\{P(b,\zeta)\}$, and the selection function $\mc{S}^\nit{rank}$. There is no finite position according to the selection function, i.e. $\mc{S}^\nit{rank}(\mc{P})=\emptyset$. The rule-assignment $(\sigma,\theta)$ with $\theta=\{x\mapsto b, y\mapsto\zeta\}$ is not applicable over $I$, because any extension $\theta'(\nit{head}(\sigma))=P(\zeta,\zeta')$ is $\emptyset$-homomorphic to $P(b,\zeta) \in I$. \ignore{\blue{Here, we use the term homomorphic for $\emptyset$-homomorphic.}}\ignore{ where $\theta'=\theta\cup\{z\mapsto \zeta'\}$ and $\zeta'$ is a fresh null. }

\ignore{
\comlb{There is no selection function here above.}
\comlb{Here below I do not understand. You defined (above) homomorphism from the head to the other atoms, and in this case, there are homomorphisms. I am lost.}
\commos{In the example, $\select^\nit{rank}({\mc{P}})=\emptyset$, $\select^\nit{rank}(\mc{P})$-homomorphism becomes $\emptyset$-homomorphism, which reduces to the usual homomorphism. And also here $\select^\nit{rank}(\mc{P}\cup D)$ is equal to $\select^\nit{rank}(\mc{P})$sine the selection function $\select^\nit{rank}$ is independent of data and syntactic.}
}

Freezing $\zeta$ in $I$ by replacing it with the constant $\zeta^f$ makes $(\sigma,\theta)$ applicable since a head extension of the form  $P(\zeta^f,\zeta')$ is not $\emptyset$-homomorphic to any of the atoms $P(a,b)$ or $P(b,\zeta^f)$ in $I$.\boxtheorem\end{example}

The technique of freezing nulls was  first used in~\citep{leone} for \qa \ over \nit{shy} \dplus \ programs. The modified chase procedure we are about to introduce is based on {\em the parsimonious chase} for Shy Programs~\citep{leone}.

We present now our new chase procedure, \qchase{\mc{Q}}{\mc{S}}. It is a modified chase that produces a finite instance. \qchase{\mc{Q}}{\mc{S}} appeals to the notions of freezing nulls and $\Pi$-homomorphism of Definition~\ref{df:freezing}, and rule applicability of Definition~\ref{df:app}. 

\begin{definition} \label{def:newCh}Given a \cq \ $\mc{Q}$ over a program $\mc{P} \cup D$ and a selection function $\mc{S}$, $\qmchase{\mc{Q}}{\mc{S}}{\mc{P} \cup D}$ is the instance $I$ that is obtained from $D$ after iteratively applying the following steps, with $I$ initially equal to $D$: (This is the \qchase{\mc{Q}}{\mc{S}} procedure.)

\vspace{-0mm}
\begin{itemize}
\item[] {\bf Step~\setword{1}{item-app}.} For every $\select(\mc{P} \cup D)$-applicable rule-assignment pair $(\sigma,\theta)$ over $I$, add $\theta'({\it head}(\sigma))$ to $I$ (c.f. Definition~\ref{df:app}). Go to Step~\ref{item-resume} if all the applicable pairs are applied, producing a possibly extended instance $I$.
\item[] {\bf Step~\setword{2}{item-resume}.} ({\em resumption step}) Freeze every null in $I$ and go to Step~\ref{item-app}.  Apply resumption $M_\mc{Q}$ times, where $M_\mc{Q}$ is the number of $\exists$-variables in $\mc{Q}$. \boxtheorem
\end{itemize}
\end{definition}

Notice that this chase does not have anything like an ``unfreezing" step. What was frozen stays frozen.

\ignore{\comlb{It is clear now. But I see there is nothing like   ``unfreezing" nulls, i.e. what was frozen at some point stays like that until the end? Am I missing something?}
\commos{We do not unfreez them. The reason for freezing is to help us know how much of the chase we need to generate for QA and when we should stop. For QA, we will treat frozen and non-frozen nulls the same way.}
}

\ignore{\commos{I fixed Step 2. Now I only use the term resumption and I applying it which I think is clear from the definition. With the new changes, it should be clear when we move on from step 1.} }

So as the usual the chase procedure\ignore{, which we reviewed in Section~\ref{sec:dpm}}, the \qchase{\mc{Q}}{\mc{S}} procedure applies a pair of rule-assignment only once. Furthermore, the \qchase{\mc{Q}}{\mc{S}} procedure applies rule-assignments in the same order as the usual chase procedure. The main difference between  \qchase{\mc{Q}}{\mc{S}} and the latter resides in the applicability condition in Step~\ref{item-app} that uses $\select$ to check $\select(\mc{P} \cup D)$-homomorphism. This requires the computation of the $\select(\mc{P} \cup D)$ positions. For computability and complexity analysis, we assume this computation is done at once by an oracle that runs $\select$ in constant time w.r.t. $D$. 

The \qchase{\mc{Q}}{\mc{S}} procedure is a partial chase procedure in the sense that the \qchase{\mc{Q}}{\mc{S}} instance is a subset of the usual chase instance modulo renaming nulls. This is because any pair of rule-assignment that is applicable in the \qchase{\mc{Q}}{\mc{S}} procedure is also applicable in the usual chase; the applicability condition in \qchase{\mc{Q}}{\mc{S}} extends the applicability condition in the usual chase. \ignore{Examples~\ref{ex:alg} and \ref{ex:algS} show two examples of running the \qchase{\mc{Q}}{\mc{S}} procedure.}

\begin{example} \label{ex:alg} Consider a query $\mc{Q}(x):\exists y\;R(x,y)$ over a program $\mc{P} \cup D$ with rules as below, and the \edb \ $D= \{P(a,b)\}$.

\ignore{\comlb{You should say why you have the variables marked here. Also, why the program is in \sch$(\select^\bot)$.}
\commos{The blue parts address this issue.} }

\vspace{-4mm}
\begin{align*}
\sigma_1:&\hspace{1cm}P(\hat{x},\hat{y})~\rightarrow~\exists z\;P(y,z),\\
\sigma_2:&\hspace{1cm}P(x,y),P(y,\hat{z})~\rightarrow~\;R(x,y).
\end{align*}

\noindent The program is sticky, because there is no repeated marked variable in a body. Then, it is in \sch$(\select^\bot)$, i.e. $\select = \select^\bot$. As a consequence, the positions we have to consider for rule applicability are those in  $\mc{S}^\bot(\mc{P})=\emptyset$.

The \qchase{\mc{Q}}{\mc{S}} runs as follows.  It starts from $I:=D$. The pair $(\sigma_1,\theta_1)$ with $\theta_1:x\!\mapsto\! a,y\!\mapsto \!b$ is applicable; and the procedure adds $P(b,\zeta_1)$ to $I$. The next applicable pair is $(\sigma_2,\theta_2)$ with $\theta_2: x\!\mapsto\!a, y\!\mapsto\!b, z\!\mapsto\!\zeta_1$ and adds $R(a,b)$ to $I$. The pair $(\sigma_1,\theta_3)$ with $\theta_3: x\!\mapsto\!b, y\!\mapsto\!\zeta_1$ is not applicable because it generates $P(\zeta_1,\zeta_2)$ that is homomorphic to $P(a,b)$. 

The query has an $\exists$-variables $y$, $M_\mc{Q}=1$. So, the procedure continues with Step~\ref{item-resume} of Definition~\ref{def:newCh}, by freezing $\zeta_1$, i.e. replacing it by the constant $\zeta_1^f$. 

As a result, $(\sigma_1,\theta_3)$ becomes applicable, and adds $P(\zeta_1^f,\zeta_2)$, that consequently makes $(\sigma_1,\theta_4)$, with $\theta_4: x\!\mapsto\!b,y\!\mapsto\!\zeta_1^f,z\!\mapsto\!\zeta_2$, applicable, and adds $R(b,\zeta_1^f)$. Note that, after $\zeta_1$ is frozen, $(\sigma_2,\theta_2)$ is not applied again.

The procedure stops since there is no applicable pair that  is not already applied, and the only allowed resumption is applied. The result of the procedure is $I=D\cup\{P(b,\zeta_1^f),R(a,b),P(\zeta_1^f,\zeta_2),R(b,\zeta_1^f)\}$.

The answer to $\mc{Q}$ over $I$ contains $a$ and $b$, i.e. $\mc{Q}(I)=\{a,b\}$. We will show in Theorem~\ref{th:correctness} that this is equal to the answer from the program $\mc{P}\cup D$, i.e. $\mc{Q}(\mc{P}\cup D)=\mc{Q}(I)$, which means the \qchase{\mc{Q}}{\mc{S}} can be used for QA over the program $\mc{P}\cup D$.\boxtheorem \end{example}

\ignore{\comlb{Shouldn't we illustrate, at least informally, what we do now with the query?}
\commos{Yes, I added the blue sentences at the end of the example.}
}

Example~\ref{ex:alg} shows running the \qchase{\mc{Q}}{\mc{S}} procedure with $\select = \select^\bot$ that determines the simplest (or better, smaller) syntactic (sticky) and semantic (\sch$(\select^\bot)$) program classes. This allowed us to easily illustrate the applicability condition, and the resumption step. In the next example, we show the \qchase{\mc{Q}}{\mc{S}} procedure with other selection functions, to show the impact of $\select$ on the procedure, and QA.

\begin{example} \label{ex:algS} Consider the  program $\mc{P} \cup D$ with rules as below and \edb \ $D= \{P(a,b), P(b,c),V(b), V(c) \}$, and the  query $\mc{Q}(x):U(x)$.
\end{example}

\vspace{-6mm}
\begin{align*}
\sigma_1:&\hspace{1cm}P(x,y),V(y)~\rightarrow~\exists z\;P(y,z), \\
\sigma_2:&\hspace{1cm}P(x,y),P(y,z)~\rightarrow~\;U(x).
\end{align*}
We consider below two different selection functions. In the first case, the program does belong to the associated semantic program class, but it the second, it does not.

\vspace{2mm}
\noindent (a) If $\select=\select^\exists$, $\select(\mc{P})$ contains every position in $\mc{P}$,  because the \edg \ of $\mc{P}$ --- a simple graph that we do not show as it only contains one node representing $z$ in $\sigma_1$ and it does not have any edges --- is cycle-free, and therefore all the positions have finite $\exists$-rank. The program is trivially in \sch$(\select)$.

\ignore{\vspace{-4mm}
\begin{align*}
\sigma_1:&\hspace{1cm}P(\hat{x},\hat{y}),V(\hat{y})~\rightarrow~\exists z\;P(y,z), \\
\sigma_2:&\hspace{1cm}P(x,\hat{y}),P(\hat{y},\hat{z})~\rightarrow~\;U(x).
\end{align*}
\begin{wrapfigure}{r}{3.5cm}
\begin{center}
\vspace{-6mm}
\includegraphics[width=3cm]{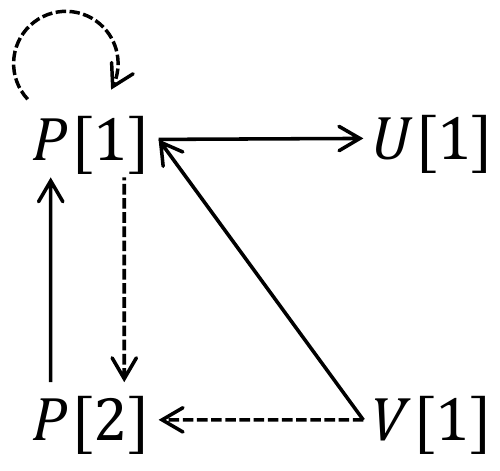}
\vspace{-3mm}
\captionof{figure}{The DG \ of
$\mc{P}$}\label{fig:dg-chase}
\vspace{-6mm}
\end{center}
\end{wrapfigure}
}

 The \qchase{\mc{Q}}{\mc{S}} runs as follows. It starts from $I:=D$. The pairs $(\sigma_1,\theta_1)$, with $\theta_1:x\!\mapsto\! b,y\!\mapsto \!c$, and $(\sigma_2,\theta_2)$, with $\theta_2:x\!\mapsto\! a,y\!\mapsto \!b,z\!\mapsto \!c$, are applicable. With them, the procedure adds $P(c,\zeta_1)$ and $U(a)$ to $I$. Notice that the pair $(\sigma_1, \{x \mapsto a, y \mapsto b\})$ is not applicable because of $P(b,c)$. 
 
 The next applicable pairs are $(\sigma_1,\theta_3)$, with $\theta_3: x\!\mapsto\!c, y\!\mapsto\!\zeta_1$, and $(\sigma_2,\theta_4)$, with $\theta_4: x\!\mapsto\!b, y\!\mapsto\!c, y\!\mapsto\!\zeta_1$.  They add $P(\zeta_1,\zeta_2)$ and $U(b)$ to $I$. Finally, the pair $(\sigma_2,\theta_5)$, with $\theta_5: x\!\mapsto\!c, y\!\mapsto\!\zeta_1, z\!\mapsto\!\zeta_2$, becomes applicable, and adds $U(c)$ to $I$.
 
There are no more applicable pairs, and we continue with Step~\ref{item-resume}. In this case, the \qchase{\mc{Q}}{\mc{S}} chase does not apply any resumptions since $\mc{Q}$ does not have any $\exists$-variables. The final instance is $I=D\cup\{P(c,\zeta_1),U(a),U(b),P(\zeta_1,\zeta_2),U(c)\}$. The instance $I$ correctly answers $\mc{Q}$, i.e. $\mc{Q}(\mc{P} \cup D)=\mc{Q}(I)=\{a,b,c\}$.

\vspace{2mm} \noindent (b) Now let us consider $\select=\select^\nit{rank}$, that, in general, determines a smaller class of programs than  $\select^\exists$. In this case,   $\select^\nit{rank}(\mc{P} \cup D)=\{V[1]\}$. More specifically, the DG of $\mc{P}$ in Figure~\ref{fig:dg-chase} includes cycles with special edges. From this, $P[1],P[2]$ and $U[1]$ have infinite rank: these cycles create paths of infinite size that end at these positions. However, there is no such paths ending with $V[1]$, and therefore, the rank of $V[1]$ is $0$. The program does not belong to  \sch$(\select^\nit{rank})$. 

We can see this more clearly as follows: If during the chase of $\mc{P}\cup D$ and while applying $\sigma_2$, the value $b$ participates in a join, $P(a,b) \wedge P(b,c)$, in positions $P[2],P[1]$, both with infinite rank, and $b$ does not appear in the result $U(a)$; and then, it does not ``stick".

\begin{wrapfigure}{r}{3.5cm}
\begin{center}
\vspace{-7mm}
\includegraphics[width=3cm]{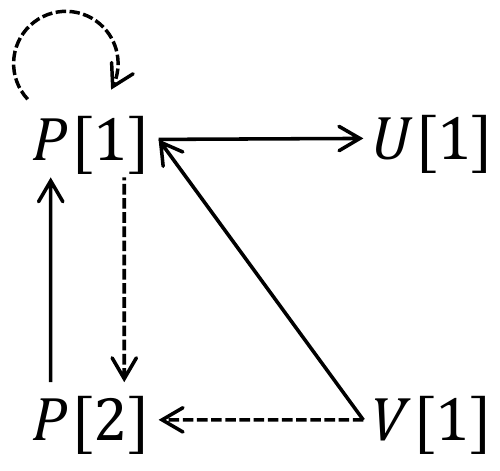}
\vspace{-2mm}
\captionof{figure}{The DG \ of
$\mc{P}$}\label{fig:dg-chase}
\vspace{-8mm}
\end{center}
\end{wrapfigure}

If, despite of this, we run the \qchase{\mc{Q}}{\mc{S}} procedure, it generates the instance $I=D\cup \{P(c,\zeta_1),U(a),U(b)\}$, after applying $(\sigma_1,\theta_1)$, $(\sigma_2,\theta_2)$ and $(\sigma_2,\theta_4)$. Notice that $(\sigma_1,\theta_3)$ is not applicable, because $P(\zeta_1,\zeta_2)$ is $\select^\nit{rank}(\mc{P})$-homomorphic to existing atoms in $I$, e.g. $P(c,\zeta_1)$. In this case, when $\select=\select^\nit{rank}$, the instance $I$ does not give all the answers to $\mc{Q}$:  $\mc{Q}(\mc{P} \cup D)=\{a,b,c\}$, but  $\mc{Q}(I)=\{a,b\}$.


This example shows that the \qchase{\mc{Q}}{\mc{S}} procedure returns an instance that may generate incomplete answers if the program is not in \sch$(\select)$.\boxtheorem

\ignore{\comlb{OLD: I am a bit lost (I think we discussed this, but it was some time ago). You seem to be suggesting that, independently from the program, you can choose any selection function to play with. Is that correct? If yes, is there anything like the ``most appropriate" selection function to use with a given program?}}

As the example above illustrates, given a program $\mc{P}\cup D$ and a query $\mc{Q}$, we can choose any selection function $\mc{S}$ to apply the QA procedure. However, there is no guarantee that the result will be correct.  Actually, as we will see in Section \ref{sec:algCorr}, the \qchase{\mc{Q}}{\mc{S}} can be guaranteed to be correct for QA if only if $\mc{P}\cup D$ is in \sch$(\select)$.
\ Still, there might be more than one (correct) selection function to use.
 For example, if the program belongs to \sch$(\select^\bot)$, then it also belongs to \sch$(\select^\exists)$. Both  \qchase{\mc{Q}}{\mc{S}^\bot} and \qchase{\mc{Q}}{\mc{S}^\exists} can be correctly used for QA with  the program. In Theorem~\ref{th:correctness} we will 
 show they run in PTIME.
 
 \ignore{++
 \comlb{Do these selection functions have to be "correct" for the class of the program at hand? I mean, to run in PTIME? This should be clarified somewhere close to that Thm.: I see the Thm does not have the hypothesis that the selection is function is correct for the program.Maybe it should.}
 
 \commos{The chase procedures for $\mc{S}^\exists$ and $\mc{S}^\bot$ run in PTIME, and they generate polynomially many tuples. Whether they can generate correct answers depends on whether the program belongs to \sch$(\select^\exists)$ or \sch$(\select^\bot)$ as you mentioned in the text.}
 ++}
\ignore{
\comlb{What can we say in general about the complexity (or run-time) of computing 
\qchase{\mc{Q}}{\mc{S}^\bot} and \qchase{\mc{Q}}{\mc{S}^\exists} for a program in both classes?}
\commos{We show they all run in PTIME on Theorem~\ref{th:correctness} in the new subsection.}  }



Having introduced the \qchase{\mc{Q}}{\mc{S}} procedure, we are not in position to formally present the \schqa \ algorithm, shown as  {\bf Algorithm~\ref{ag:qaa}} below. Its main component is the chase procedure for \qa \ over programs in $\sch(\mc{S})$.

\ignore{
\comlb{Next paragraph, first to lines in red below: You should clarify/refresh why the algorithm may have to run several times, and why with possibly different queries. Maybe start with this point? At the beginning I thought these were different instantiations of the same query, but looking at the example below, it seems you want to reuse computations, for other queries. Even more need to clarify the point. I wonder if we really need to introduce this reuse as a part of the algorithm or we should do it simple, and indicate later how we could do the reusing.  }
}

First, the  \schqa \ algorithm returns the error message ``not in the class'' if the input program is not in $\sch(\select)$ (Line~\ref{line:checking}). Otherwise, it runs the \qchase{\mc{Q}}{\mc{S}} procedure to generate the corresponding instance $I$, and uses it to answer the given query $\mc{Q}$, as usual. If $\mc{Q}$ has free variables, the answers to $\mc{Q}$ are those in $\mc{Q}(I)$ that do not contain any nulls. Notice that some tuples in $\mc{Q}(I)$ may contain nulls but they cannot serve as certain query answers. \newline

\ignore{\comlb{The last statement above, in blue,  can be interpreted in two ways: there are no answers with nulls or those that may have them are not considered. It would be better to clarify this. From what I read below, the latter seems to be the case.}
\commos{NEW (2021/7/13): Yes, the latter case is true. Those that have nulls wont be part of the answer because the nulls are not certain vales. I added the last sentence.} }

\ignore{\comlb{If a query has free variables, will  all the answers be obtained from the same $I$, right? If yes, it may be good to say it like this.}
\commos{That is true. I added the blue sentence above.}
}

\begin{algorithm}[H]
\caption{\schqa \ algorithm: parameter is a selection function $\select$}\label{ag:qaa}
\KwIn{A program $\mc{P} \cup D \in $ \sch$(\select)$ and a \cq \ $\mc{Q}$ over $\mc{P} \cup D$.}
\KwOut{$\mc{Q}(\mc{P} \cup D)$.}

\lIf{$\mc{P} \cup D \not\in \sch(\select)$}{ \KwRet{``not in the class''}}\label{line:checking}
\textbf{run the} \qchase{\mc{Q}}{\mc{S}} \textbf{procedure and store the result in} $I$\\

\KwRet{\textbf{\em the tuples in} $\mc{Q}(I)$ \textbf{\em that do not have any nulls.}}\label{line:return}
\end{algorithm}\vspace{3mm}

\subsection{ Correctness of the \schqa \ Algorithm} \label{sec:algCorr}

The correctness of \schqa \ algorithm relies on the correctness of the \qchase{\mc{Q}}{\mc{S}} procedure for answering $\mc{Q}$. This means that, first, it always terminates; and second, the resulting instance can be correctly used for \qa \ (i.e. returning all and only answers). In the following we explain in intuitive terms these properties and why they should hold. They are formally stated and proved in Theorem~\ref{th:correctness}.

The \qchase{\mc{Q}}{\mc{S}} procedure always terminates, and then returning a finite instance $I$, because of the modified applicability condition in Definition~\ref{df:app}, whose  Step~\ref{item-app} allows to create only finitely many atoms during all iterations of the procedure. This is because the applicability condition does not allow \qchase{\mc{Q}}{\mc{S}} to add two $\select(\mc{P} \cup D)$-homomorphic atoms to $I$, and there are only finitely many atoms that satisfy this condition. New applicable rule-assignments may appear after each resumption and allow adding more atoms. However, we show in Theorem~\ref{th:correctness} (Property~\ref{item-finite}) that the number of such atoms is also finite.

\ignore{
\comlb{I do not quite get the part in red. Mainly the ``More". Maybe there is something missing. I do not know what you want to say and what for.}
\commos{NEW (2021/7/13): I think instead of 'More' I should have said 'New' as I updated it in blue. I wanted to emphasise that resumptions can cause new applicable pairs and applying those new pairs will add new atoms to the chase instance. The number of these new pairs and the atoms they will add is limited. I am not sure if here was the right place to say this. This will be discussed later in the proof of the theorem.}
}

\ignore{
\comlb{It is not enough to add finitely many objects at each step if the iteration does not terminate. Should we say that finite termination does not depend on the membership of the program to the associated class?}\\
\commos{It is not only about one iterations of Step 1. My initial sentence was not clear and I updated it. In all iterations of Step 1, the procedure can only add finitely many atoms because of the applicability condition. The condition allows finitely many atoms. I explained it in blue.}
}

When $\mc{P} \in \sch(\select)$, the generated finite instance contains only and all the answers to the query $\mc{Q}$ on $\mc{P} \cup D$. \ The first part of this correctness claim is 
the {\em soundness property}, which tells us  that any tuple in $\mc{Q}(I)$ that does not include nulls is an answer in $\mc{Q}(\mc{P} \cup D)$. This is because $I$ is a subset of the usual chase instance modulo renaming nulls, as we explained in the previous section. Actually, for the soundness property,
the program does not have to belong to the class associated to the selection function at hand.

\ignore{
\comlb{You see in red above? It seems that there may be answers with nulls. Is this fine? If yes, could we say why this can happen? I have lost track a bit: Is this shown in an example somewhere?}
\commos{NEW (2021/7/13): Yes, you are right. There might be nulls but those tuples with nulls can not serve as certain answers. There is no example about this. The fact that nulls cannot be in the answer immediately follow from the certain answer semantics, which we discuss in sec 2.2. Please let me know if we should add an example, and where it should go.} }

The second part of the correctness claim is the {\em completeness property}.  It tells us that any answer in $\mc{Q}(\mc{P} \cup D)$ can be found in $\mc{Q}(I)$, i.e. $\mc{Q}(\mc{P} \cup D) \subseteq \mc{Q}(I)$. This is proved by showing that $I$ contains a large enough portion of the possibly infinite (usual and query independent) chase instance -modulo renaming of nulls- to generate all the answers to $\mc{Q}$.

\begin{theorem} \label{th:correctness} Consider a  \cq \ $\mc{Q}$, possibly with free variables, over a \dplus \ program $\mc{P} \cup D$ with schema $\schema$. For every computable selection function $\mc{S}$ over $\schema$, the \qchase{\mc{Q}}{\mc{S}} procedure has the following properties: 

\begin{itemize}
  \item[\setword{(a)}{item-finite}] The \qchase{\mc{Q}}{\mc{S}} terminates with a finite instance $I$. \ignore{ after XXX applications of Step XXX in Definition XXX, where XXX is XXX.}  
  
 \ignore{ \comlb{Can we say something about the number of steps depending on the query? We had a long and detailed analysis in the previous (main) section, but I guess it did not refer to the modified chase.}
 \\
  \commos{$M_\mc{Q}$ is the only parameter from the query that affects \qchase{\mc{Q}}{\mc{S}}. This is stated in the definition of \qchase{\mc{Q}}{\mc{S}} and is used in the proof of (b) and (c). I am not sure how we can add that to the theorem as a property.} }
  
  \item[\setword{(b)}{item-complete}] Every answer in $\mc{Q}(I)$ that does not contain nulls belongs to $\mc{Q}(\mc{P} \cup D)$.
  \item[\setword{(c)}{item-complete2}] If $\mc{P} \cup D \in \sch(\select)$, then every answer in $\mc{Q}(\mc{P} \cup D)$ is also in $\mc{Q}(I)$.
  \item[\setword{(d)}{item-time}] When $\mc{S}(\mc{P} \cup D)$ can be computed in polynomial time in the size of $D$, the \qchase{\mc{Q}}{\mc{S}} procedure runs in polynomial time in the size of $D$.  
  \boxtheorem \end{itemize}
\end{theorem}

\ignore{
\comlb{Why, what for, you have the second part in 4.? Comment? It looks weird beside the first claim, that has a clear purpose.}\\
\commos{You are right, the second part looked weird. I removed it. It is used to prove the first claim and is in the proof. About $D$, the result is in terms of $D$ because the selection function $\mc{S}$ can depend on data and we can only guarantee tractability if $\mc{S}$ is also tractable. However, $\mc{S}^\bot,\mc{S}^\exists,\mc{S}^\nit{rank}$ are all data independent.}\\
\comlb{To make the proof below more clear it would be more clear to have letters, i.e. (a) ..., instead of 1. ..., saying: Proof of (a): ... I do not want to change without knowing if you need those numbers later. Apart from this, you have huge paragraphs below.}\\
\commos{I updated as you suggested. I also revised the paragraphs to shorter paragraphs.} }

\ignore{
\comlb{You are not uniform in the letters/numbers you use for cases. Sometimes you use letter, sometimes numbers, and in each class, of different kinds. This does not help the reader.}

\comlb{Later in the paper you use ``Step 1", etc. It is not clear if they are steps for  def. 4.8 above (where you also have ``steps") or for the algorithm. I think you should straighten our the use of cases and references.}

\commos{I fixed it as I explained before.}
}


\noindent {\bf Proof of  \ref{item-finite}:} {Step~\ref{item-app} in Definition \ref{def:newCh} can only add finitely many atoms to $I$ during all iterations of the algorithm. Before the first resumption, the applicability condition does not allow adding two $\select(\mc{P} \cup D)$-homomorphic atoms in $I$, which means any pair of atoms in $I$ differ by at least a constant or a null in a position in $\select(\mc{P} \cup D)$. Therefore, an upper bound on the size of $I$ before any resumptions is the number of atoms that can be generated with the finite constants and nulls in the positions of $\select(\mc{P} \cup D)$. 

Each resumption freezes all nulls in $I$, and allows the procedure to add new atoms to $I$. The number of the frozen nulls and the number of the new atoms after each resumption are finite. Since there are $M_\mc{Q}$ resumptions, we can conclude that the total number of atoms in $I$ is finite and the procedure always terminates.

\noindent {\bf Proof of  \ref{item-complete}:} It follows from the fact that $I$ is a subset of $\nit{chase}(\mc{P}\cup D)$ modulo renaming of nulls. To prove this consider any applicable pair of rule-assignment $(\sigma,\theta)$ over $I$ during the \qchase{\mc{Q}}{\mc{S}} procedure. There is a corresponding rule-assignment $(\sigma,\theta')$ in $\nit{chase}(\mc{P}\cup D)$ where $\theta'(\nit{body}(\sigma))$ and $\theta'(\nit{head}(\sigma))$ are respectively equal to $\theta(\nit{body}(\sigma))$ and $\theta(\nit{head}(\sigma))$ modulo renaming of nulls. This is because the only difference between  \qchase{\mc{Q}}{\mc{S}} and $\nit{chase}(\mc{P}\cup D)$ is in their applicability conditions, where \qchase{\mc{Q}}{\mc{S}} imposes a more restricted condition.

\ignore{
\comlb{Not very technical the ``discussed before". Where exactly? And was a proof or hand waiving?}
\commos{I added the text in blue. Do you have suggestions on how I can make it more clear or formal?}
}

\noindent \nit{\bf Proof of ~\ref{item-complete2}:} Consider an answer $a \in \mc{Q}(\mc{P} \cup D)$ with a proof-tree schema $\mc{T}$. We will show that $a$ is also in $\mc{Q}(I)$ by building a proof-tree schema $\mc{T}^I$ for $a$ that is mapped to $I$. 

Let $I_0$ be the instance $I$ before the first resumption, and let $I_i, i>0$ be the instance $I$ after the $i$-th resumption. For any path $\pi$ from a leaf node to the root in $\mc{T}$, we build a similar path $\pi^I$ in $\mc{T}^I$ that is mapped to $I$. We do that by iterating over the nodes $v$ in $\pi$ from its leaf to its root and building $\pi^I$ by adding corresponding nodes $v^I$. We assume $v$ is mapped to an atom $a_v$ in $\nit{chase}(\mc{P}\cup D)$, and consider the following possible scenarios. 

\begin{itemize}
    \item[i.] If $v$ is a leaf in $\pi$, then $a_v \in D$, and $\pi^I$ has the same leaf node, $v^I=v$, which is also mapped to $I_0$, because $D\subseteq I_0$.
    \item[ii.] Considering the first non-leaf node $v$ in $\pi$, there are the following possibilities:
    
\ignore{\comlb{It would be good to remind (refer) the reader to where those variable types were introduced.}
      \commos{NEW (2021/7/13): The blue numbers, e.g., \ref{subitem:one}, are hyperlinks and take the reader to where the types are explained. I also added the footnote to clarify.} }
    
    \begin{enumerate}
        \item If $v$ only has variables of sub-type~(\ref{subitem:one}), i.e. variables that appear in two different paths but only in $\mc{S}$-finite positions, then $\pi^I$ also has the same node $v$.\footnote{\ We introduced the variable types~(\ref{item:I}) and~(\ref{item:II}) in Section~\ref{sec:intuitive}, and the sub-types~(\ref{subitem:one}) and~(\ref{subitem:two}) in Section~\ref{sec:intuitiveS}.} This is because $a_v \in I_0$ as every position of $v$ is in $\mc{S}$, and therefore, there is no possible $\mc{S}$-homomorphic atom to $a_v$.
        \item If $v$ has variables of type~(\ref{item:I}), then it is possible to have an atom $a'_v$ in $I_0$ that is $\mc{S}(\mc{P}\cup D)$-homomorphic to $a_v$, and prevents $a_v$ from appearing in $I_0$ due to the applicability condition. However, in this case, we can map $v$ to $a'_v$, because $v$ does not have any variable that appears in other paths in $\pi$. This might require changing the path from $v$ to the leaf node in $\pi^I$.
        \item The last case occurs when $v$ has some variables of sub-type~(\ref{subitem:two}). In this case, $a_v$ might not be in $I_0$ due to an $\mc{S}(\mc{P}\cup D)$-homomorphic atom $a'_v$ in $I_0$. However, we claim $a_v$ is always in $I_1$. In fact, after the first resumption, the nulls in $a'_v$ are frozen, $a'_v$ is not $\mc{S}(\mc{P}\cup D)$-homomorphic to $a_v$, and $a_v$ is added to $I_1$.
        
        Therefore, we can add the same $v$ to $\pi^I$. This means that, as we build $\pi^I$, we can add a variable of sub-type~(\ref{subitem:two}) to $\pi^I$ if there is one more resumption. This means the number of resumptions must be at least equal to the number of variables of sub-type~(\ref{subitem:two}). As we discussed in Section~\ref{sec:intuitiveS}, the number of variables of sub-type~(\ref{subitem:two}) is limited by the number of variables in the root query atom due to the $\mc{S}$-stickiness property\ignore{ that requires these variables to continue to appear in the atoms in $\pi^I$ including the root atom}. 
        
        A tighter upper bound is $M_{\mc{Q}}$, i.e., the number of $\exists$-variables in the root query atom, because only the $\exists$-variables can be mapped to nulls, and the resumptions are needed only if the variables are mapped to nulls. The reason that only the $\exists$-variables can be mapped to nulls is because the nulls cannot appear in the query answer. Therefore, the number of required resumptions to guarantee all variables of sub-type~(\ref{subitem:two}) are added to $\mc{T}^I$ is $M_\mc{Q}$.
        
        \ignore{\red{We discussed before} that the number of variables of sub-type~(\ref{subitem:two}) is limited by $M_\mc{Q}$, which is due to the $\mc{S}$-stickiness property.}
    \end{enumerate}
\end{itemize}

\ignore{\comlb{Difficult to parse above. The ``because" if for the preceding claim of what follows? Also, where before?} 
\commos{NEW (2021/7/14): I rephrased it. Actually, it was my mistake. This is the only place we discuss that the number of variables of sub-type~(\ref{subitem:two}) is $M_\mc{Q}$. As I mentioned above, this is based on the fact that nulls cannot appear in the query answer, which we argued in the previous comments (that clarification is also useful here).} }

\noindent The discussion around the first non-leaf node can be inductively extended to the other nodes between the leaf node and the root. This means, we can build $\mc{T}^I$ by building a path $\pi^I$ for every path $\pi$ in $\mc{T}$.

\ignore{
\comlb{For the last part below, is there anything from Section \ref{sec:intuitiveS} that can be reused? This long last part of the paper feels somehow repetitive.}
\commos{All these properties, except maybe (c), are intuitively explained before. The point of the theorem is to have a rather formal proof and that is why I repeated the whole discussion and added some more detail.}
}

\noindent \nit{\bf Proof of ~\ref{item-time}:} We will  prove that the size of $I_{M_\mc{Q}}$ is polynomial in $D$. First, we start by showing that the number of atoms in $I_0$ is a polynomial function of $d_0$ and $s$, where $d_0$ is the number of constants in $D$ and $s$ is the number of terms (constants and nulls) in the $\select(\mc{P} \cup D)$-finite positions during the chase of $\mc{P}\cup D$. This holds because the \qchase{\mc{Q}}{\mc{S}} procedure can generate at most $p\times (d_0 + s + 1)^r$ atoms in $I_0$ with $d_0$ constants in $D$, $s$ terms in the $\select(\mc{P} \cup D)$-finite positions, and a placeholder $\star$ that represents the nulls in the non-$\select(\mc{P} \cup D)$-finite positions.\footnote{ \ We explained the use of the placeholder $\star$ in Section~\ref{sec:intuitive} (see Example~\ref{ex:rproof}).} Here, $p$ and $r$ are, respectively, the number of predicates in $\mc{P}$ and the maximum arity of the predicates. 

\ignore{\comlb{Maybe refer to where the placeholder was introduced.}
\commos{NEW (2021/7/14): I added the footnote.} }

To extend the above upper bound to $I_1$, note that the nulls in $I_0$ will be frozen and considered as constants in $I_1$. Since the number of these constants is proportional to the number of atoms in $I_0$ and is polynomial in $d_0$ and $s$, the maximum number of atoms in $I_1$ will be also polynomial in $d_0$ and $s$. This will extend to $I_{M_\mc{Q}}$, and since $M_\mc{Q}$ is independent of $D$ and $\mc{P}$, we can conclude that the number of atoms in $I_{M_\mc{Q}}$ is also polynomial in $d_0$ and $s$, which proves\blue{~\ref{item-time}} The polynomial upper bound only holds if $\select(\mc{P} \cup D)$ can be computed in polynomial time. \ignore{ as stated in the \red{last item}.} \boxtheorem}

From Theorem~\ref{th:correctness}, we conclude that \schqa \ runs in polynomial time for programs in \sch$(\select^\bot)$ \ and its syntactic subclass, Sticky. Due to the fact that there are polynomially many values in finite-rank positions during the chase of a \dplus \ program (cf.~\citep[Theorem 3.9]{fagin}), we can also claim that the algorithm is tractable for programs in \sch$(\select^\nit{rank})$ including those in \ws. We can conclude that \schqa \ runs in polynomial time for programs in $\select^\exists(\mc{P})$ when $\mc{S}=\mc{S}^\exists$ which means \qa \ is tractable for \jws \ programs.

\begin{corollary} \label{cr:ptime} For a selection function $\mc{S} \in \{\select^\bot, \select^\exists, \select^\nit{rank}\}$, \schqa runs in polynomial time w.r.t. the size of $D$.\boxtheorem\end{corollary}

It is implicit in the construction of the query-related chase in Definition  ~\ref{def:newCh} and the proof of Theorem \ref{th:correctness}(d) that we can reuse the same instance $I$ obtained from a run for query $\mc{Q}$ to answer other  queries  $\mc{Q}'$. This is formally established later on, in Corollary \ref{cr:q}. The idea is as follows:  If $M_\mc{Q}$ and $M_{\mc{Q}'}$ are the numbers of $\exists$-variables in $\mc{Q}$ and $\mc{Q}'$, resp. (cf. Definition~\ref{def:newCh}), and $M_{\mc{Q}'} > M_{\mc{Q}}$, the algorithm for $\mc{Q}'$ does not need to run the chase from scratch:  it can resume  the procedure, starting from the already generated instance $I$ for $\mc{Q}$, $(M_{\mc{Q}'}-M_{\mc{Q}})$ times.   If $M_{\mc{Q}'} \le M_{\mc{Q}}$,  $I$ has already a large enough portion of the chase to correctly answer $\mc{Q}'$, and no resumption is needed.

\begin{example} (ex.~\ref{ex:algS} cont.) \label{ex:algorithm} Let us  run the \schqa \ algorithm with program $\mc{P}\cup D$, $\select=\select^\exists$, and the \cq \ $\mc{Q}(x)\!:U(x)$. \  As the program is in \sch$(\select^\exists)$, it passes the test in Line~\ref{line:checking}. If this is  the first query to be answered, it initializes $I=\emptyset$; the \schqa \ algorithm runs the \qchase{\mc{Q}}{\mc{S}} procedure that generates the instance $I=D\cup\{P(c,\zeta_1),U(a),U(b),$ \ $P(\zeta_1,\zeta_2),U(c)\}$. Next, the query is posed to this instance,  returning \ $\mc{Q}(I)=\{a,b,c\}=\mc{Q}(\mc{P} \cup D)$. 

Now, if we want to answer the query  $\mc{Q}':\exists y\;(P(x,y) \wedge U(y))$, we need to run \qchase{\mc{Q}}{\mc{S}} in the algorithm with one resumption since $M_{\mc{Q}'}=1$. However, the algorithm does not need to run the \qchase{\mc{Q}}{\mc{S}} procedure from scratch. It can start from the previous instance $I$, and resume only once to answer $\mc{Q}'$, because $(M_{\mc{Q}'} - M_{\mc{Q}}) = 1-0 =1$. With this additional resumption, we obtain an instance $I=D\cup\{P(c,\zeta_1),U(a),U(b),P(\zeta_1,\zeta_2),U(c),P(\zeta_2,\zeta_3),U(\zeta_1)\}$. The instance can be used to answer any query $\mc{Q}''$ with $M_{\mc{Q}''}=M_{\mc{Q}'}$, e.g. the query $\mc{Q}''(z)\!: \exists y( P(x,z)\wedge V(z))$.

If we do not resume the algorithm sufficiently many times, $I$ may return incomplete answers, e.g. the answer to $\mc{Q}'$ without resumption is $\{a,b\}$ while the complete answer obtained after one resumption is $\{a,b,c\}$.

$M_\mc{Q}$ is an upper bound for the number of necessary resumptions. This means for some query $\mc{Q}$, it might be possible to answer it on $I$ after fewer than $M_\mc{Q}$ resumptions. For example, we can answer $\mc{Q}''(x):\exists y\;P(x,y)$ on $I$ above obtained without any resumptions although $M_{\mc{Q}''}=1$.
\boxtheorem \end{example}

\begin{corollary} \label{cr:q} Consider CQs $\mc{Q}$ and $\mc{Q}'$ over a program $\mc{P} \cup D \in \sch(\mc{S})$. Let $I$ be the result of the \qchase{\mc{Q}}{\mc{S}} procedure. If $M_{\mc{Q}'} \le M_{\mc{Q}}$, i.e. $\mc{Q}'$ has equal or fewer $\exists$-variables than $\mc{Q}$, then $\mc{Q}'(I)=\mc{Q}'(\mc{P} \cup D)$ (the query $\mc{Q}'$ can be answered on the result of the \qchase{\mc{Q}}{\mc{S}} procedure).\boxtheorem\end{corollary}

Corollary~\ref{cr:q} follows from Theorem~\ref{th:correctness}, and implies that we can run the \qchase{\mc{Q}}{\mc{S}} procedure with $n$~resumptions to answer queries with up to $n$ $\exists$-variables. If a query has more than $n$ variables, we can incrementally retake the already-computed instance $I$, adding the required number of resumptions.

\section{Magic-Sets Query Optimization \ignore{for \dplus \ } and JWS Programs}\label{sec:mg}

As we saw in the previous section, the fact that a same instance generated by a partial chase can be used to answer a multitude of queries, even when they do not have the same subschema, is an indication that we are generating more facts than needed to answer a particular query. This situation was investigated long ago in the context of Datalog: computing bottom-up the minimal model of a program to answer a particular can be very and unnecessarily expensive. For this reason, the {\em magic-sets technique} was invented for Datalog programs, to answer queries by following still a bottom-up approach, but restricting the generation of facts according to and as guided by the query at hand \citep{beeri-ms,ceri}.

More specifically, magic-sets (MS) is a general query answering technique based on rewriting logical rules, so that they can be applied in a  bottom-up manner, but avoiding the generation of irrelevant facts. The advantage of doing bottom-up query answering with the rewritten rules resides in the use of the structure of the query and the data values in it, and so optimizing the data generation process. It turns out that magic-sets can be extended to \dplus \ programs~\citep{alviano12-datalog}. This technique,  denoted by \ms, is introduced in the rest of this section. We slightly adapt it to our setting. Furthermore, we show that when the program under optimization is a JWS program, then the optimized program also belongs to this class.

\ms \ takes a \dplus \ program and rewrites it, starting from a given query, into a new \dplus \ program. It departs in two ways from the MS technique for classical Datalog as presented  in~\citep{ceri}, due to the need to: (a) work with $\exists$-variables in \tgd s, and (b) consider predicates that may have both extensional and  intentional data defined by the rules. For (a), we apply the solution proposed in~\citep{alviano12-datalog}. However, we still have to accommodate  (b), which do below.  \ignore{is particularly relevant for \dplus \ programs that allow predicates with both extensional and intentional data, and we address it in \ms.}

\ignore{\comlb{In classic Datalog one assumes a partition of predicates between extensional and intentional, which can be easily achieved introducing new predicates. Is that the issue or there is something deeper?}
\commos{Yes, that is the issue and the change is simple. \citep{alviano12-datalog} is similar to classic Datalog and distinguishes between extensional predicates and intentional predicates.}
}

To present \ms, and so as for classical Datalog, we first introduce {\em adornments}, a convenient way for representing binding information for intentional predicates~\citep{ceri}.

\begin{definition} \label{df:adornement} Let $P$ be a predicate of arity $k$ in a program $\mc{P}$. An adornment for $P$ is a string $\alpha=\alpha_1...\alpha_k$ over the alphabet $\{b,f\}$ (for ``bound" and ``free"). The $i$-th position of $P$ is considered {\em bound} if $\alpha_i=b$, or {\em free} if $\alpha_i=f$. For an atom $A = P(a_1,...,a_k)$ and an adornment $\alpha$ for $P$, the magic atom of $A$ wrt. $\alpha$ is the atom $\nit{mg}\!\_P^\alpha(\bar{t})$, where $\nit{mg}\!\_P^\alpha$ is a predicate not in $\mc{P}$, and $\bar{t}$ contains all the terms in $a_1...a_k$ that correspond to bound positions according to $\alpha$.\boxtheorem\end{definition}

\begin{example}  \label{ex:adornment} If ``\nit{bfb}" is a possible adornment for ternary predicate $S$, then $\nit{mg}\!\_S^{bfb}(x,z)$ is the magic atom of $S(x,y,z)$ wrt. ``\nit{bfb}".\boxtheorem\end{example}

Binding information can be propagated in rule bodies according to a {\em side-way information passing strategy}~\citep{beeri-ms}.

\begin{definition} \label{df:sips} Let $\sigma$ be a \tgd \ and $\alpha$ be an adornment for the predicate of $P$ in $\nit{head}(\sigma)$. A {\em side-way information passing strategy} ({\em \sips}) for $\sigma$ wrt. $\alpha$ is a pair $(\prec^\alpha_\sigma,f^\alpha_\sigma)$, where:

\begin{itemize}
  \item $\prec^\alpha_\sigma$ is a strict partial order over the set of atoms in $\sigma$, such that if $A=\nit{head}(\sigma)$ and $B \in \nit{body}(\sigma)$, then $B \prec^\alpha_\sigma A$.
  \item $f^\alpha_\sigma$ is a function assigning to each atom $A$ in $\sigma$, a subset of the variables in $A$ that are bound after processing $A$. $f^\alpha_\sigma$ must guarantee that if $A=\nit{head}(\sigma)$, then $f^\alpha_\sigma(A)$ contains only and all the variables in $\nit{head}(\sigma)$ that correspond to the bound arguments of $\alpha$.\boxtheorem
\end{itemize}
\end{definition}
The {\em default sips} is obtained from the partial order of the atoms as they appear in rule bodies, from left to right no matter in which direction the arrow points. Despite having a linear order, we only need to compare atoms that share variables. Accordingly, we basically have a partial order. To explain and illustrate \ms, we will use this default \sips. However, our results in Theorem~\ref{th:closed-mg} holds for arbitrary \sips.

Now, we present \ms, illustrating the technique with a running example, namely Example~\ref{ex:mg}.

\begin{example}  \label{ex:mg} Let $\mc{P}$  be a program with $D=\{U(b_1)$, $R(a_1,b_1)$,
$U(b_2)$, $R(a_2,b_2)$,
...,
$U(b_n)$, $R(a_n,b_n)\}$, and the rules

\vspace{-4mm}
\begin{align}
R(x,y),R(y,z) ~&\rightarrow~ P(x,z),\label{frm:m1}\\
U(y),R(x,y) ~&\rightarrow~ \exists z\;R(y,z),\label{frm:m2}
\end{align}

\noindent and consider the \cq \ $\mc{Q}:\exists x\;P(a_1,x)$ posed to  $\mc{P}$. The program is \jws, because every position in the program is in $\finiteExists$. The \edg \ of $\mc{P}$ does not have any cycles, because $B_x\not\subseteq T_z$ and $B_y\not\subseteq T_z$ in Rule~(\ref{frm:m2}). This means that the null values generated by $z$ do not appear in $x$ or $y$ during the chase of $\mc{P}$ (see Section~\ref{sec:ja} for the definitions of  \edg, $B_x,B_y$ and $T_z$). 

We will show below that the program resulting from applying \ms \ on $\mc{P}$ is also \jws.\boxtheorem\end{example}

The \ms \ rewriting technique takes a \dplus \ program $\mc{P}$ with \edb \ $D$ and a \cq \ $\mc{Q}$ of schema $\mc{R}$, and returns a program $\mc{P}_m$ with the same \edb \ $D$ and a \cq \ $\mc{Q}_m$ of schema $\mc{R}_m \subseteq \mc{R}$, such that $\mc{Q}(\mc{P} \cup D)=\mc{Q}_m(\mc{P}_m \cup D)$. It has the following steps:

\paragraph*{1. Generation of adorned rules:} \ms \ starts from $\mc{Q}$ and generates adorned predicates by annotating predicates in $\mc{Q}$ with strings of $b$'s and $f$'s in the positions that contain constants and variables, resp. For every newly generated adorned predicate $P^\alpha$, \ms \ finds every rule $\sigma$ with the head predicate $P$ and it generates an adorned rule $\sigma'$ as follows and adds it to $\mc{P}_m$. According to the predetermined, default {\em \sips}, \ms \ replaces every body atom in $\sigma$ with its adorned atom and the head of $\sigma$ with $P^\alpha$. The adornment of the body atoms is obtained from the \sips and its function $f_\sigma^\alpha$. This possibly generates new adorned predicates for which we repeat this step.

\begin{example}  (ex.~\ref{ex:mg} cont.) \label{ex:mstep1} Starting from the \cq \ $\mc{Q}:\exists x\;P(\blue{a_1},x)$, \ms \ generates the \cq \ $\mc{Q}_m:\exists x\;P^{bf}(\blue{a_1},x)$ and creates the new adorned predicate $P^{bf}$. The adornment $\nit{bf}$ shows that the first position in $P^{bf}(a_1,x)$ is bounded to a constant, namely $a_1$, and the second position is free as $x$ can take any values. \ms \ considers $P^{bf}$ and (\ref{frm:m1}) and generates the rule,

\vspace{-4mm}
\begin{align*}
R^{bf}(x,y),R^{bf}(y,z) ~\rightarrow~ P^{bf}(x,z),
\end{align*}

\noindent and adds it to $\mc{P}_m$. This makes new adorned predicate $R^{bf}$. \ms \ generates the adorned rule,

\vspace{-4mm}
\begin{align*}
U(y),R^{fb}(x,y) ~\rightarrow~ \exists z\;R^{bf}(y,z),
\end{align*}

\noindent and adds it to $\mc{P}_m$. Here, (\ref{frm:m2}) is not adorned wrt. $R^{fb}$, because this bounds the position $R[2]$ that holds the $\exists$-variable $z$. The following are the resulting adorned rules:

\vspace{-3mm}
\begin{align}
R^{bf}(x,y),R^{bf}(y,z) ~&\rightarrow~ P^{bf}(x,z).\label{frm:adrn1}\\
U(y),R^{fb}(x,y) ~&\rightarrow~ \exists z\;R^{bf}(y,z).\label{frm:adrn2}
\end{align}

\noindent In this example, we used the default \sips, which applies the partial order of the atoms in (\ref{frm:m1}) and (\ref{frm:m2}). According to this \sips, \ for $\sigma_1$ in (\ref{frm:m1}) with $\alpha=\nit{bf}$, we have $R(y,z)\prec^\nit{bf}_{\sigma_1}R(x,y)\prec^\nit{bf}_{\sigma_1}P(y,z)$, because  $P(y,z)$ appears in the head and $R(x,y)$ appears before $R(y,z)$ in the body of $\sigma_1$, and  $f^\nit{bf}_{\sigma_1}(P(x,z))=\{x\}$, $f^\nit{bf}_{\sigma_1}(R(x,y))=\{x,y\}$, and $f^\nit{bf}_{\sigma_1}(R(y,z))=\{x,y,z\}$. All this specifies the bound variables, while generating the adorned rule for $\sigma_1$.
\boxtheorem
\end{example}

\ignore{\comlb{I reformulated it a bit. Please, check. The part in red is not too clear. You mean this processing you just performed? What do you want to say?}
\commos{It is fine. "processing" refers to the processing needed for generation of adorned rules based on a given sips, which must follow the sips, i.e. the order and bound/free variables to generate adornments for the next atoms. I updated the blue phrase.}}


\ignore{\comlb{Could you explain the last point in more detail? How exactly the adornments are influenced by the sips? This is the only detailed example we have. It seems, according to your description,  that the sips is used in Steps 1 and 2. Is it in 2.? If yes, it should be highlighted and explained in Example 5.4. The sips is not mentioned. }
\commos{Yes, I added the additional blue sentences to the end of example. To show the point about total order vs partial order we needed a different example as this example is too simple. E.g. for a rule $R(x,y),S(x,z),T(y,w) \rightarrow P(x,w)$ with $\alpha=\nit{bf}$, we only have a partial order in the default sips as $S(x,z)$ and $T(y,w)$ are not comparable. I was not sure if this is too much for here to add to the example.}
}

\paragraph*{2. Adding magic atoms to the adorned rules}  Let $\sigma$ be an adorned rule in $\mc{P}_m$ with the head predicate $P^\alpha$ (which was obtained   using the predetermined {\em sips} in Step 1). \ms \ adds magic atom $\nit{mg}\!\_P^\alpha$ of $\nit{head}(\sigma)$ (cf. Definition~\ref{df:adornement}) to the body of $\sigma$.

\begin{example}  (ex.~\ref{ex:mstep1} cont.) \label{ex:step-2} Adding the magic atoms to the adorned rules \blue{(\ref{frm:adrn1}) and (\ref{frm:adrn2}),} we obtain the following rules:

\vspace{-4mm}
\begin{align}
\nit{mg}\!\_P^{bf}(x),R^{bf}(x,y),R^{bf}(y,z) ~\rightarrow&~ P^{bf}(x,z).\label{frm:msfirst}\\
\nit{mg}\!\_R^{bf}(y),U(y),R^{fb}(x,y) ~\rightarrow&~ \exists z\;R^{bf}(y,z).\label{frm:ext}
\end{align}

\noindent We add the magic atom $\nit{mg}\!\_P^{bf}(x)$ to the body of (\ref{frm:msfirst}) due to the head atom $P^{bf}(x)$. \ignore{According to the adornment in (\ref{frm:msfirst}), we only need to add the adorned rule (\ref{frm:ext}) for \red{the only adorned predicate in the body} of (\ref{frm:msfirst}). Note that there is no adorned rule for $R^{fb}(x,y)$ that appears in the body of  (\ref{frm:ext}), because we cannot bound the  $\exists$-variable in (\ref{frm:m2}), which is the only definition rule for $R$.}\boxtheorem
\end{example}

\ignore{\comlb{In red above. I thought that ``adorned" means having ``b" and ``f"s. In the body all the atoms have them. What do you mean?}
\commos{The sentence was not relevant since adornment is already done in step 1. I removed it. We also already mentioned this point in the prev example.} }

\paragraph*{3. Generation of magic rules} \ignore{Additionally, it generates magic rules as follows.} For every occurrence of an adorned predicate $P^\alpha$ in the body of an adorned rule $\sigma$, \ms \ generates a magic rule $\sigma'$ that defines $\nit{mg}\!\_P^\alpha$ (a magic predicate might have more than one definition).\ignore{\red{We apply a particular and fixed {\em sips} \ignore{that is obtained from the atoms in the same} that corresponds to the  order in which the atoms appear rule bodies.} \ignore{they appear in the body of the rules but} \red{However,  our results are obtained with arbitrary sips.} \ignore{\red{We assume that the atoms in $\sigma'$ are ordered according to the partial order in the {\em \sips} \ of $\sigma$ and $\alpha$}} } If the occurrence of $P^\alpha$ is in atom $A$, and there are the body atoms $A_1,...,A_n$ on the left hand side of $A$ in $\sigma$, in this order (which coincides here with the the order induced by the predetermined {\em sips}), the body of $\sigma'$ contains $A_1,...,A_n$, and the magic atom of the head of $\sigma$. \blue{Notice that the atoms that appear in the body of this new rule are determined by the {\em sips}.}

\ignore{\comlb{Check my modification in blue above. If I understand correctly, the atoms that appear in the body of this new rule are determined by the sips. So, it might not be the same as from left to right.}
\commos{Yes, sips decides what atoms appear in the magic rule. As it is explained in the text, each body atom A should be considered and leads to a magic rule. The atoms in the body of these rules is decided by the order of the atoms in the sips and in particular atom A.}
}

We also create a {\em seed} for the magic predicates, in the form of a fact, obtained from the query. Seed facts correspond to the constants  in the bounded positions in the query, and act as the extensional data for the magic predicates.

\ignore{\comlb{If you see my preceding comment, and it is correct, the role of the sips here would be more than indirect. It would determine which atoms go in the body. Maybe I am wrong, which shows -maybe- the need to clarify this point.}
\commos{Yes, you are not wrong. Sips is directly involved in the new step 3. I hope it is clear now. Step 2 only depends on adorned atoms and step 1 and is indirectly influenced by sips.} }

\ignore{\commos{I think it is the other way around, we use the sips that is based on the default join order. I updated the text, please let me know if I am wrong.}
\comlb{I do not know. Now I notice that I read it possibly wrong the last time. I thought $A$ was the head atom. So, better you decide how to leave it. It would be good to illustrate this issue in the example. Actually, I do not see the sips issue illustrated explicitly in the examples.}
\commos{The sips is fixed in step 1 and in this step we use it to generate the magic rules. I added sentences to Ex~\ref{ex:mstep1} to clarify the use of sips.}
}

\ignore{\comlb{This should be better explained. In particular, why that atom in red below? Why $a_1$? Some other comments follow on this below.}
\commos{I added the blue paragraph above for better explanation, and I explain why $a_1$ is used is explained in Ex~\ref{ex:step-2}.} }

\begin{example}  (ex.~\ref{ex:step-2} cont.) \label{ex:step-3} We generate the following {\em magic rules} that define the magic predicates:

\vspace{-5mm}
\begin{align}
\nit{mg}\!\_P^{bf}(x) ~\rightarrow&~ \nit{mg}\!\_R^{bf}(x).\label{frm:magic1}\\
\nit{mg}\!\_P^{bf}(x),R^{bf}(x,y) ~\rightarrow&~ \nit{mg}\!\_R^{bf}(y).\label{frm:magic2}\\
\nit{mg}\!\_R^{bf}(x),R^{bf}(y,z) ~\rightarrow&~ \nit{mg}\!\_R^{fb}(y).\label{frm:magic3}
\end{align}

We add (\ref{frm:magic1}) for the adorned atom $R^{bf}(x,y)$ in (\ref{frm:adrn1}). The head of (\ref{frm:magic1}) is the magic atom of $R^{bf}(x,y)$, i.e. $\nit{mg}\!\_R^{bf}(x)$), and its body only contains the magic atom of the head of (\ref{frm:adrn1}), i.e. $\nit{mg}\!\_P^{bf}(x)$. There is no other atom in the body of (\ref{frm:magic1}), because, according to the default {\em sips}, there is no atom on the left of $R^{bf}(x,y)$ in the body of (\ref{frm:adrn1}). 

Similarly, we add (\ref{frm:magic2}) for the adorned atom $R^{bf}(y,z)$ in (\ref{frm:adrn1}). (\ref{frm:magic2}) has $R^{bf}(x,y)$ in its body, because $R^{bf}(x,y)$ is on the left of $R^{bf}(y,z)$ in the body of (\ref{frm:adrn1}). we finally generate and add (\ref{frm:magic3}) for the adorned atom $R^{fb}(y,z)$ in (\ref{frm:adrn2}).

It is always the case that magic rules do not have $\exists$-variables. We also add the seed fact $\nit{mg}\!\_P^{bf}(a_1)$ because of $a_1$ that appears in $P^{bf}(a_1,x)$ in the query $\mc{Q}_m$.
\boxtheorem\end{example}



\paragraph*{\blue{4. Adding} rules to load extensional data:} This step applies only if $\mc{P}$ has intentional predicates with extensional data in $D$. The \ms \ algorithm adds rules to load the data from $D$ when such a predicate gets adorned. 



\begin{example} (ex.~\ref{ex:step-3} cont.) \label{ex:step-4}
$R$ is an intentional predicate that is adorned and has extensional data $R(a_1,b_1),R(a_2,b_2),...$ (see \edb \ $D$ in Example~\ref{ex:mg}). \ms \ adds the following rules to load its extensional data for $R^\nit{bf}$, $R^\nit{fb}$, and $P^\nit{bf}$:

\vspace{-4mm}
\begin{align}
\nit{mg}\!\_R^{bf}(x),R(x,y) ~\rightarrow&~ R^{bf}(x,y).\\
\nit{mg}\!\_R^{fb}(y),R(x,y) ~\rightarrow&~ R^{fb}(x,y).\\
\nit{mg}\!\_P^{bf}(x),P(x,y) ~\rightarrow&~ P^{bf}(x,y).\label{frm:mslast}
\end{align}
\boxtheorem\end{example}

Example~\ref{ex:ms-result} below demonstrates that the resulting program $\mc{P}_m$ from \ms, which contains Rules~(\ref{frm:msfirst})-(\ref{frm:mslast}), gives the same answer to $\mc{Q}_m$ as the initial program to $\mc{Q}$, i.e. $\mc{Q}(\mc{P}\cup D)=\mc{Q}_m(\mc{P}_m\cup D)$. The example also shows that  program $\mc{P}_m$ also remains in \jws. Furthermore, the example shows the optimization gain of \ms \ during the data generation process.

\begin{example}\label{ex:ms-result} (ex.~\ref{ex:step-4} cont.) Running the chase procedure on the programs before and after \ms, i.e. on $\mc{P}$ and $\mc{P}_m$, generates the following instances $I$ and $I_m$:

\vspace{-5mm}
\ignore{
\begin{align*}
    I= D\cup \{&R(b_1,\zeta_1),R(b_2,\zeta_2),...,R(b_n,\zeta_n),\\
   &    P(a_1,\zeta_1),P(a_2,\zeta_2),...,P(a_n,\zeta_n)\}\\
    I_m= D\cup \{&\nit{mg}\_P^\nit{bf}(a_1),\nit{mg}\_R^\nit{bf}(a_1), R^\nit{bf}(a_1,b_1),\\
    &\nit{mg}\_R^\nit{fb}(b_1), R^\nit{bf}(b_1,\zeta_1),  P^\nit{bf}(a_1,\zeta_1)\}
\end{align*}
}
\begin{eqnarray*}
    I&=& D\cup \{R(b_1,\zeta_1),R(b_2,\zeta_2),...,R(b_n,\zeta_n),
      P(a_1,\zeta_1),P(a_2,\zeta_2),...,\\&&~~~~~~~~P(a_n,\zeta_n)\}\\
    I_m&=& D\cup \{\nit{mg}\_P^\nit{bf}(a_1),\nit{mg}\_R^\nit{bf}(a_1), R^\nit{bf}(a_1,b_1),
    \nit{mg}\_R^\nit{fb}(b_1), R^\nit{bf}(b_1,\zeta_1),\\ && ~~~~~~~~P^\nit{bf}(a_1,\zeta_1)\}
\end{eqnarray*}
Answering $\mc{Q}:\exists x\;P(a_1,x)$ and $\mc{Q}_m:\exists x\; P^{bf}(a_1,x)$, respectively on $I$ and $I_m$ we obtain the same answer, i.e. true. However, with a large value for $n$, $I_m$ contains much fewer atoms than $I$. \ $I_m$ contains only the  atoms that are relevant for answering $\mc{Q}_m$. Although $I_m$ includes the additional magic atoms, instance $I$ still may contains many more atoms than $I_m$.

Note that $\mc{P}_m$ remains \jws \ because the \edg \ of $\mc{P}_m$ does not have any cycles. The values generated by $z$ in Rule~(\ref{frm:ext}) cannot appear in the variables $x$ and $y$ in the body of the rule to make a cycle.\boxtheorem\end{example}

\ms \ slightly differs from the rewriting algorithm in~\citep{alviano12-datalog} in that we have the additional  Step~4, due to the fact \ignore{In particular, in the latter Step~4 is not needed since, unlike the former, it assumes the} we allow intentional predicates in $\mc{P}$ and adorned predicates in $\mc{P}_m$ \ignore{do not} to have extensional data. The correctness of \ms, i.e. that $\mc{Q}(\mc{P} \cup D)=\mc{Q}_m(\mc{P}_m \cup D)$ holds, follows from both the correctness of the rewriting algorithm in~\citep{alviano12-datalog} and Step~4.

\ignore{\comlb{Last claim. Why is this relevant?}
\commos{Reading it more carefully I think the whole paragraph is irrelevant here. I commented the paragraph.} }

It is worth showing that, when applying \ms \ to a \ws \ program $\mc{P}$, the resulting program,  $\mc{P}_m$, may not necessarily be \ws \ or belong to \sch$(\select^\nit{rank})$.

\begin{example}  \label{ex:notclosed} Consider \bcq \ $\mc{Q}:\exists x\;R(x,a)$ over program $\mc{P}$ with extensional database $D=\{R(a,b),U(b)\}$ and rules:

\vspace{-3mm}
\begin{align}
R(x,y) ~\rightarrow~& \exists z\;R(y,z).\label{eq:mc1}\\
R(x,y) ~\rightarrow~& \exists z\;R(z,x).\label{eq:mc2}\\
R(x,y),R(y,z),U(y) ~\rightarrow~& R(y,x).\label{eq:mc3}
\end{align}

\noindent $\mc{P}$ is \ws \ since the only repeated marked variable, $y$ in (\ref{eq:mc3}), appears in $U[1] \in \finiteRank(\mc{P})$. Note that every body variable is marked. The result of the magic-sets rewriting $\mc{P}_m$ contains the adorned rules:

\vspace{-3mm}
\begin{align}
\nit{mg}\!\_R^\nit{fb}(y),R^\nit{fb}(x,y) &\rightarrow \exists z\;R^\nit{bf}(y,z).\label{eq:m1}\\
\nit{mg}\!\_R^\nit{bf}(x),R^\nit{bf}(x,y) &\rightarrow \exists z\;R^\nit{fb}(z,x).\label{eq:m2}\\
\nit{mg}\!\_R^\nit{bf}(x),R^\nit{bf}(x,y),R^\nit{bf}(y,z),U(y) &\rightarrow R^\nit{fb}(y,x).\label{eq:m3}\\
\nit{mg}\!\_R^\nit{bf}(y),R^\nit{fb}(x,y),R^\nit{bf}(y,z),U(y) &\rightarrow R^\nit{bf}(y,x).\label{eq:m4}
\end{align}

\noindent and the magic rules:

\vspace{-3mm}
\begin{align}
&\nit{mg}\!\_R^\nit{fb}(a).\label{eq:m5}\\
\nit{mg}\!\_R^\nit{bf}(x),R^\nit{bf}(x,y) ~\rightarrow~& \nit{mg}\!\_R^\nit{fb}(y).\label{eq:m6}\\
\nit{mg}\!\_R^\nit{fb}(y),R^\nit{fb}(x,y) ~\rightarrow~& \nit{mg}\!\_R^\nit{bf}(x).\label{eq:m7}
\end{align}

Here, every body variable is marked. Note that according to the description of \ms, the magic predicates $\nit{mg}\!\_R^\nit{fb}$ and $\nit{mg}\!\_R^\nit{bf}$ are equivalent and so we replace them with a single predicates, $\nit{mg}\!\_R$.

$\mc{P}_m$ is {\em not} \ws, since $R^\nit{fb}[1], R^\nit{fb}[2], R^\nit{bf}[1], R^\nit{bf}[2],\nit{mg}\!\_R^\nit{fb}[1],\nit{mg}\!\_R^\nit{bf}[1]$ are not in $\finiteRank(\mc{P}_m)$; and (\ref{eq:m1}), (\ref{eq:m2}), (\ref{eq:m6}) break the syntactic property of \ws \ since in each rule there is a join variable that only appears in these infinite-rank positions. The program is {\em not} in \sch$(\select^\nit{rank})$ either because the chase of $\mc{P}_m$ includes a chase step of (\ref{eq:m6}), which applies the join between $\nit{mg}\!\_R^\nit{bf}(a)$ and $R^\nit{bf}(a,b)$, where the value ``$a$" replaces variable $x$ that appears only in infinite-rank positions $\nit{mg}\!\_R^\nit{bf}[1]$ and $R^\nit{bf}[1]$. The rewriting introduces new join variables between the magic predicates and the adorned predicates, and these variables might be marked and appear only in the infinite-rank positions. That means the joins may break the $\select^\nit{rank}$-stickiness as in this example. This proves that \sch$(\select^\nit{rank})$ and \ws \ are not closed under \ms.

\ms \ does not break $\select$-stickiness for finer selection functions, such as $\select^\exists$. The actual reason why \ms \ might break  $\select^\nit{rank}$-stickiness is due to the fact that $\select^\nit{rank}$ may decide that some finite positions of $\mc{P}_m$ are infinite positions. The positions of the new join variables are always bounded and are finite, and therefore \ms \ does not break $\select$-stickiness if we consider a finer selection function $\select$. For example, $\mc{P}_m$ is \jws \ and in \sch$(\select^\exists)$, because $R^\nit{fb}[2], R^\nit{bf}[1]$ are in $\finiteExists(\mc{P}_m)$, and every repeated marked variable appears at least once in one of these two positions.\boxtheorem\end{example}

We show in Theorem~\ref{th:closed-mg} that the class of \sch$(\select^\exists)$ and its syntactic subclass \jws \ are closed under \ms. This is due to the use by both classes of the $\select^\exists$ selection function, which better specifies finite positions compared to $\select^\nit{rank}$.

\begin{theorem} \label{th:closed-mg}  Let $\mc{P}$ and $\mc{P}_m$ be the input and the result programs of \ms, resp. If $\mc{P}$ is \jws, then $\mc{P}_m$ is \jws. \boxtheorem\end{theorem}

\hproof{To prove $\mc{P}_m$ is in \jws, we show every repeated marked variable in $\mc{P}_m$ appears at least once in a position of $\finiteExists(\mc{P}_m)$. The repeated variables in $\mc{P}_m$ either: (a) are in adorned rules and correspond to the repeated variables in $\mc{P}$,  or (b) appear in magic predicates. For example, $y$ in $\nit{mg}\!\_R(x),R^\nit{bf}(x,y),R^\nit{bf}(y,z) \rightarrow R^\nit{fb}(y,x)$ is of type (a) since it corresponds to $y$ in $R(x,y),R(y,z) \rightarrow R(y,x)$. $x$ is a variable of type (b), because it appears in the magic predicate $\nit{mg}\!\_R$.

The bounded positions in $\mc{P}_m$ are in $\finiteExists(\mc{P}_m)$. That is because an $\exists$-variable never gets bounded during \ms, and if a position in the head is bounded the corresponding variable appears in the body only in the bounded positions. As a result, a bounded position is {\em not} in the target of any $\exists$-variable, so it is in $\finiteExists(\mc{P}_m)$.

The join variables in (a) do not break the $\select^\exists$-stickiness property since they correspond to join variables in $\mc{P}$ and $\mc{P}$ is \jws. This follows two facts: first, a variable in $\mc{P}_m$ that corresponds to a marked variable in $\mc{P}$ is marked, second, variables in $\mc{P}_m$ that correspond to variables in $\finiteExists(\mc{P})$ are in $\finiteExists(\mc{P}_m)$. As a result if a repeated variable is not marked or appears at least once in a $\finiteExists(\mc{P})$, its corresponding variable in $\mc{P}_m$ also has these  properties. The join variables in (b), also satisfy the \jws \ syntactic condition, because they appear in positions of the magic predicates that are in $\finiteExists(\mc{P})$.}

Theorem~\ref{th:closed-mg} ensures that we can correctly apply \ms \ to optimize \schqa \ for the  \jws \ class and its sticky and \ws \ subclasses. With this we have established that \jws \ has the desirable properties formulated at the beginning of Section~\ref{sec:stk}: It extends \ws \ programs, and allows the application of the proposed bottom-up \qa \ algorithm, \schqa. Now, we have obtained the remaining property:   \schqa \ for \qa \ under \jws \ programs can be optimized through magic-sets rewriting.

\section{Conclusions and Future Research}\label{sec:conc}

We have defined a framework for the analysis, classification, and comparison of classes of \dpm programs in relation to their associated {\em selection functions} and the behaviour of the chase with respect to the latter. Selection functions determine some positions in a program's predicate as finite, i.e. that they take finitely many values during the chase. The property that is studied is that of stickiness of values that appear in them and joins in rule bodies.

Selection functions provide a useful abstraction and elegant tool to analyse the behavior of program classes in relation to the chase. New classes could be introduced and investigated following our approach. Selection functions can be quite general. In this work we have considered a range of them, including non-computable ones, which do occur as we have shown, and are both natural and of scientific interest.

Several already studied classes of programs, in their semantic and syntactic versions, fit in this framework, e.g. the classes of sticky and weakly-sticky programs. A new syntactic class, that of {\em Join Weakly-Sticky} programs (\jws), that extends the last two, was identified and investigated in this work. We proposed a practical, polynomial time, bottom-up QA algorithm, \schqa, for these programs. We introduced a magic-set rewriting technique, \ms, to optimize \schqa. The  \jws \ class turns out to be closed under the proposed \ms \ rewriting, which may not hold for sticky or weakly-sticky programs. 

\begin{figure}[h]
\begin{center}
\includegraphics[width=9cm]{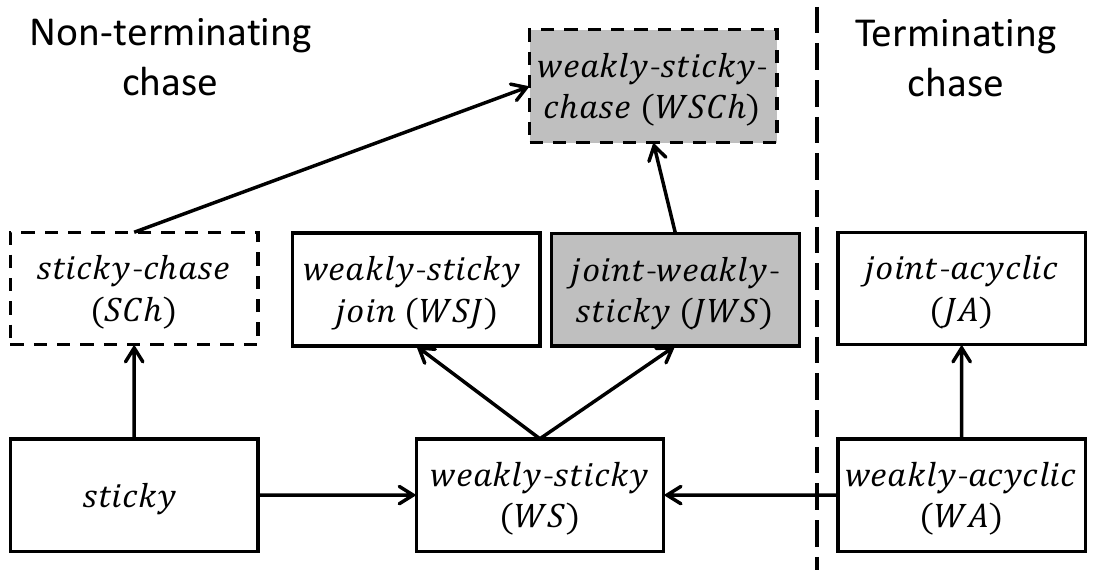}
\end{center}
\vspace{-0.3cm}
\caption{Generalization relationships between program classes. Those depicted with dotted lines are semantic. Those highlighted in grey have been introduced in this work.}
\label{fig:general}
\end{figure}

Figure~\ref{fig:general} shows the introduced class of \jws \ and other discussed program classes in this work, with their inclusion relationships.

Several research directions are part of our ongoing and future work. Among them we find the following:

\begin{enumerate}
  \item The investigation of  the application the magic-set rewriting for \dpm \ in the presence of program constraints, i.e. negative constraints and equality generating dependencies. 
  \item The implementation of \schqa \ and \ms \  and experiments on large real-world data.
  
  We find particularly interesting in this direction trying out an {\em in-database approach}, that is, the implementation of our QA algorithm inside the database, as opposed to having it as an application program running in interaction with the database.
  \item We want to study the application of \qa \ ideas in this paper (semantic generalization of program classes, freezing nulls and chase resumption) for \qa \ over programs in different classes of \dpm \ programs.

      \item The problem of representing and reasoning about Datalog with numerical and set aggregates has received considerable interest~\citep{bertossi18datalog,bellomarini,mohapatra,zaniolo}. We intend to study the \schqa \ algorithm for \dpm \ with aggregation. Numerical aggregations have been recently introduced for {\em Warded Datalog}, a different class of \dpm \ programs~\citep{bellomarini}.

      \item As mentioned in Section \ref{sec:intr}, the motivation for our work had origin in applications of \dpm \ to problems of quality data  extraction \citep{bertossi17}. Now, we would like to investigate the application of the QA algorithm and its optimization in that scenario. 
\end{enumerate}




\vspace{2mm} \noindent {\bf Acknowledgements:} \ This work was supported by NSERC Discovery Grants 2016-06148 and 2021-04120, the NSERC Strategic Network on Business Intelligence (BIN), and by ANID - Millennium Science Initiative Program - Code ICN17002.

\end{document}